\documentclass[twocolumn]{aastex61}
\pdfoutput=1
\usepackage[latin1]{inputenc}
\usepackage{amsmath}
\usepackage{amsfonts}
\usepackage{amssymb}
\usepackage{chngcntr}

\usepackage{graphicx}
\usepackage[varg]{txfonts}
\usepackage{mwe}  
\usepackage{url}
\usepackage{longtable}
\usepackage{float}

\accepted{October 10, 2017} 
\submitjournal{ApJ}

\shorttitle{AGN populations in large volume X-ray surveys}
\shortauthors{Ananna et al.}

\begin{document}
\title{AGN populations in large volume X-ray surveys: Photometric redshifts and population types found in the Stripe~82X survey}
	
\author{Tonima Tasnim Ananna}
\affiliation{Department of Physics, Yale University, PO BOX 201820, New Haven, CT 06520-8120}
\affiliation{Yale Center for Astronomy and Astrophysics, P.O. Box 208121, New Haven, CT 06520, USA 0000-0002-5554-8896}
\email{tonimatasnim.ananna@yale.edu}
	
\author{Mara Salvato}
\affiliation{MPE, Giessenbachstrasse1, 85748, Garching, Germany}
	
\author{Stephanie LaMassa}
\affil{Space Telescope Science Institute, 3700 San Martin Dr, Baltimore, MD 21218}
	
\author{C. Megan Urry}
\affiliation{Department of Physics, Yale University, PO BOX 201820, New Haven, CT 06520-8120}
\affiliation{Yale Center for Astronomy and Astrophysics, P.O. Box 208121, New Haven, CT 06520, USA 0000-0002-5554-8896}
\affiliation{Department of Astronomy, Yale University, P.O. Box 208101, New Haven, CT 06520, USA}

\author{Nico Cappelluti}
\affiliation{Department of Physics, Yale University, PO BOX 201820, New Haven, CT 06520-8120}
\affiliation{Yale Center for Astronomy and Astrophysics, P.O. Box 208121, New Haven, CT 06520, USA 0000-0002-5554-8896}
\author{Carolin Cardamone}
\affiliation{Department of Math \& Science, Wheelock College, Boston, MA 02215, USA}

\author{Francesca Civano}
\affiliation{Harvard-Smithsonian Center for Astrophysics, 60 Garden Street, Cambridge, MA 02138, USA}

\author{Duncan Farrah}
\affiliation{Department of Physics, Virginia Tech, Blacksburg, VA 24061, USA}

\author{Marat Gilfanov}
\affiliation{Max-Planck Institut fuer Astrophysik, Karl-Schwarzschild-Str. 1, Postfach 1317, D-85741 Garching, Germany}
\affiliation{Space Research Institute of Russian Academy of Sciences, Profsoyuznaya 84/32,117997 Moscow, Russia}

\author{Eilat Glikman}
\affiliation{Department of Physics, Middlebury College, Middlebury, VT 05753, USA 0000-0003-0489-3750}

\author{Mark Hamilton}
\affiliation{Yale University}

\author{Allison Kirkpatrick}
\affiliation{Department of Physics, Yale University, PO BOX 201820, New Haven, CT 06520-8120}
\affiliation{Yale Center for Astronomy and Astrophysics, P.O. Box 208121, New Haven, CT 06520, USA 0000-0002-5554-8896}

\author{Giorgio Lanzuisi}
\affiliation{Dipartimento di Fisica e Astronomia, Università di Bologna, Via Gobetti 93/2, 40129 Bologna, Italy}
\affiliation{INAF - Osservatorio Astronomico di Bologna, Via Gobetti 93/3, 40129 Bologna, Italy}

\author{Stefano Marchesi}
\affiliation{Department of Physics \& Astronomy, Clemson University, Clemson, SC 29634, USA}

\author{Andrea Merloni}
\affiliation{MPE, Giessenbachstrasse 1, 85748, Garching, Germany}

\author{Kirpal Nandra}
\affiliation{MPE, Giessenbachstrasse 1, 85748, Garching, Germany}

\author{Priyamvada Natarajan}
\affiliation{Department of Astronomy, Yale University, P.O. Box 208101, New Haven, CT 06520, USA}
\affiliation{Yale Center for Astronomy and Astrophysics, P.O. Box 208121, New Haven, CT 06520, USA 0000-0002-5554-8896}
\affiliation{Department of Astronomy, Yale University, P.O. Box 208101, New Haven, CT 06520, USA}

\author{Gordon T. Richards}
\affiliation{Department of Physics, Drexel University, 32 S. 32nd Street, Philadelphia, PA 19104, USA}
	
\author{John Timlin}
\affiliation{Department of Physics, Drexel University, 32 S. 32nd Street, Philadelphia, PA 19104, USA}

\begin{abstract}
		
Multi-wavelength surveys covering large sky volumes  are necessary to obtain an accurate census of rare objects such as high luminosity and/or high redshift active galactic nuclei (AGN). Stripe~82X is a 31.3 deg$^2$ X-ray survey with {\it Chandra} and {\it XMM}-Newton observations overlapping the legacy Sloan Digital Sky Survey (SDSS) Stripe~82 field, which has a rich investment of multi-wavelength coverage from the ultraviolet to the radio. The wide-area nature of this survey presents new challenges for photometric redshifts for AGN compared to previous work on narrow-deep fields because it probes different populations of objects that need to be identified and represented in the library of templates. Here we present an updated X-ray plus multi-wavelength matched catalog, including {\it Spitzer} counterparts, and estimated photometric redshifts for 5961 (96\% of a total of 6181) X-ray sources, which have a normalized median absolute deviation, {$\sigma_{\rm nmad}$ = 0.06 and an outlier fraction, $\eta$ = 13.7\%}. The populations found in this survey, and the template libraries used for photometric redshifts, provide important guiding principles for upcoming large-area surveys such as {\it eROSITA} and 3{\it XMM} (in X-ray) and the Large Synoptic Survey Telescope (LSST; optical).
		
\end{abstract}
	
 \section{Introduction} \label{sec:intro}
Over the last two decades, X-ray surveys have been a major tool for advancing our understanding of galaxies and black hole growth (e.g., \citealp{brandthasinger2005}). The most massive black holes in the most massive galaxies are particularly interesting because, although they are rare, population synthesis models suggest that they may account for more than half of the total mass enclosed in black holes (e.g., \citealp{Treister2009, Ueda2014, buchner2015}). Massive galaxies and black holes are also likely to have formed early and to have grown rapidly, at or above the Eddington rate \citep{yu2002, hopkins2006, vandokkum2010, du2015}. It is therefore important that any complete census of black hole growth include high luminosity and/or high-redshift AGN---i.e., quasars---which means surveying a large enough volume to find these relatively rare objects.
	
To date, most quasars have been found in wide-area, large-volume optical surveys, such as the Sloan Digital Sky Survey (SDSS), which preferentially selects AGN that are unobscured by circumnuclear or galactic-scale dust. Soft X-ray surveys like {\it ROSAT} (0.1-2.4 keV) cover even larger areas but are also insensitive to obscured AGN. In contrast, infrared and hard X-ray (2-10 keV) observations can recover obscured or reddened AGN missed by optical surveys \citep{brandthasinger2005,Cardamone2008,donley2012,mendez2013,kirkpatrick2015,delmoro2016}. However, while infrared surveys can efficiently select the most luminous obscured AGN \citep{lacy2004,assef2013,kirkpatrick2013,stern2012}, at fainter fluxes, star-forming galaxies are a significant contaminant and a large fraction of AGN can be missed altogether (e.g., \citealp{Cardamone2007,donley2010,mendez2013}). 
	
Hard X-ray surveys efficiently find both unobscured and obscured AGN at a range of luminosities.
The moderate-luminosity AGN found in deep and medium-deep X-ray surveys like the Chandra Deep Fields \citep{alexander2003, xue2011, luo2017, lehmer2005, comastri2011}, COSMOS \citep{hasinger2007, cappelluti2009, Brusa2010, Civano2016, Marchesi2016} and All-Wavelength Extended Groth Strip International Survey (AEGIS; \citealp{nandra2015}) have an intrinsic obscured-to-unobscured ratio of roughly 3:1 \citep{tresiter2004, buchner2015}, and the higher energy {\it Swift}/BAT \citep{swiftbat} X-ray survey shows a similar obscured fraction, $\sim$70\%, in the local universe \citep{ricci2015}, suggesting obscured black hole growth may dominate overall. 
{Theories of black hole growth in quasars also imply an extended phase of obscured accretion \citep{sanders1988, hopkins2006}.}
	
However, this mix of information leaves unclear the question of whether there is a substantial population of obscured AGN at high luminosity and/or redshift, because small volume X-ray surveys do not sample high luminosity quasars.
{Fortunately, large-volume hard X-ray surveys sensitive enough to yield large numbers of quasars, and with sufficient multi-wavelength data to study black hole accretion and star formation rate in the host galaxy, are now becoming available}
%
{(e.g., $Chandra$ Bo\"{o}tes, 10~deg$^2$, \citealp{xbootes, stern2012}; $XMM$-XXL, $\sim50$~deg$^2$, \citealp{pierre2016, menzel2016, Georgakakis2017}; $XMM$-LSS, 11.1~deg$^2$, \citealp{xmm-lss, xmmlssmultiwavelength2013}; Stripe~82X, 31.3~deg$^2$, $XMM$+$Chandra$, \citealp{steph2013a,steph2013b,steph2016}).}
	%
Stripe~82X is among the largest of these hard X-ray surveys,
with the most extensive multi-wavelength data 
(millimeter, ACT, \citealp{act}; radio, FIRST, VLA; \citealp{first, hodge2011}; ultraviolet, $GALEX$, \citealp{galex};  optical, SDSS, \citealp{annis2014, jiang2014, fliri2016}; 
near-infrared, VHS, UKIDSS, \citealp{vhs,lawrence2007}; mid-infrared, \textit{WISE, Spitzer}, \citealp{stern2012, spies, sheladata};  far-infrared, \textit{Herschel}, \citealp{hers}). {It also has extensive spectroscopy 
\citep{sdssIspectroscopy2002, sdssIII2012, dr13_2016, sdssIII2012, ahn2014, 2slaqspectra2009, deep2-2013, coil2011, garilli2008, jones2004, jones2009, drinkwater2010}, including our own work with WIYN, Palomar and Keck (\citealp{steph2016}, hereafter LM16).} 
Nearly one third of Stripe~82X sources have spectroscopic redshifts used in this work, and the SDSS-IV eBOSS program \citep{sdssdr14} will raise the spectroscopic completeness to 43\%(LaMassa et al., submitted.).
	
\begin{figure*}[th]
\centering
\includegraphics[height=7cm, width=16cm]{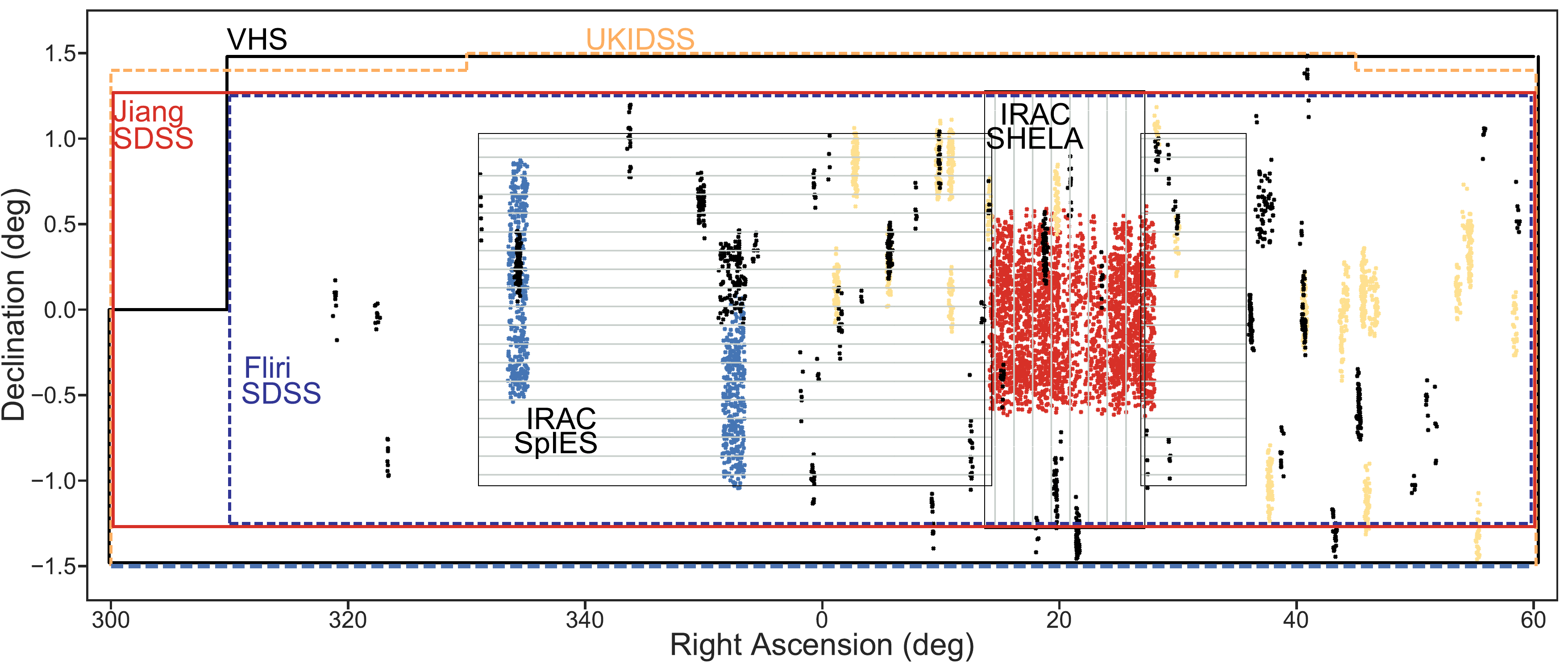}
\caption{Map of the Stripe~82X survey, along with footprint of multi-wavelength coverage. Note that the vertical extent of the plot is $\sim2.5$~deg and the horizontal extent is 120~deg. {The dots represents X-ray sources}. \textit{Red dots}: AO13; \textit{blue dots}: AO10; \textit{yellow dots}: archival {\it XMM}-Newton sources; \textit{black dots}: $Chandra$ sources. The \textit{solid red line} and the \textit{dark blue dashed line} encompass areas covered by two SDSS co-added optical catalogs, \citet{jiang2014} and \citet{fliri2016}, respectively. The \textit{solid black line} delineates the area covered by the near-infrared (NIR) Vista Hemispherical Survey (VHS) \citep{vhs} and the \textit{yellow dashed line} is the area covered by the United Kingdom Infrared Telescope (UKIRT) Infrared Deep Sky Survey \citep[UKIDSS, NIR]{lawrence2007}. 
The patches with horizontal lines and vertical lines show areas covered by the mid-infrared \textit{Spitzer} Infrared Array Camera Equatorial Survey \citep[SpIES]{spies} and \textit{Spitzer}-Hetdex Exploratory Large Area survey \citep[SHELA]{sheladata} respectively. 
Not explicitly shown in this map are the Wide-field Infrared Survey (WISE) and Galaxy Evolution Explorer ({\it GALEX}), which cover the whole sky. 
As explained later, we do not use {\it GALEX} or UKIDSS data to identify multi-wavelenth counterparts of X-ray sources, although we do add data from these surveys to construct spectral energy distributions (SED).} 
\label{fig:vhs_map} 
\end{figure*} 
	
Previously, accurate photometric redshifts have been obtained for X-ray samples from small, deep surveys like {\it XMM}-COSMOS (2.13 deg$^2$; \citealp{MS09}, hereafter S09); {\it Chandra} Deep Field South (CDFS, 0.13 deg$^2$; \citealp{luo2010,hsu2014}); the Extended {\it Chandra} Deep Field-South (ECDFS, 0.3 deg$^2$; \citealp{Cardamone2010,hsu2014}); {\it Chandra}-COSMOS (0.90 deg$^2$; \citealp{MS11}, hereafter S11); 
{\it Chandra} COSMOS-Legacy (2.2 deg$^2$; \citealp{Civano2016, Marchesi2016}); 
the Lockman Hole (0.20 deg$^2$; \citealp{fotopoulou2012}); and AEGIS (0.67 deg$^2$; \citealp{nandra2015}). But the population probed in these narrow-deep surveys were different from the X-ray bright sample of wider and shallower surveys such as Stripe~82X.

In this paper, we identified a list of templates appropriate for wider and shallower X-ray survey samples with higher luminosity AGN, with which we obtain photometric redshifts with accuracy comparable to some of the deeper surveys. 
Not only does this make the Stripe~82X sample valuable, but it also provides an initial calibration for surveys of comparable depth, such as Bo\"{o}tes, $XMM$-XXL and the upcoming {\it eROSITA} All-sky survey \citep{merloni2012}. This work will also be relevant in calculating photometric redshifts for AGN in large-scale non X-ray surveys such as Large Synoptic Survey Telescop (LSST; \citealp{ivezic2008}).
	
The paper is organized as follows: In Section~\ref{sec:assoc}, we describe the Stripe~82X data; we also explain the multi-wavelength catalog matching, with details about how ambiguous associations were resolved. The procedures used to calibrate the photometric redshifts, which closely follow S09 and S11, are described in \S~\ref{sed_fitting}.
In \S~\ref{sec:results_discussion}, we present the final photometric redshifts along with preliminary characterization of the sources, and some conclusions about calculating photometric redshifts for a large volume X-ray survey using broad-band data.
In \S~\ref{sec:conclusions}, we summarize the key results of this work. 
Appendix A describes the final Stripe~82X multi-wavelength catalog and redshifts. Appendix B describes our process for selecting the set of templates used in fitting for photometric redshifts with an example and Appendix C details how we calculated X-ray-to-optical flux ratios. 
	
\section{{Identifying Multi-Wavelength Counterparts of Stripe~82 X-Ray Sources}}\label{sec:assoc}

\subsection{Characteristics of the Stripe~82X Population}\label{sec:data_available}
	
Figure~\ref{fig:vhs_map} shows the regions covered by the Stripe~82X survey. There are two regions with contiguous {\it XMM}-Newton coverage awarded to our team in \textit{XMM}-Newton cycles 10 and 13 (AO10 and AO13), and the rest are from archival {\it XMM}-Newton and {\it Chandra} pointings, as described in \citet{steph2013a, steph2013b, steph2016}. The AO10 and AO13 surveys have slightly different exposure times, 6-8 ks for AO13 and 4-6 ks for AO10. The non-uniform exposure times occur due to mosaicking. Key ancillary data sets are indicated in the figure (see caption for details).
	
\begin{figure}[th]
\centering
\includegraphics[width=1.0\linewidth]{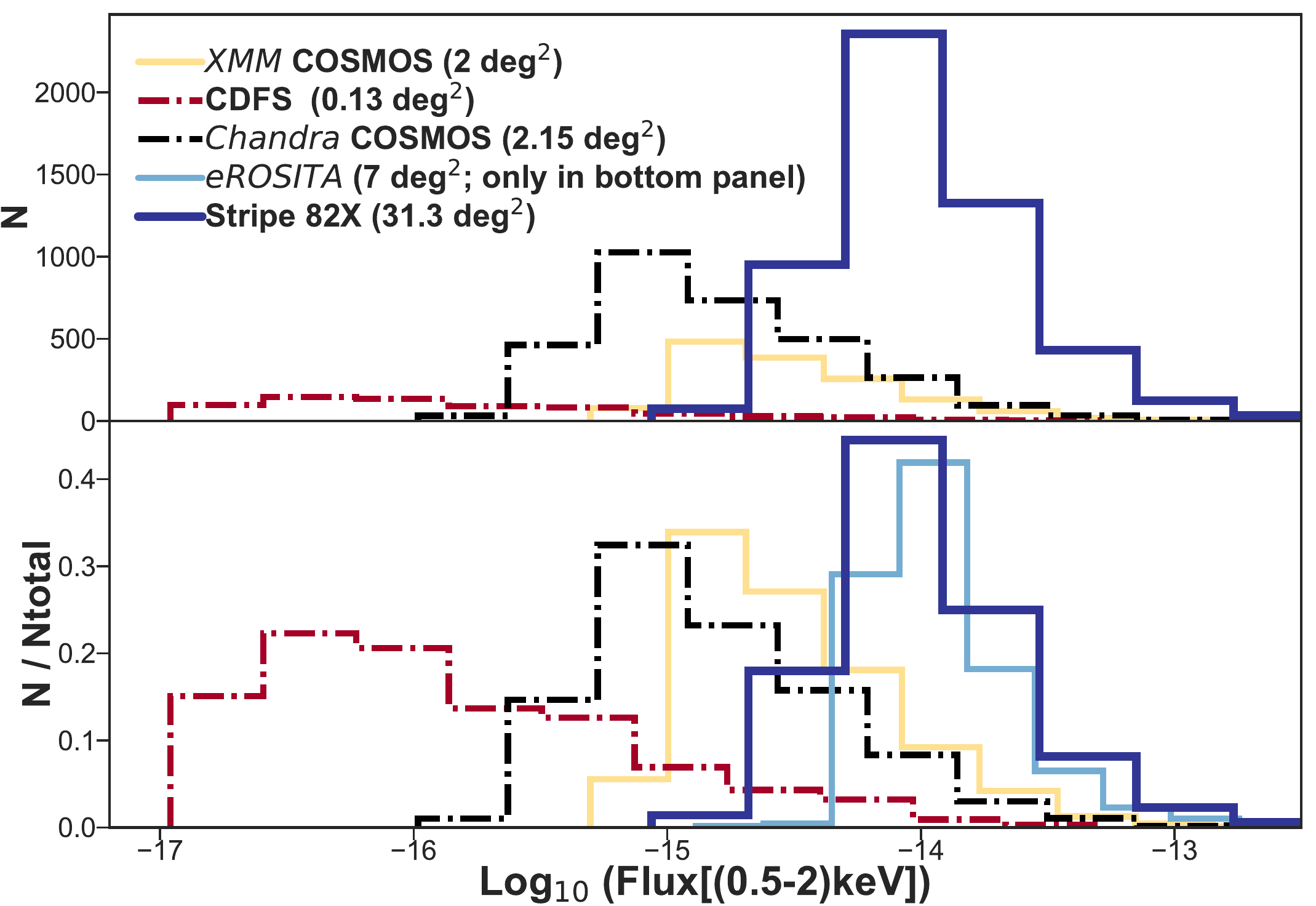}
\caption{Soft X-ray flux distributions for point-like sources in several X-ray surveys. \textit{Top panel:} absolute number of sources in each survey. 
\textit{Bottom panel:} Each survey is normalized to the total number of sources in that survey.
The bright sources that dominate Stripe~82X are barely present in the deep pencil-beam surveys. This is why the library of spectral energy distribution (SED) templates for computing photometric redshifts must be optimized. The \textit{eROSITA} histogram in the bottom panel was estimated using contiguous 7 deg$^2$ $XMM$-XXL data \citep{menzel2016,liu2016,Georgakakis2017}, cut at the fluxes expected for \textit{eROSITA}.} 
\label{fig:all_xray_histo}
\end{figure}

The brighter flux limit of Stripe~82X presents several challenges. First, the flux distribution of AGN population sampled by Stripe~82X is not well represented in pencil-beam surveys for which photometric redshifts have been computed in previous works. Figure~\ref{fig:all_xray_histo} shows histograms of soft X-ray flux (0.5-2 keV) for deeper, smaller-volume surveys compared to Stripe~82X. The bright AGN typically found in Stripe~82X are rare in COSMOS and practically absent in the CDFS. 

{Conversely, the bottom panel of Figure~\ref{fig:all_xray_histo} shows that \textit{eROSITA} All-sky survey is expected to have a flux limit similar to our {\it XMM}-Newton AO10 and AO13 surveys. Photometric redshifts for other large-volume X-ray surveys such as the 3$XMM$ Serendipitous Catalog (880 deg$^2$, \citealp{xmm_ra_dec_err}), and the {\it Chandra} Source Catalog are not available yet, but the object types will likely be similar to Stripe~82X.} 
	
{We resolve the first challenge by selecting a reduced set of templates that represents the population of Stripe~82X, as described in \S~\ref{sec:library_selection} and Appendix B. These templates should also be applicable to the other wide shallow fields such as 3$XMM$ and \textit{eROSITA}. In fact, the template library we compiled for this work has already been used successfully in \citet{Georgakakis2017} to calculate photometric redshifts for {\it XMM}-XXL sources (50 deg$^2$, \citealp{menzel2016}) with X-ray fluxes (soft band) F$_X < 10^{-13}$ erg s$^{-1}$ cm$^{-2}$.}

The second challenge is that Stripe~82X contains both 
archival observations of varying depths --- 6~deg$^2$ from {\it Chandra} and 7.4~deg$^2$ from {\it XMM}-Newton \citep{steph2013a, steph2013b, steph2016} --- and relatively uniform {\it XMM}-Newton imaging awarded to our team in AO10 and AO13 (PI: Urry), \citet{steph2013b}, LM16.
After removing periods of high background, the latter allocations resulted in 4.6~deg$^2$ and 15.6~deg$^2$ of independent area, respectively.
We dealt with the varying depths by calibrating each sub-sample separately for each field, i.e., optimizing the spectral energy distribution (SED) libraries independently, but in any case we converged on the same set of templates for all four fields (\S~\ref{sec:library_selection}).
	
The third challenge is that for areas covering more than a few square degrees with a variety of { observations taken by independent teams with different instruments}, multi-wavelength { catalogs are assembled  by  simple matches in position. Instead, for smaller fields, such as COSMOS (S09, S11, \citealp{hsu2014}), CANDELS GOODS-S \citep{candelsmultiwavelength} and ECDF-S MUSYC \citep{Cardamone2010}, the multi-wavelength catalogs are obtained in a homogeneous manner including the registration of the images at the same reference and taking into account the individual point spread function (PSF)}. For Stripe~82X, {we had to reach a compromise while dealing} with catalogs that have different types of fluxes or magnitudes (e.g., aperture corrected, Petrosian) and very different spatial resolution, which is discussed further in \S~\ref{sec:systematicshifts}.

\subsection{{Generating a New Stripe~82X Multi-Wavelength Catalog}}
\label{ssec:multi}
 
A common challenge for all the X-ray surveys  is finding the
correct {\it and} consistent multi-wavelength counterpart of each X-ray source. 
LM16 presented a combined multi-wavelength catalog for all Stripe~82 X-ray fields, with identifications done pair-wise using a statistical Maximum Likelihood Estimator (MLE - described in detail in \S~\ref{sec:running_mle}) algorithm \citep{SutherlandandSaunders1992}.
That is, each ancillary dataset ({\it GALEX}, SDSS, VHS, UKIDSS,  {\it AllWISE}) was matched separately to the master X-ray catalog. Once that the counterpart was found, a match in coordinates was performed for retrieving radio and infrared information.

\begin{figure*}[th]
\centering
\includegraphics[width=0.31\linewidth]{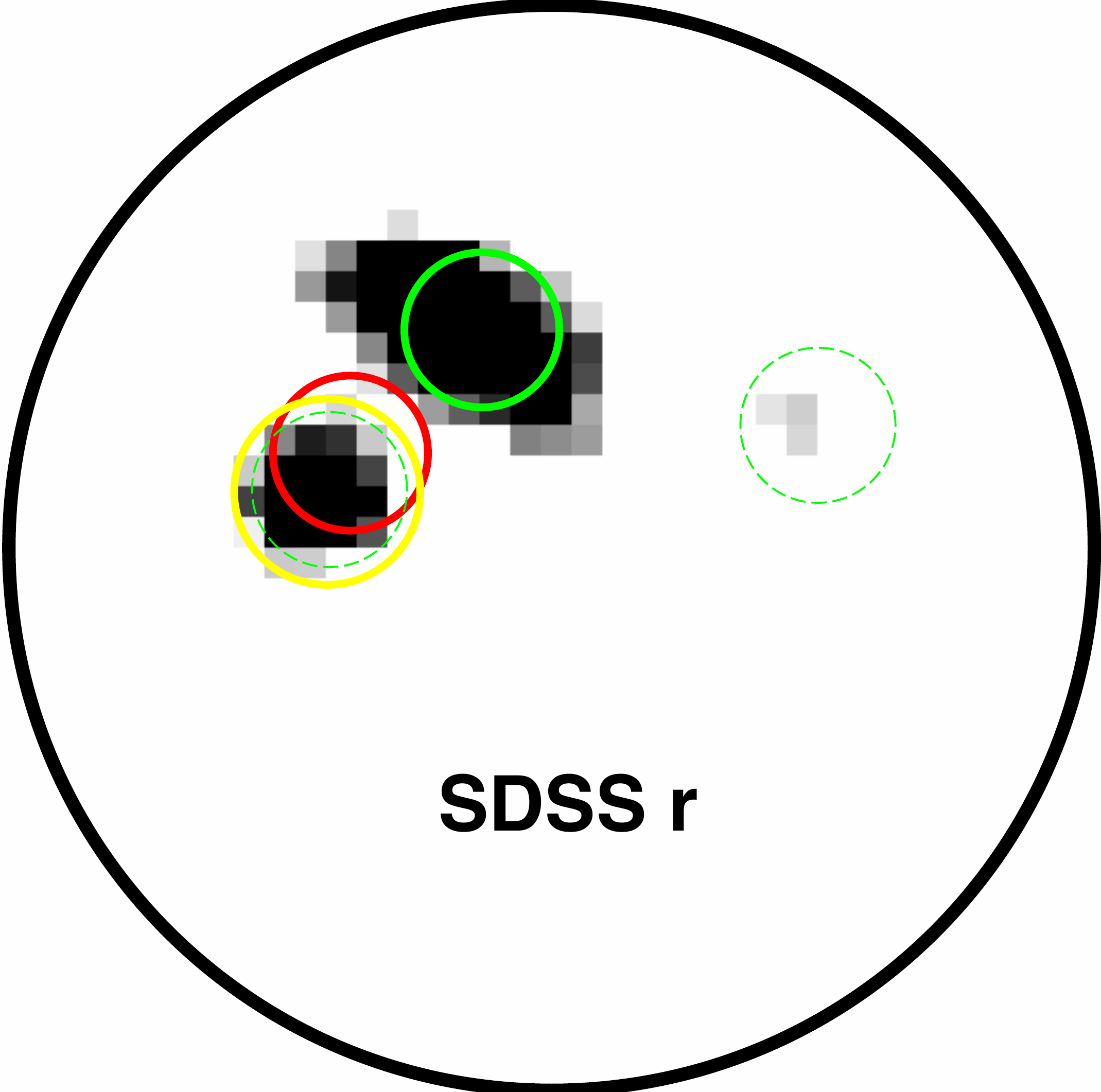}~
\includegraphics[width=0.3\linewidth]{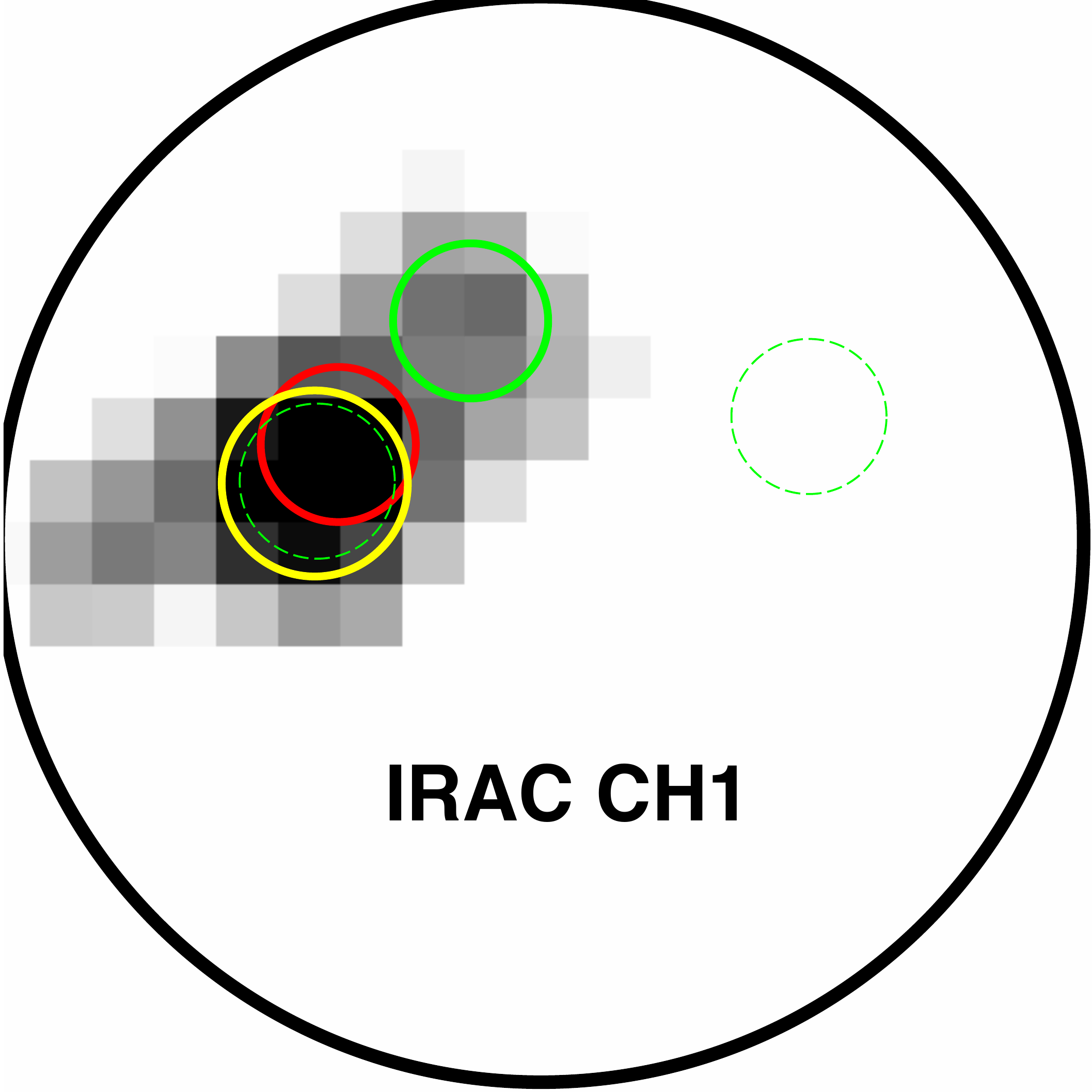}~ 
\includegraphics[width=0.29\linewidth]{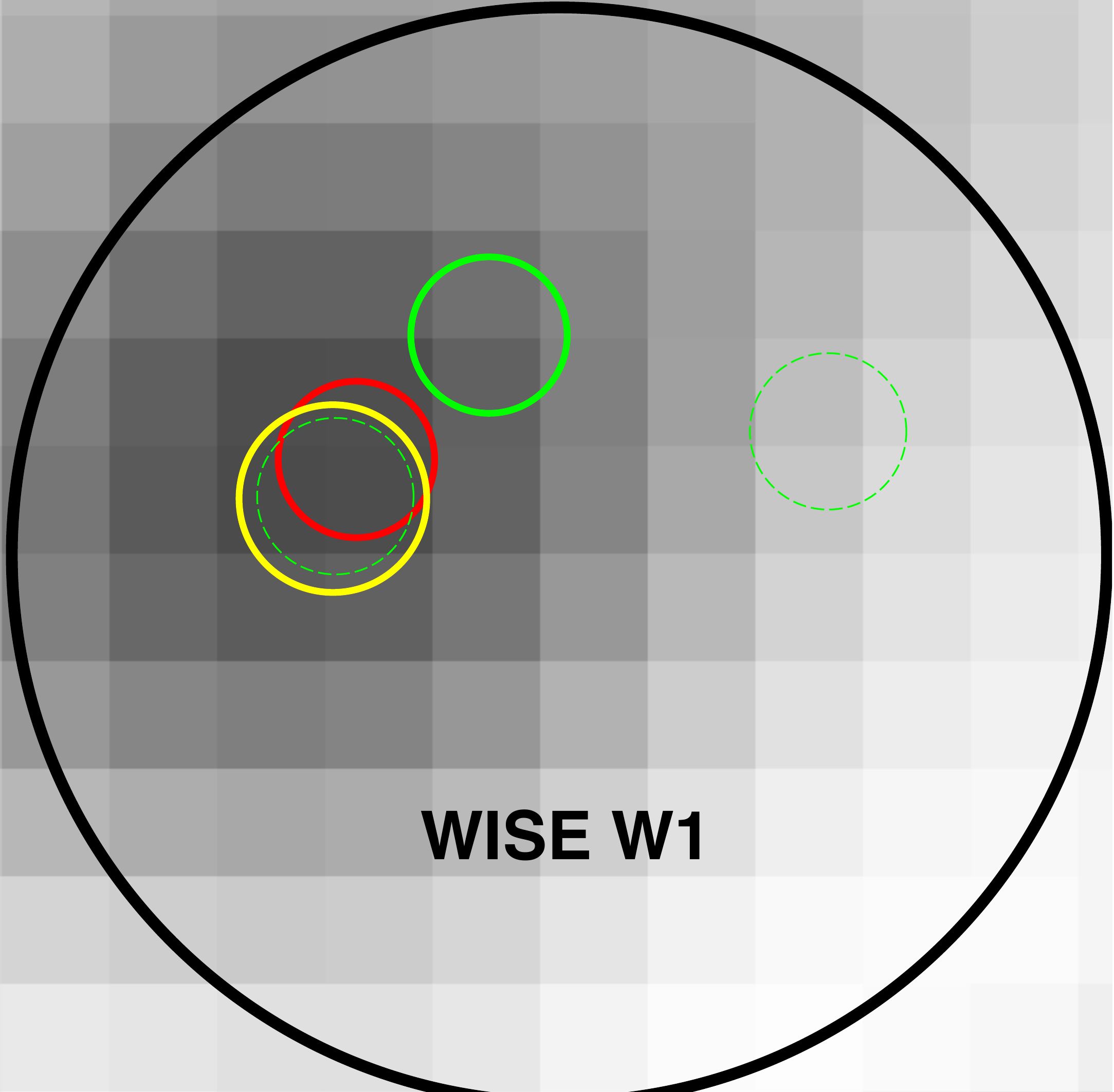}
\caption{From left to right: The field of Stripe~82X $XMM$-Newton source ID 2794 from LM16 ({\it solid black circle} with radius 7$^{\prime\prime}$) at optical SDSS $r$ band, near-infrared IRAC Ch1 (3.6~$\mu$m) and {\it AllWISE} W1 (3.4~$\mu$m). The positional errors for SDSS, VHS and IRAC are only a few milli-arcseconds; for clarity, we use bigger circles to indicate the location of source in each band. {\it Dashed green circle} with radius 1$^{\prime\prime}$ around SDSS $r$ band, {\it solid yellow circle} with radius 1.2$^{\prime\prime}$ for VHS $K$ band and {\it solid red circle} with radius 1$^{\prime\prime}$ for IRAC CH1 {source positions}. 
In this example, the most likely optical counterpart from the MLE analysis ({\it left image}) is the bright source at the top ({\it solid green circle}) but the infrared images suggest a more reliable counterpart below and to the left. The dashed green circles indicate other nearby optical sources that have a lower reliability match. {We identify the source circled in yellow as the correct counterpart, which is the most likely counterpart according to VHS and IRAC, but more accurately pin-pointed by VHS.}}
\label{fig:AO13_2794} 
\end{figure*}

In LM16, running MLE on each band returned the same counterpart across multi-wavelength catalogs for approximately  4943 (80\%) cases. 
If we only consider catalogs/wavebands that this work and LM16 have in common --- $GALEX$, SDSS, VHS, UKIDSS and {\it AllWISE} --- 487 (8\%)
X-ray sources in LM16 have conflicting associations (i.e. distinctly different sources are chosen as counterparts in different bands, see Fig.~\ref{fig:AO13_2794} for an example). 
In such cases, LM16 left it to the reader to decide which ancillary band provides the correct X-ray counterpart.
We note that LM16 found no associations for 12\% of the sample because the potential counterparts were below an empirically determined reliability threshold (10\%), or because no candidate counterparts  were found within the search radius around the X-ray source (2\%).
	
In this paper we update the multi-wavelength catalog, based in part on deeper and more homogeneous ancillary data than were previously available. 
Specifically, using new SDSS co-added catalogs from \citeauthor{fliri2016} (\citeyear{fliri2016}; hereafter FT16) and MIR data from {\it Spitzer} \citep{spies, sheladata}, we repeated the 
MLE matching (for a detailed description of the process, see \S~\ref{sec:running_mle}).
Then, in cases of conflicting counterparts at different wavelengths, we established a procedure for identifying the same, correct counterpart at each wavelength (see \S~\ref{ssec:correct_assoc}), which is necessary if we are to use the SED to calculate photometric redshifts. 
	 
The optical data are from SDSS, which repeatedly imaged 300-deg$^2$ area in $u$, $g$, $r$, $i$ and $z$ broad-band filters, with 70-90 exposures at each location. The SDSS single-epoch images were combined to create deeper co-added catalogs (FT16; \citealp{jiang2014}, hereafter J14) that reach $\sim 2.5$~mag deeper than the single-epoch data (or the depth of the larger SDSS survey) and increases the likelihood of finding an optical association for an X-ray source. The FT16 catalog is 0.2 magnitudes deeper in SDSS $i$-band than the earlier J14 co-added catalog because it eliminates data of poor quality. We use FT16 when available, and J14 when the X-ray source is outside the footprint of the FT16 catalog.
	
We use near-infrared (NIR) data from the Vista Hemispherical Survey (VHS; \citealp{vhs}), which primarily covers the southern hemisphere but includes Stripe~82 (the equatorial coverage goes up to a declination of +1.5 degrees). VHS has a 5$\sigma$ $K_{AB}$ depth of 20.3~mag and provides broad-band data in $J$, $H$ and $K$ filters. We also use United Kingdom Infrared Telescope (UKIRT) Infrared Deep Sky Survey (UKIDSS; \citealp{lawrence2007}) NIR data when constructing SEDs, but not in the phase of counterpart identification (explained in \S~\ref{ssec:sed_construction}).

MIR data comes from the Wide-field Infrared Survey {\it AllWISE} catalog and the \textit{Spitzer} Space Telescope Infrared Array Camera (IRAC). These two instruments have slightly different broad-band filters that cover approximately the same wavelengths in the first two channels. {\it AllWISE} has a resolution of 1.875$^{\prime\prime}$/pixel and has 95\% completeness at a depth of 19.8~mag (AB) at 3.4~$\mu$m\footnote{\url{http://wise2.ipac.caltech.edu/docs/release/allwise/expsup/sec2_4a.html}}. This means nearby objects are often blended together, and the positional accuracy is lower than for VHS and SDSS. IRAC has much better spatial resolution than {\it AllWISE} (0.8$^{\prime\prime}$/pixel in CH1, at 3.6~$\mu$m). 

The \textit{Spitzer} IRAC Equatorial Survey (SpIES) catalog \citep{spies} has a 
95\% completeness at $\sim$ 20.34~mag (AB) in CH1. In the same band, {\it Spitzer}-HETDEX Exploratory Large Area Survey (SHELA) catalog \citep{sheladata} has a 
95\% completeness at 20.0~mag (AB).
Because of their depth and resolution, for determining the counterparts, we use IRAC data rather than {\it AllWISE} when possible.

\subsection{{Running MLE on Optical and IR data}}
\label{sec:running_mle}
	
\begin{deluxetable*}{llllll}[th]
\tablewidth{0pt}
\tablecaption{\label{tab:exposure_time} \textsc{Number of X-ray Sources with Counterparts Identified in Ancillary Catalogs.}}
\tablehead{ \colhead{} & \colhead{\textsc{AO13}} & \colhead{\textsc{AO10}} & \colhead{\textsc{Archival $XMM$}} &
\colhead{\textsc{Archival $Chandra$}} & \colhead{\textsc{Total}}}
\startdata
Area (deg$^2$) & 15.6 & 4.6 & 7.4 & 6 & 31.3\tablenotemark{1} \\ 
X-ray Exposure time range (ks) & 6-8 & 4-8 & 17-65 & 5.7-71.2 & \\ 
Number of X-ray sources\tablenotemark{2} & 2862 & 751 & 1607 & 1146 & 6366 \\
Unique sources & 2820 & 721 & 1496 & 1144 & 6181 \\
Total number of multi-wavelength associations\tablenotemark{3} & 2812 (99\%) & 716 (99\%) & 1453 (97\%) & 1063 (93\%) & 6044 (97.8\%) \\
Number of associations common with LM16\tablenotemark{4} & 2555 (91\%) & 633 (88\%) & 1230 (82\%) & 911 (80\%) & 5329 (86\%) \\
Changed Association\tablenotemark{5} & 23 (0.8\%) & 5 (0.7\%) & 7 (0.5\%) & 5 (0.4\%) & 40 (0.6\%)\\
New associations\tablenotemark{6} & 172 (6\%) & 57 (8\%) & 165 (11\%) & 113 (10\%) & 507 (8\%) \\
Sub-threshold new associations & 54 (2\%) & 17 (2\%) & 48 (3\%) & 32 (3\%) & 151 (2\%) \\
Number with spectra & 820 (29\%) & 270 (37\%) & 435 (29\%) & 361 (32\%) & 1886 (30\%) \\
\enddata
\tablenotetext{1}{Non-overlapping area.}
\tablenotetext{2}{The number of sources detected in different X-ray bands at a 4.5$\sigma$ level are given in Table~3 of LM16.}
\tablenotetext{3}{Among unique sources.}
\tablenotetext{4}{LM16 reports the best match of the X-ray sources to each multi-wavelength band, but for 8\% of those counterparts, not all bands agree on a single counterpart (e.g., the best NIR counterpart is different from the optical or ultraviolet counterpart). We resolve these ambiguities (\S~\ref{ssec:correct_assoc}), so here we report the number of our counterparts that lie within 1$^{\prime\prime}$ of an LM16 counterpart in \textit{any} band. We are only considering SDSS, VHS and {\it AllWISE} associations from LM16 for this comparison, as these are the catalogs in which we look for associations.}
\tablenotetext{5}{Compared to LM16.}
\tablenotetext{6}{Not present in LM16. In this row, we ignore the sub-threshold associations in this catalog, as LM16 ignores sub-threshold associations.}
\end{deluxetable*}

The Maximum Likelihood Estimator (MLE) has been used to identify the correct association between sources from different catalogs, in particular for counterparts to point-like X-ray sources (e.g. \citealp{Brusa2007, Rovilos2011, xue2011, GeorgakakisNandra2011, Civano2012, steph2013b, nandra2015, steph2016, Marchesi2016}). In this work, we highlight the major steps of the matching procedure while the reader can refer to the more detailed description in \citet{Naylor2013}.
	
We performed MLE matching for six different optical, NIR and MIR catalogs; the total number of matches for each catalog are given in Table~\ref{tab:exposure_time}. The MLE method considers three factors of the ancillary waveband to determine correct counterparts to the X-ray sources: (i) the area-normalized density of background sources in the field, as a function of magnitude (or flux), (ii) the area-normalized density of sources in the vicinity of the X-ray source, per unit  magnitude, after subtracting the background distribution, and (iii) the positional errors associated with each catalog (astrometric offsets between catalogs must be corrected in advance).
	
The most critical point is the estimate of the magnitude distribution of the background sources, so that we can distinguish between background sources and X-ray counterparts. For example, an over-subtraction of sources in point (ii) will reduce the likelihood that a faint source is identified as the right counterpart to an X-ray source, with the effect being stronger for shallow X-ray surveys like Stripe~82X. 
Also, many of the Stripe~82 counterparts will be stars, bright nearby objects, or quasars, which will outshine some fainter sources, reducing the effective depth (i.e., this part of the sky will appear shallower than other, random locations in the sky lacking bright objects). To put it another way, the background near our X-ray objects has fewer visible faint sources than the field as a whole, so the background estimated from elsewhere has more faint objects than the region of interest.
This was already noticed by \citet{Rutledge2000} and \citet{Brusa2007}, and is further explained in \citet{Naylor2013}.
	
To mitigate the inaccuracy at faint magnitudes in Stripe~82, we use the following process. First, we find all the objects close to the X-ray positions, i.e., within circles of radius 5$^{\prime\prime}$ for $Chandra$ fields or 7$^{\prime\prime}$ for $XMM$-Newton fields; this includes the true counterparts plus the background. The area normalized histogram of the magnitudes of these objects is what we call the \textbf{total magnitude distribution}. Then we estimate a preliminary area normalized background tallying all the objects in a 2~deg$^2$ X-ray surveyed area. 
The total magnitude distribution includes fewer sources, is noisier, and has fewer objects than the estimated background at faint magnitudes. Therefore, we replace the total magnitude distribution at $r > 23.5$~mag with the estimated background at those faint magnitudes. Using these two distributions, we tentatively identify the counterparts of all the X-ray sources using the MLE method. We then refine the background estimate by removing a circle of 2$^{\prime\prime}$ radius around each counterpart in the 2~deg$^2$ area (a few hundred sources), and repeat the counterpart identification. The magnitude distribution of these counterparts will be correct at the bright end, {but does not include objects fainter than $r = 23.5$~mag because they are considered background objects}.
	
We still need to determine the background at faint magnitudes. We define a control sample of 18,000-20,000 objects with the same magnitude distribution as the bright counterparts ({this is the distribution of counterpart magnitudes with $r<23.5$~mag identified in the previous step}) but located far enough away from the X-ray source positions to not be counterparts (though still in the X-ray surveyed region). We then measure the background near these objects within an annulus of inner radius 1$^{\prime\prime}$ and outer radius of 5$^{\prime\prime}$ (for $Chandra$ fields) or 7$^{\prime\prime}$ (for $XMM$-Newton fields) around these objects. This empirically measures the background in regions where faint objects are masked by bright objects, similar to what occurs around X-ray sources. This corrected background agrees with the previously estimated background at bright magnitudes but no longer exceeds the total magnitude distribution at faint magnitudes (at $r \gtrsim 23$~mag). We refer to \S~2.2 in \citet{Brusa2007} for a demonstration. 
	
We then repeat the identification process using the corrected background {and the original total magnitude distribution}. In addition, we make an astrometric correction by finding the median offset between the X-ray and ancillary catalogs, and correcting the X-ray position after each iteration of the MLE identification (following \citealp{hsu2014}). Typical offsets between X-ray and SDSS coordinates are $\sim 1^{\prime\prime}$ for \textit{XMM}-Newton and $\sim 0.02^{\prime\prime}$ for $Chandra$. In the final output catalog, we provide the original Right Ascension and Declination of the X-ray sources without the offsets applied.

As discussed above, we use the FT16 co-added SDSS catalog ($r < 25$~mag) where possible, otherwise we use the J14 co-added catalog (u, r, z bands).
For the NIR, we used VHS $K$ band \citep{vhs} as it is deeper than UKIDSS and has better photometric precision. 
For the MIR, we predominantly used the SpIES and SHELA {\it Spitzer} IRAC surveys (3.6 $\mu$m), since they have better spatial resolution than {\it {\it AllWISE}} and deeper coverage.
However, for archival pointings outside the SpIES or SHELA footprints, we used {\it {\it AllWISE}} to find infrared counterparts (W1).
	
\subsection{{Error Radius for Positional Matching}}\label{sec:error_rad}
	
To determine whether or not we have identified the same source in two different catalogs, we consider the separation between the two positions.
For separations $<1^{\prime\prime}$ (for SDSS and VHS) and $<2^{\prime\prime}$ (for IRAC), discrete sources cannot be resolved, and are identified  as a single object in their respective catalogs.
There could be accidental alignments with a random faint source, because they have the largest space density. 

\begin{figure}[th]
\centering
\includegraphics[width=0.85\linewidth]{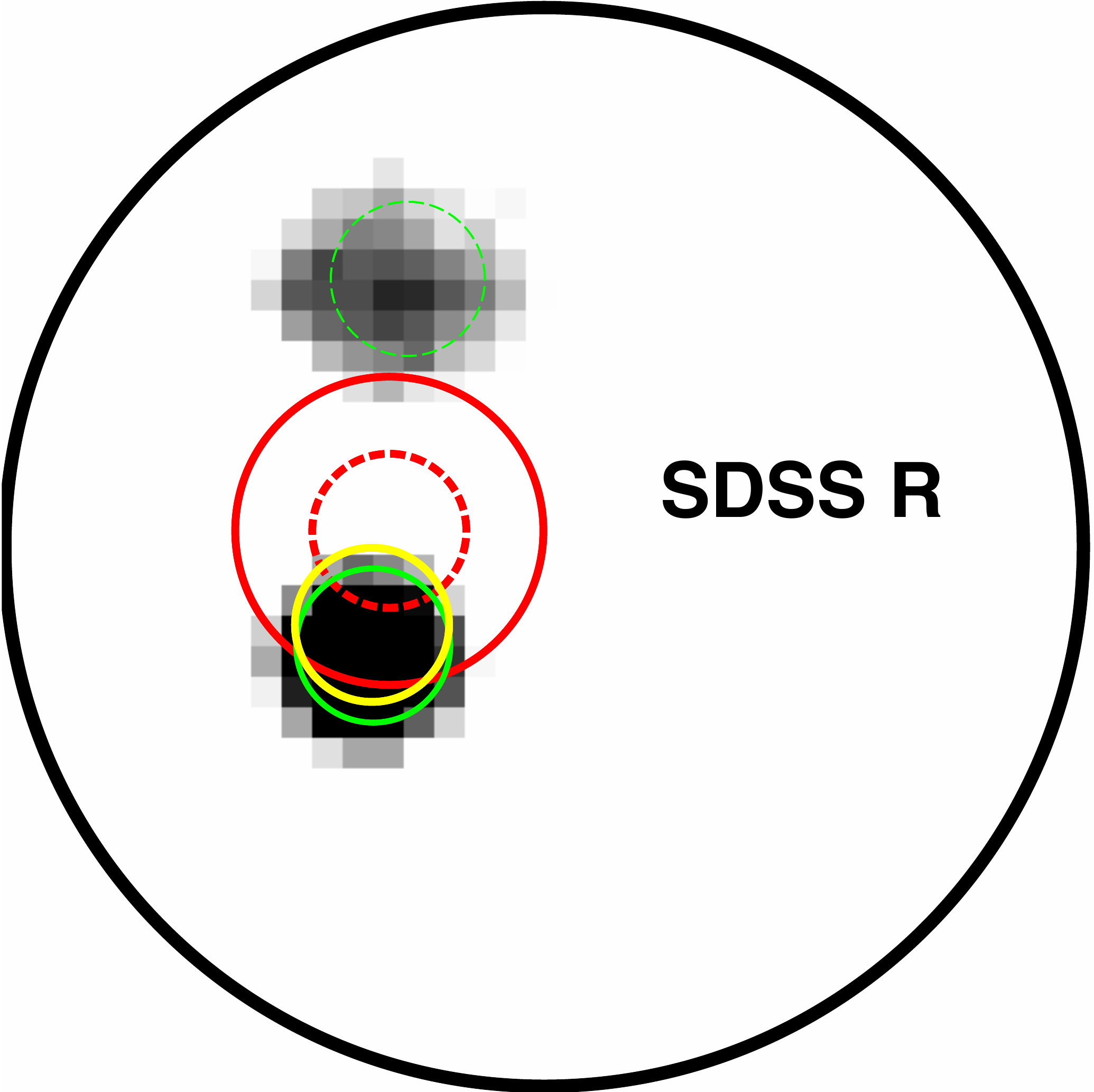}
\includegraphics[width=0.85\linewidth]{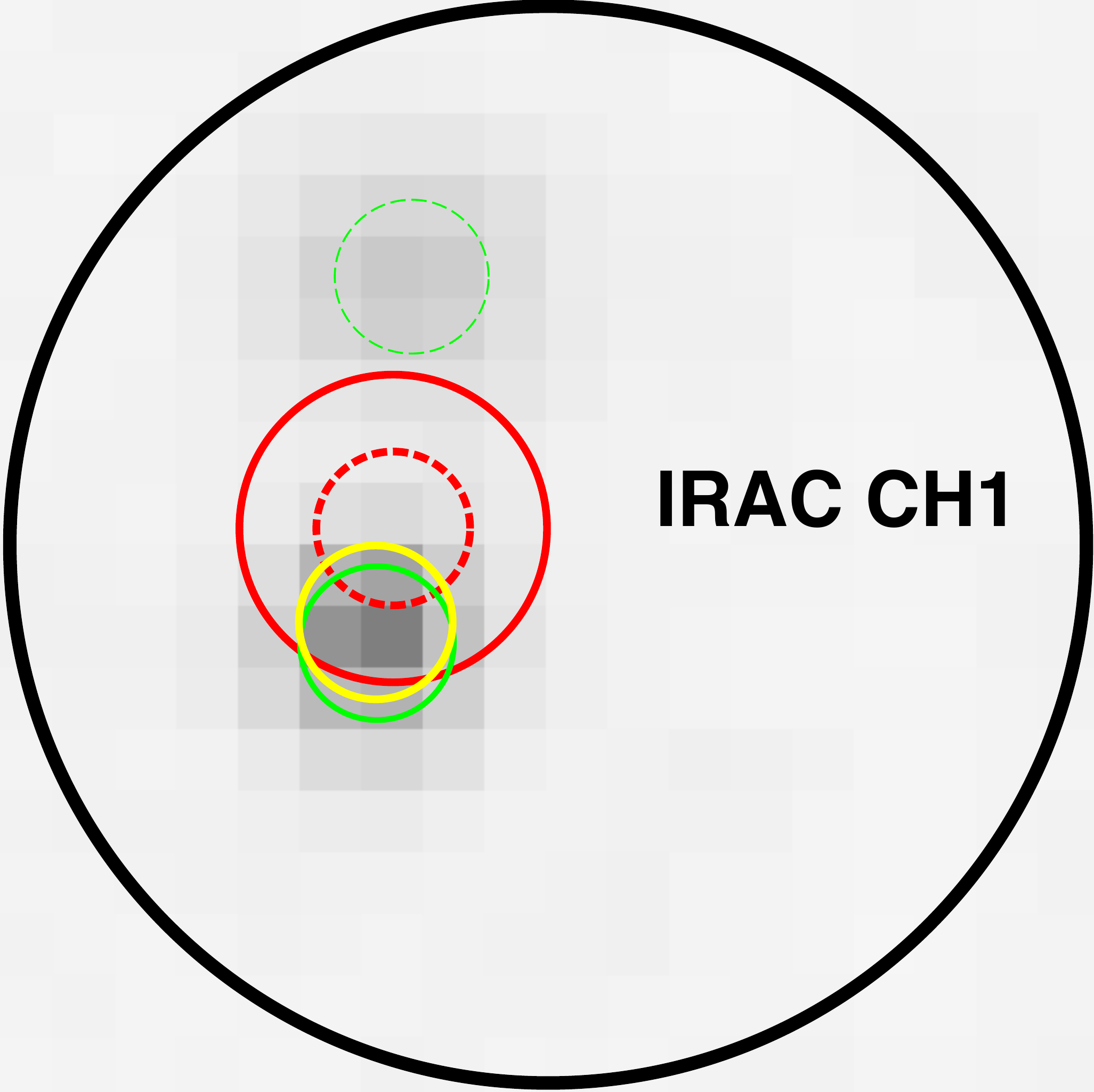}
\caption{{An example of how source positions are affected by nearby sources:} 
{\it (top)} SDSS $r$ band image, {\it (bottom)} IRAC CH1 image. 
According to our criteria, the correct counterpart to the X-ray object is the lower source because SDSS source position {\it (solid green circle)} and VHS source position {\it (yellow circle)} agree on this as the most likely counterpart. (The {\it dashed green circle} shows another possible SDSS match for this X-ray object, but with much lower reliability, so we ignore it).
Note that even though the lower source is brighter in the IRAC image, IRAC source position {\it (at the center of the red circles)} points slightly away as it is affected by blending with the adjacent faint object. We use two circles of different radii to show the IRAC source position: \textit{dashed red circle} is 1$^{\prime\prime}$ in radius, and the \textit{solid red circle} is 2$^{\prime\prime}$. While comparing source positions between catalogs, if we had used a 1$^{\prime\prime}$ matching radius, we would have incorrectly concluded that IRAC disagrees with SDSS and VHS. Therefore, to account for blending, we allow a 2$^{\prime\prime}$ matching radius for IRAC. To avoid contamination, we visually checked all sources like these where IRAC source position falls between 1-2$^{\prime\prime}$ from SDSS/VHS, and added comments in our final catalog.} 
\label{fig:irac_blended} 
\end{figure}

For SDSS and VHS, at the faintest magnitudes, the closest pairs lie at $\sim1.2^{\prime\prime}$, so we are safe in assuming that counterparts within 1$^{\prime\prime}$ are the same object, and those with larger separations are distinct objects. 
For IRAC, the smallest separation between faint sources is $\sim2.3^{\prime\prime}$, so an error radius of $2^{\prime\prime}$ is appropriate; that is, IRAC sources should be within $2^{\prime\prime}$ of the catalogued SDSS or VHS position.
Source blending within the error radius can make coordinates inaccurate, as illustrated by Figure~\ref{fig:irac_blended}. Even though SDSS, VHS and IRAC all identify the same bright counterpart, the IRAC position is less accurate because it is affected by blending with the object above. Therefore, we allow a bigger matching radius for IRAC (2$^{\prime\prime}$) to account for blending. 
	
As an extra precaution to avoid contamination, we visually inspected all 157 cases (2\%) where the separation between SDSS/VHS and IRAC counterparts is between 1$^{\prime\prime}$ and 2$^{\prime\prime}$, and added comments in our final catalog, Appendix A. Not surprisingly, we found that these cases have very crowded fields which result in blending, or are extended sources where each band points at a different part of the same source, or are very bright sources that cause saturation, leading to errors in pinpointing the correct position of the source.
For crowded fields, where several objects are too close to each other, we cannot obtain clean photometry and this will affect the accuracy of the photometric redshifts. Thus, the ``nearby\_neighbor\_sextractor" and the ``manual\_check" columns in our final catalog are provided to assist the user. The description of all columns in the final catalog are given in Appendix A.

\begin{figure}[th]
\centering
\includegraphics[width=0.95\linewidth]{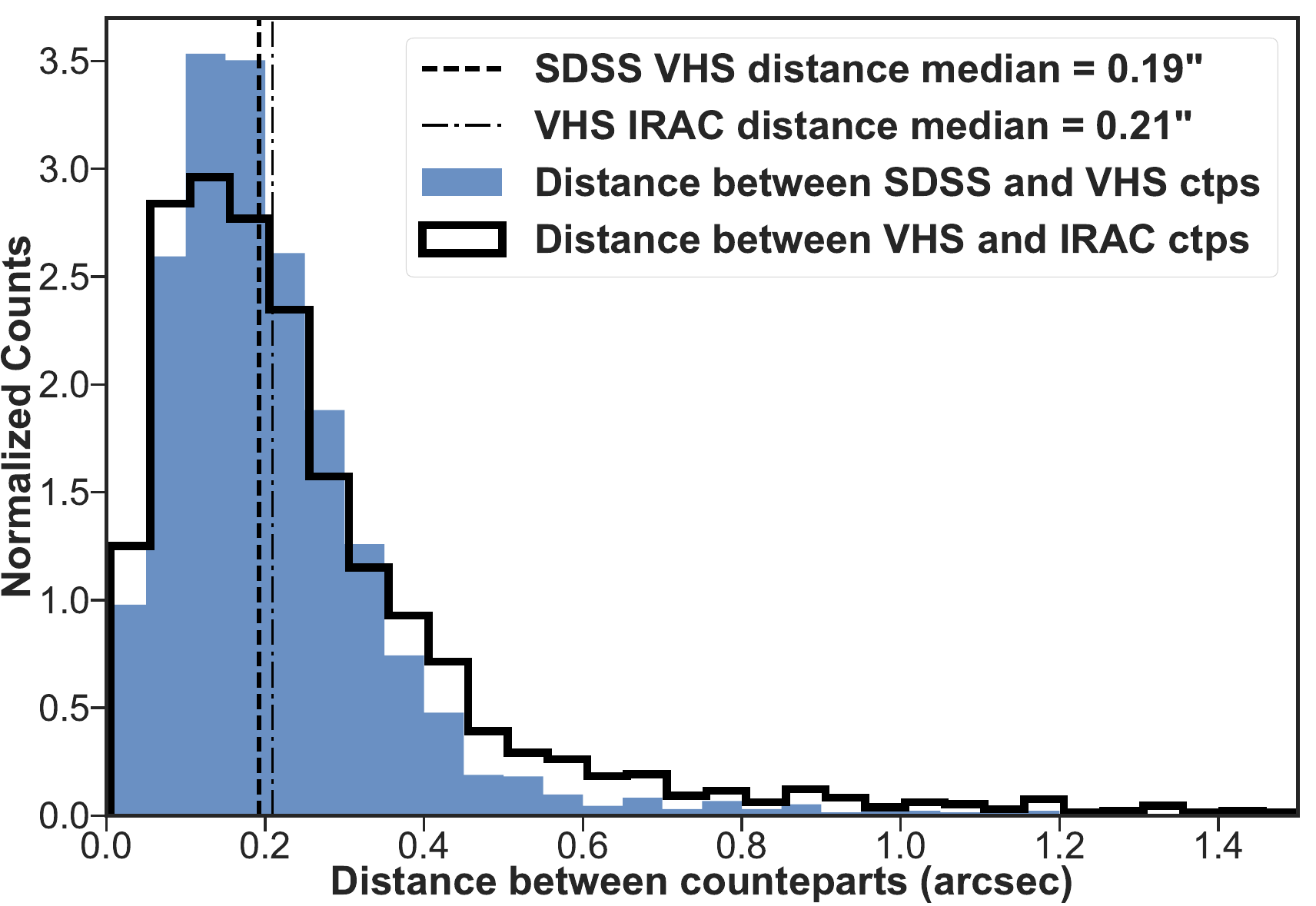}
\caption{An area-normalized histogram of distances between SDSS $r$ band and VHS $K$ band sources for \textit{XMM} AO10 and AO13 fields. The median distance between SDSS and VHS sources is 0.19$^{\prime\prime}$ and 86\% of SDSS and VHS counterparts fall within 1$^{\prime\prime}$ of each other. A distance greater than 1$^{\prime\prime}$ indicates that the two counterparts are likely different.} 
\label{fig:sdss_vhs_dist} 
\end{figure}

Comparing the positions of all SDSS and VHS counterparts for the \textit{XMM} AO10 and AO13 X-ray sources, we found that 86\% are within 1$^{\prime\prime}$ of each other and the median distance between them is 0.19$^{\prime\prime}$ (Fig.~\ref{fig:sdss_vhs_dist}). Because the offsets between SDSS and VHS coordinates are so small, and as both SDSS and VHS catalogs have sufficiently high resolution to spatially distinguish between sources that are $\sim1.2^{\prime\prime}$ apart, we conclude that if the counterparts chosen by SDSS and VHS are more than 1$^{\prime\prime}$ apart, they must be two different objects.

Due to the vastly different resolution of {\it AllWISE}, those counterparts are not considered unless we do not have match for a source in any other catalog {(66 cases)}. We do not discard an association completely if it is only detected in one band because very obscured objects may only be identifiable in the MIR, and can potentially be more interesting, e.g., obscured and/or high-redshift AGN, than bright objects identified in all bands. We provide quality flags and reliability classes to indicate level of uncertainty in counterpart identification (\S~\ref{ssec:correct_assoc}).
	
\subsection{{Resolving Multiple Associations}}\label{ssec:correct_assoc}
	
We find at least one possible counterpart within the search radius of an X-ray source, either in the optical or in the NIR and MIR wavelength, for 97.8\% of the Stripe~82X sample. However, not all the associations are correct because a fraction of them will be chance associations. We use the Likelihood Ratio (LR) results from MLE to guide us as to whether a counterpart is reliable. The LR is the probability that the correct counterpart is found within the search radius divided by the probability that the association is a chance coincidence with a background source. We determined a critical threshold ($LR_{\rm th}$) for each band, below which we assume a counterpart is not reliable (these are identified in the master catalog as having ``sub-threshold" reliability class), following the method elaborated in \citet{Civano2012} and \citet{Marchesi2016}, which we summarize below.
	
The $LR_{\rm th}$ is the LR value at which (reliability+ completeness)/2 is maximum, where reliability is defined by R $= N_{ID}/N_{LR > L_{th}}$ (ratio of sum of all the reliabilities of the candidate counterparts and total number of sources above threshold) and completeness C $= N_{ID}/N_{X}$ (ratio of sum of reliabilities of all the sources identified as possible counterparts and the total number of X-ray sources) \citep{Civano2012, Marchesi2016}. After determining $LR_{\rm th}$, we look closely at cases with more than one reliable counterpart identified in any given band (about 10\% of the X-ray sources have multiple counterparts above this threshold in at least one band). 
For each band, we look at the ratio of LR values for the most and the next most reliable counterparts, i.e., $LR_{\rm 12}$ = $LR_1$/$LR_2$. If that ratio is above the median value for all $LR_{\rm 12}$ in that band, we assign it a ``secure" reliability class.
If the ratio falls below the median, we put that source in the ``ambiguous" reliability class. Note that this classification is done among counterparts identified within a single band of multi-wavelength data, for each band on which we ran MLE.
	
Next we compare results of the analysis described above among different bands and find that for 12\% of X-ray sources, different bands choose different objects as the most reliable counterpart (\S~\ref{sec:error_rad}). 
If the two counterparts have different reliability classes (secure, ambiguous, or sub-threshold), we choose the more secure case.
If both counterparts have the same reliability class, we choose the one with a higher likelihood ratio.
Table~\ref{tab:spurious_ambiguous} lists the reliability class for all counterparts.
	
As we resolve conflicting associations, we assign quality flags to reflect the confidence with which we identify the correct counterpart.
			
\noindent{\bf Quality Flag (QF) 1:} 
4524 X-ray sources (73.2\%) have a unique counterpart in at least two bands and their positions agree, so we consider the identifications unambiguous. For 28 of these sources, the LR is sub-threshold in all bands but the same counterpart is identified in more than one band; therefore, we consider these unambiguous as well. 

\noindent{\bf Quality Flag 1.5:} 
174 sources (2.8\%) have ambiguous counterparts in at least one band but the other bands help resolve the ambiguity, so these associations are also considered relatively secure.

\noindent{\bf Quality Flag 2:} 
190 sources (3.0\%) have counterparts that disagree, of which one is either sub-threshold or very close to the threshold ($LR < LR_{\rm th} + 2$) while the other is well above the threshold ($LR > LR_{\rm th} +5$). We choose the latter case, regardless of band, but flag the association as not quite as secure.

\noindent{\bf Quality Flag 3:} 
416 sources (6.7\%) have different counterparts in multiple bands with comparable $LR$ but different reliability classes (this occurs because $LR_{\rm th}$ and $LR_{\rm 12}$ varies between bands). Here we choose the association with the more secure reliability class, assigning it QF 3 and the secondary counterpart {QF -1}.

\noindent{\bf Quality Flag 4:}
740 sources (12\%) have a counterpart in only one waveband. Of these, 631 sources have $LR > LR_{\rm th}$, {which are considered reliable.} 109 sources are below the LR threshold and therefore possibly spurious matches.

\noindent{\bf Quality Flag 0, -1:}
A QF 0 indicates that we found no association to that X-ray source in any multi-wavelength catalog, or that these are sources taken from LM16. There are 137 (2.2\%) such sources. QF -1 is an alternate to association given for Quality Flag 3. This means that in the final catalog, one X-ray source can be listed twice, QF 3 being the more secure counterpart and QF -1 is the less secure alternative according to our MLE analysis.

\begin{deluxetable}{lcccc}[th]
\tablewidth{0pt}
\tablecaption{\label{tab:spurious_ambiguous} \textsc{Summary of Matches within Each Band.}}
\tablehead{\colhead{\textsc{Band}} &  \colhead{\textsc{Total}} & \colhead{\textsc{LR $>$ LR$_{\rm th}$}} &\colhead{\textsc{Secure}} &\colhead{\textsc{Ambiguous}}}
\startdata
SDSS \tablenotemark{1} & 5825 (94\%) & 5518 (89\%) & 5358 (87\%) & 160 (2.6\%)  \\
VHS $K$ \tablenotemark{2} & 4253 (69\%) & 4123 (67\%)  & 3853 (62\%) & 270 (4.4\%) \\
	IRAC CH1 \tablenotemark{3} & 4728 (76\%) & 4354 (70\%) & 4102 (66\%) & 252 (4.1\%) \\
	\enddata
	\tablenotetext{1}{FT16 $r$ band has $LR_{\rm th}$ of $\approx$0.6-0.7 and a median $LR_{12} \approx$ 3-4.}
	\tablenotetext{2}{VHS $K$ band has $LR_{\rm th}$ of $\approx$1-1.5 for $XMM$ AO10 and AO13, and $\approx$0.1-0.2 for Archival $XMM$ and Chandra. Median $LR_{12} \approx$1.5-4.5.}
	\tablenotetext{3}{IRAC Ch1 band has $LR_{\rm th}$ of $\approx$0.8-2.0 and a median $LR_{12}$ $\approx$ 1-2.5 ($LR_{12}$ $>$ $LR_{\rm th}$ in each field).}
\end{deluxetable}

173 X-ray sources only have matches below threshold (QF 1: 28, QF 2: 1, QF 3: 35, QF 4: 109).
This totals 6181 X-ray sources to which we assigned a quality flag.

To summarize, 
of the 6181 unique X-ray sources, 
4524 (73\%) have unambiguous counterparts in one or more bands (QF 1),
780 (12\%) have different counterparts in different bands that we resolve with three levels of confidence (QF 1.5, 2, 3),
740 (12\%) have counterparts in only one band (QF 4).  {These are considered reliable when they fall above LR$_{\rm th}$ (631 cases)},
and	137 (2.2\%)  have no counterparts from our association analysis (QF 0).
173 ($\sim$ 3\%)  have sub-threshold counterparts (at various QF values).
Other than the 29 sub-threshold sources, we take QF 1, 1.5 and 2 sources to be secure.

\begin{figure*}[th]
	\centering
	\includegraphics[width=0.7\linewidth]{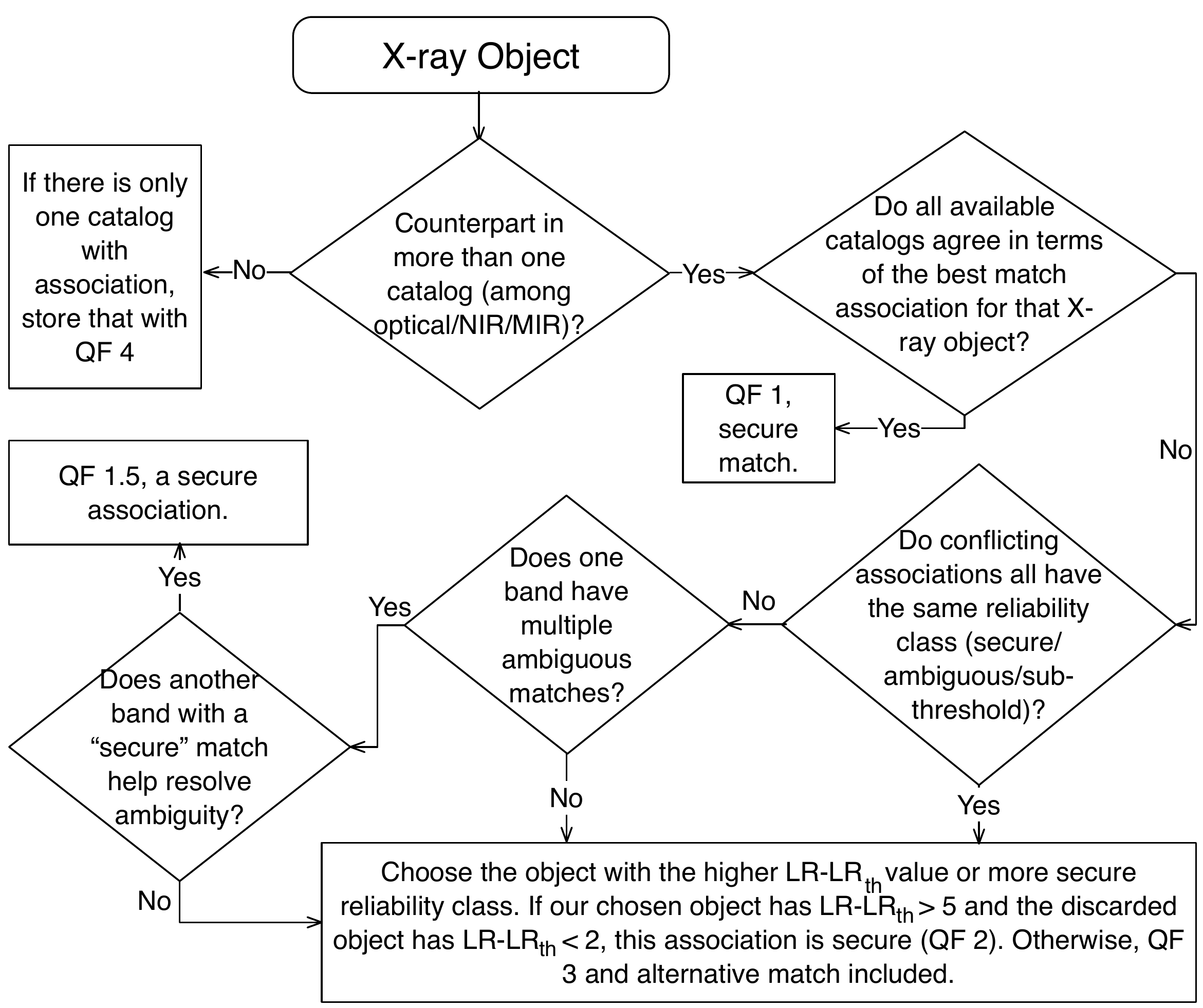}
	\caption{Flow chart showing how we assign counterpart(s) to each X-ray source and how the Quality Flag is defined (see \S~\ref{ssec:correct_assoc}). Quality Flag values 1, 1.5 and 2 are secure; 3 and 4 are problematic; and -1 refers to the discarded, second-best candidate.}
		\label{fig:assoc_flowcharts} 
	\end{figure*}

The flow chart in Figure~\ref{fig:assoc_flowcharts} summarizes the steps in the identification process.
We note that some identifications may be improved when deep Subaru HyperSuprimeCam optical imaging will become available.
In our final output catalog (Appendix A), we report the multi-wavelength associations and photometry for all objects, along with their reliability classes and quality flags.

	\subsection{{Duplicate X-ray Observations}}\label{ssec:duplicate}
	
There is some overlap between the X-ray observations (between all fields except AO10 and AO13 and AO10 and archival $Chandra$, as shown in Fig.~\ref{fig:vhs_map}), resulting in 6181 unique sources and 185 duplicates. The identification of duplicates was done in LM16. 
We have determined, independently, the counterparts  for all 185 duplicates and found that even though in most cases duplicate X-ray observations were assigned the same multi-wavelength counterpart, in {11 cases (23 observations - 10 of these sources have two duplicate observations each, but one source has three different observations), objects identified as the same X-ray source were assigned two different counterparts}. This occurs due to slight differences in X-ray coordinates, which in turn changes the surrounding area in which MLE looks for ancillary counterparts.
	
	In order to determine which X-ray source to rely on in case of disagreement of counterparts, we choose the $Chandra$ X-ray source over the \textit{XMM} X-ray source due to the high positional accuracy of the former (20 observations of 10 sources). {When deciding between duplicate $Chandra$ observations (2 or 3 observations of the same source), we choose the source with lowest positional uncertainty}. In the final output catalog, we use a ``duplication\_flag" to indicate which source we preferred, and sources for which counterparts disagreed. The flag is described in Appendix A.

	\subsection{{Summarizing Differences with LM16 Multi-wavelength Catalog}}
	
	There are some differences between this catalog and LM16 because we use a deeper, updated optical catalog (FT16) and new IRAC catalogs, and because we use a different calculation of the background magnitude distribution used for MLE matching. For 5329 (86\%) cases, there is at least one band in which our chosen association agrees with LM16 within 1$^{\prime\prime}$. For 40 X-ray sources (0.6\%), our chosen counterpart is different from the previous catalog in all bands. There are also 658 new associations (11\%) in our catalog due to deeper data. Of these, 507 are above $LR_{\rm th}$ in at least one band. The LM16 catalog purposely did not include any sub-threshold sources, whereas our catalog does (with reliability class marked as sub-threshold) because they are used to resolve conflicting counterparts.
	
	In addition, the LM16 catalog has 20 SDSS counterparts that were not detected in the co-added catalogs that we used.
	Seven of these sources were not reported in SDSS DR13 (LM16 used DR12), suggesting they are likely spurious, and we found no counterparts in other bands. We did not include them in our final catalog.
	Thirteen of the 20 are in the SDSS DR13 catalog because they were detected in one or more bands at one or more epochs. 
	We include these 13 in our final catalog (``Association" field says ``LM16"). However, two of these are not detected in any other multi-wavelength catalogs (VHS, IRAC, {\it AllWISE}, co-added SDSS), and so should be considered with caution.
	
	Another 5 have detections in only one of those wavebands, which is better but also requires caution.
	Finally, 5 have detections in five or more wavebands and have co-added SDSS data as well, { but after the offset correction that we applied, they  fall outside the X-ray error radius (5-7$^{\prime\prime}$), so our MLE analysis ignored these possible associations.} 
	There is 1 source from LM16 for which we get two bands of data, in VHS $J$ and $H$ band, but as we do not run MLE in these bands, we missed this association.
	We annotate these 13 objects with appropriate ``manual\_check" in our final catalog.

	\subsection{{SED Construction}}\label{ssec:sed_construction}
	
	Before constructing broad-band SEDs, we identify NIR counterparts in the UKIDSS catalog \citep{lawrence2007} and ultraviolet (UV) counterparts in the {\it GALEX} catalog \citep{galex}, using nearest neighbor matching. These catalogs are not used for association analysis, but once we have identified a counterpart with an appropriate QF, nearest neighbor matching with an error radius is sufficient to identify that counterpart in ancillary catalogs. Table~\ref{tab:error_radius} shows the error radius allowed between different ancillary catalogs to construct the final multi-wavelength catalog (Appendix A). 
	For the nearest neighbor match, the counterpart coordinate comes from the catalog with the most accurate astrometry (as in \S~\ref{ssec:correct_assoc}); 
	for example, if SDSS and IRAC counterparts for an X-ray source agree, we choose the SDSS coordinates because of their better positional accuracy.
	
 \begin{deluxetable}{lcc}[th]
	\tablewidth{0pt}
	\tablecaption{\label{tab:error_radius} \textsc{Search Radius Allowed for Nearest Neighbor Matches between Ancillary Catalogs.}}
	\tablehead{\colhead{\textsc{Catalog}} &\colhead{\textsc{SDSS and VHS}} & \colhead{\textsc{IRAC}}}
	\startdata
	SDSS DR12 redshifts & 1.0$^{\prime\prime}$ & 2.0$^{\prime\prime}$ \\
	{\it GALEX} & 2.0$^{\prime\prime}$ & 2.0$^{\prime\prime}$ \\
	SDSS (FT16) & 1.0$^{\prime\prime}$ & 2.0$^{\prime\prime}$ \\
	SDSS (J14) & 1.0$^{\prime\prime}$ & 2.0$^{\prime\prime}$ \\
	VHS & 1.0$^{\prime\prime}$ & 2.0$^{\prime\prime}$ \\
	UKIDSS & 1.0$^{\prime\prime}$ & 2.0$^{\prime\prime}$ \\
	SPIES & 2.0$^{\prime\prime}$ & 1.0$^{\prime\prime}$ \\
	SHELA & 2.0$^{\prime\prime}$ & 1.0$^{\prime\prime}$ \\
	AllWISE & 1.0$^{\prime\prime}$ & 1.0$^{\prime\prime}$ \\
	\enddata		
\end{deluxetable}	
	
\begin{figure}[th]
		\centering
		\includegraphics[width=0.95\linewidth]{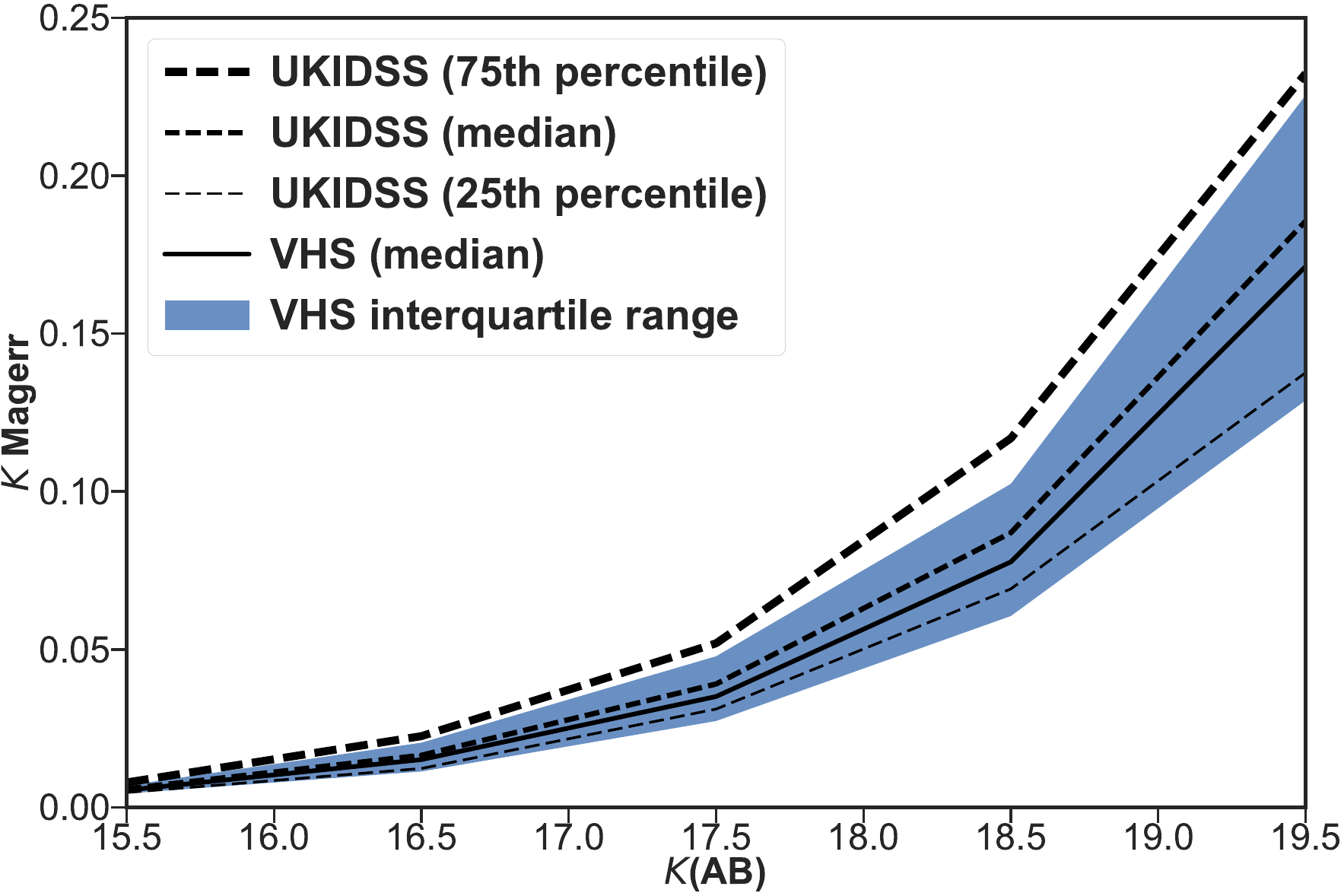}
		\caption{VHS and UKIDSS K band magnitude errors plotted against K band magnitude. VHS magnitude errors are slightly smaller than UKIDSS magnitude errors, and VHS is slightly deeper, so we only use VHS for MLE matching. However we add UKIDSS data in after finding associations, and use it to construct SEDs.}
		\label{fig:vhs_ukidss_k}
	\end{figure}

	Like VHS, UKIDSS provides broad-band data in $J$, $H$ and $K$ filters but is shallower (and thus with larger photometric errors, see Fig.~\ref{fig:vhs_ukidss_k}), with a 5$\sigma$ $K_{AB}$ depth of 20.1~mag. 
	We still include UKIDSS when constructing SEDS (and nor for counterpart identification) to check if having additional NIR data leads to more robust photometric redshifts, and discuss the results in \S~\ref{ssec:ukidss_wise}.
	Roughly 58.7\% of the X-ray sources have UKIDSS counterparts.
	Because the {\it GALEX} survey of Stripe~82 is relatively shallow, we find UV counterparts for only 20\% of the X-ray sources.

	We correct UV, optical and NIR data for extinction using the Galactic extinction values from SDSS DR13 \citep{dr13_2016} and color excess from VHS \citep{vhs}. 
	The corrected magnitudes can then be compared to template SEDs. 
	(The transformation of template SEDs to the observed photometric system is discussed in \S~4.1 of S09.) 
	
	{To summarize, the SEDs consist of far- and near-ultraviolet (FUV and NUV, respectively) from $GALEX$, $u$, $g$, $r$, $i$ and $z$ from SDSS, $J$, $H$ and $K$ from VHS/UKIDSS, CH1 and CH2 from IRAC and W1 and W2 from $AllWISE$. We also provide $AllWISE$ W3 and W4 magnitudes in our final catalog, but it is not used for SED construction. The bands used for SED fitting are discussed in \S~\ref{ssec:ukidss_wise}.}

	\section{{Photometric Redshifts}}\label{sed_fitting}
	
	{We present photometric redshifts for
	a total of 96.4\% (5961) of the sources of Stripe~82X; missing are the 2.2\% (137) without counterparts in other wavebands and the 1.4\% (83) with data in only one or two bands, which is insufficient for template fitting. In this section we will discuss the procedure to calculate photometric redshifts.}
	
	\subsection{{Method of Fitting SEDs}}

	\begin{deluxetable*}{lcccc}[th]
	\tablewidth{0pt}
	\tablecaption{\label{tab:comparison_object_type_other_fields} \textsc{Population type comparison between different fields\tablenotemark{1}.}}
	\tablehead{\colhead{\textsc{Field}} & \colhead{\textsc{Source count\tablenotemark{2}}} & \colhead{\textsc{Extended source count\tablenotemark{3}}}& \colhead{\textsc{Point-like source count\tablenotemark{4}}} & \colhead{\textsc{F$_{\rm 0.5-2 keV} > $ 1e$^{-14}$ ergs/cm$^2$ s}}}
	\startdata
	S82X & 6044\tablenotemark{5} & 2486 (41\%) & 3289 (54\%) & 2304 (38\%) \\
	Lockman Hole \citep{fotopoulou2012}  & 388 & 134 (34.5\%)\tablenotemark{6} & 140 (36\%) & 19 (5\%)  \\ 
	CDFS \citep{hsu2014}  & 744 & 591 (79\%) & 153 (21\%) & 7 (1\%) \\  				
	ECDFS \citep{hsu2014} & 1207 & 974 (81\%) & 233 (19\%) & 19 (2\%) \\ 				
	$Chandra$-COSMOS \citep{Marchesi2016}  & 4016 \tablenotemark{7} & 2023 (50\%) & 1726 (43\%) & 221 (5.5\%)  \\ 
	$XMM$-COSMOS (S09) & 1542 & 464 (30\%) & 1032 (67\%) & 179 (11.6\%) \\
	\enddata
	\tablenotetext{1}{More details about each of these fields is given in Table~\ref{compare_fields}. This table does not include fields for which photozs were only calculated for point-like sources.}
	\tablenotetext{2}{Sources that had associations or were categorized as extended and point-like sources.}
	\tablenotetext{3}{Extended sources tend to be nearby and galaxy dominated.}
	\tablenotetext{4}{Point-like sources tend to be AGN-dominated and variable.}
	\tablenotetext{5}{6044 out of 6181 have associations, and 269 could not be categorized a point-like or extended.}
	\tablenotetext{6}{103 are too faint to resolve and 12 are blended with nearby sources.}
	\tablenotetext{7}{Not all objects could be categorized as point-like of extended due to lack of optical counterpart.}
	\end{deluxetable*}
	
	We computed photometric redshift via the SED fitting technique, using the code LePhare \citep{Arnouts1999, Ilbert2006}, which has been extensively used and tested both for normal galaxies and AGN. LePhare can accommodate user-specified SED templates, extinction law(s) and extinction value(s). In addition, it allows the use of {luminosity} priors to reject unlikely redshifts. Finally, LePhare corrects for intergalactic absorption, an important effect for high redshift sources. 
	
	Since the mix of active nucleus and host galaxy can be different in every object, the fitting process is more complicated than for normal, inactive galaxies.
	This added complexity manifests as degeneracy in the results, as small parameter shifts cause more than one template to be a good fit to the photometric SED. 

\begin{figure*}[th]
	\centering
	\includegraphics[width=0.5\linewidth]{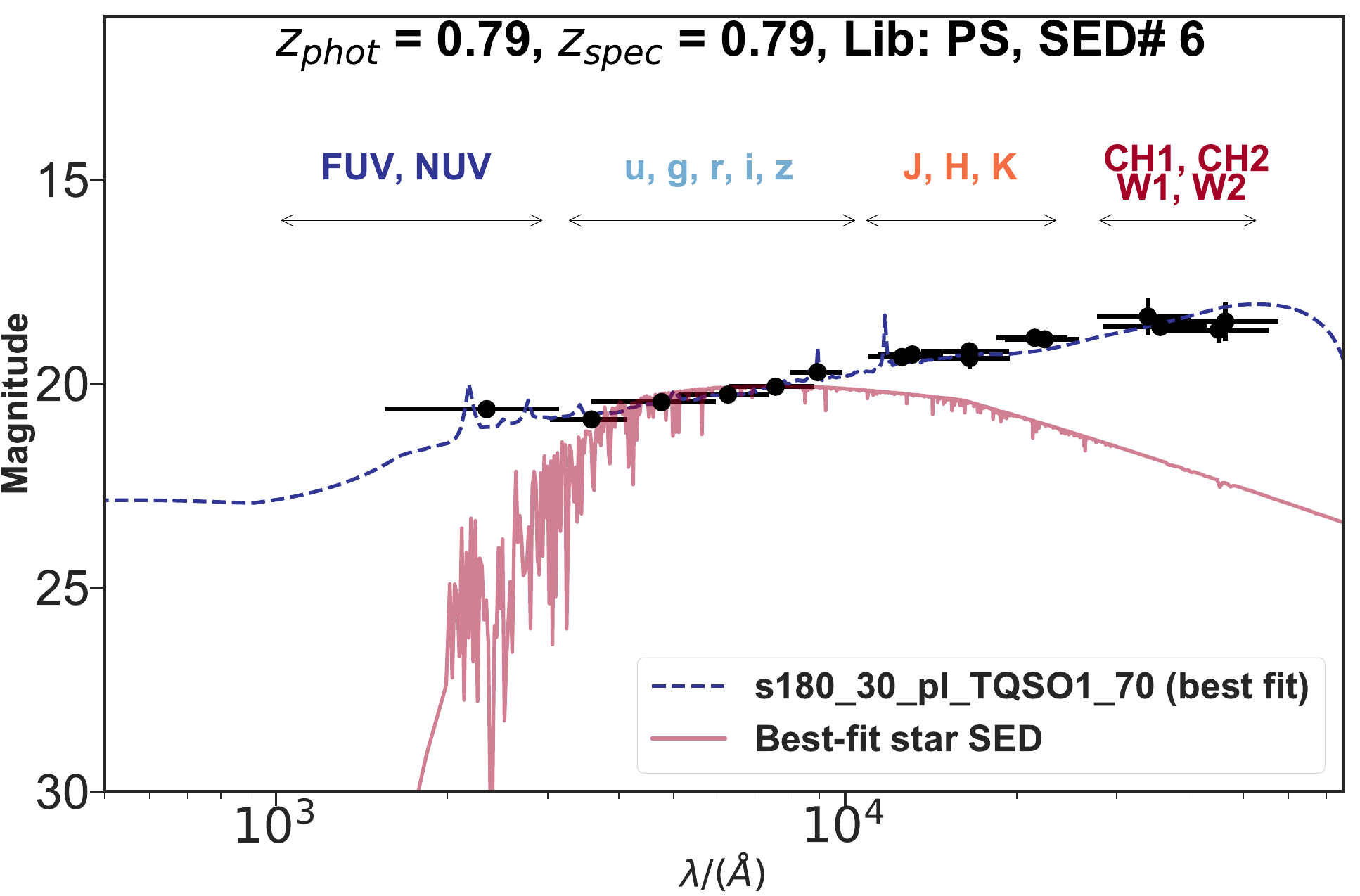}~
	\includegraphics[width=0.5\linewidth]{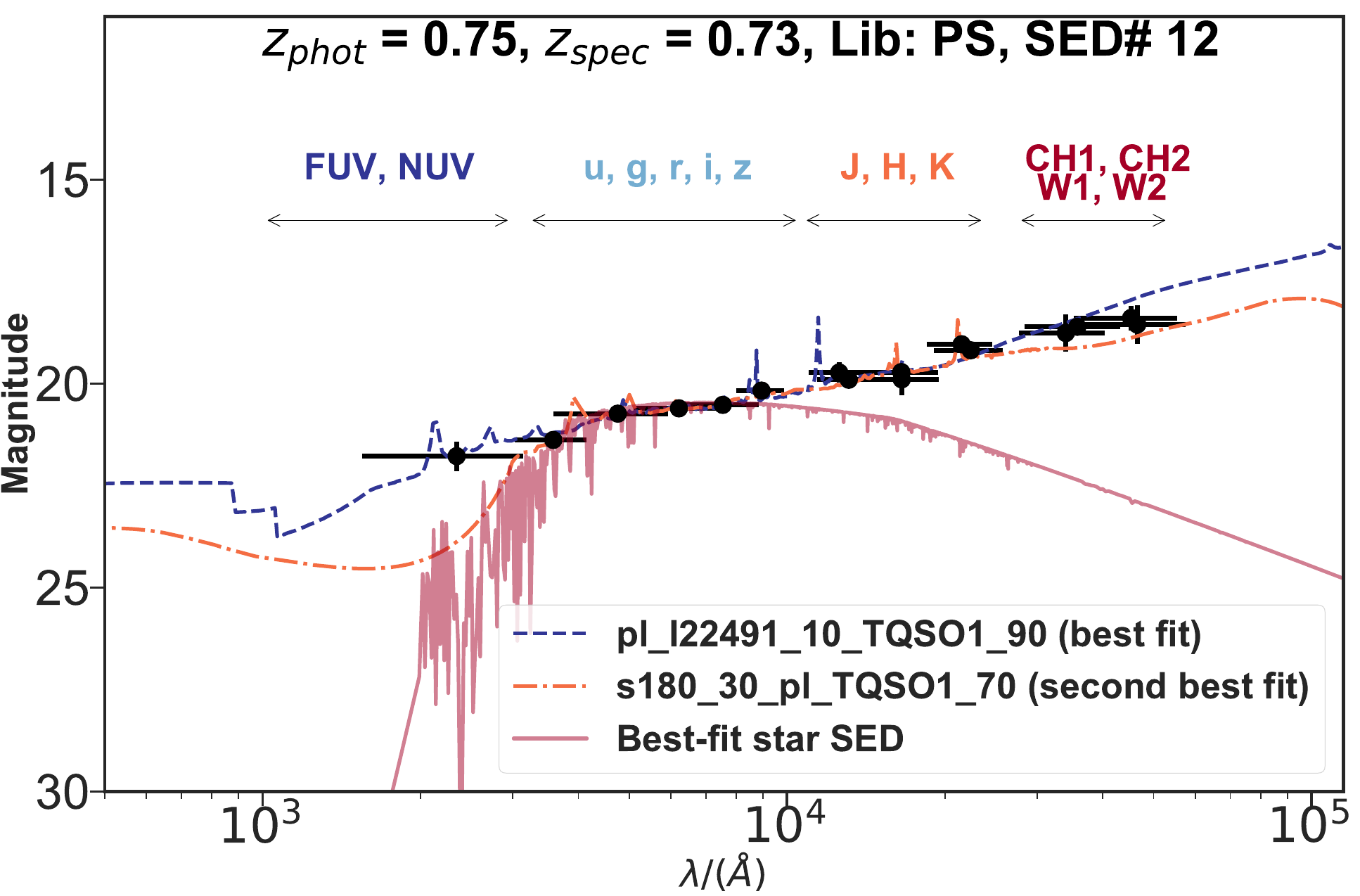}
	\includegraphics[width=0.5\linewidth]{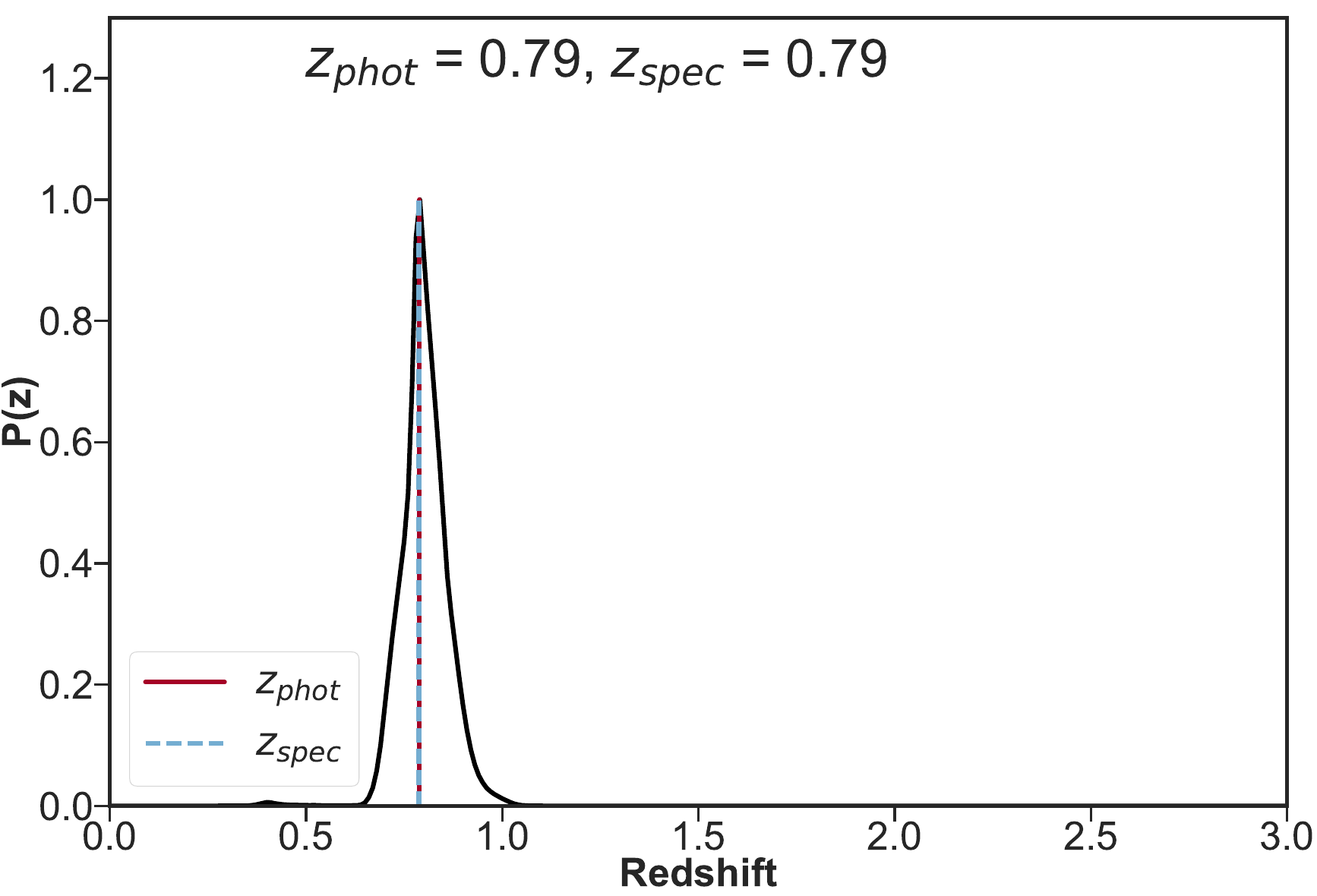}~
	\includegraphics[width=0.5\linewidth]{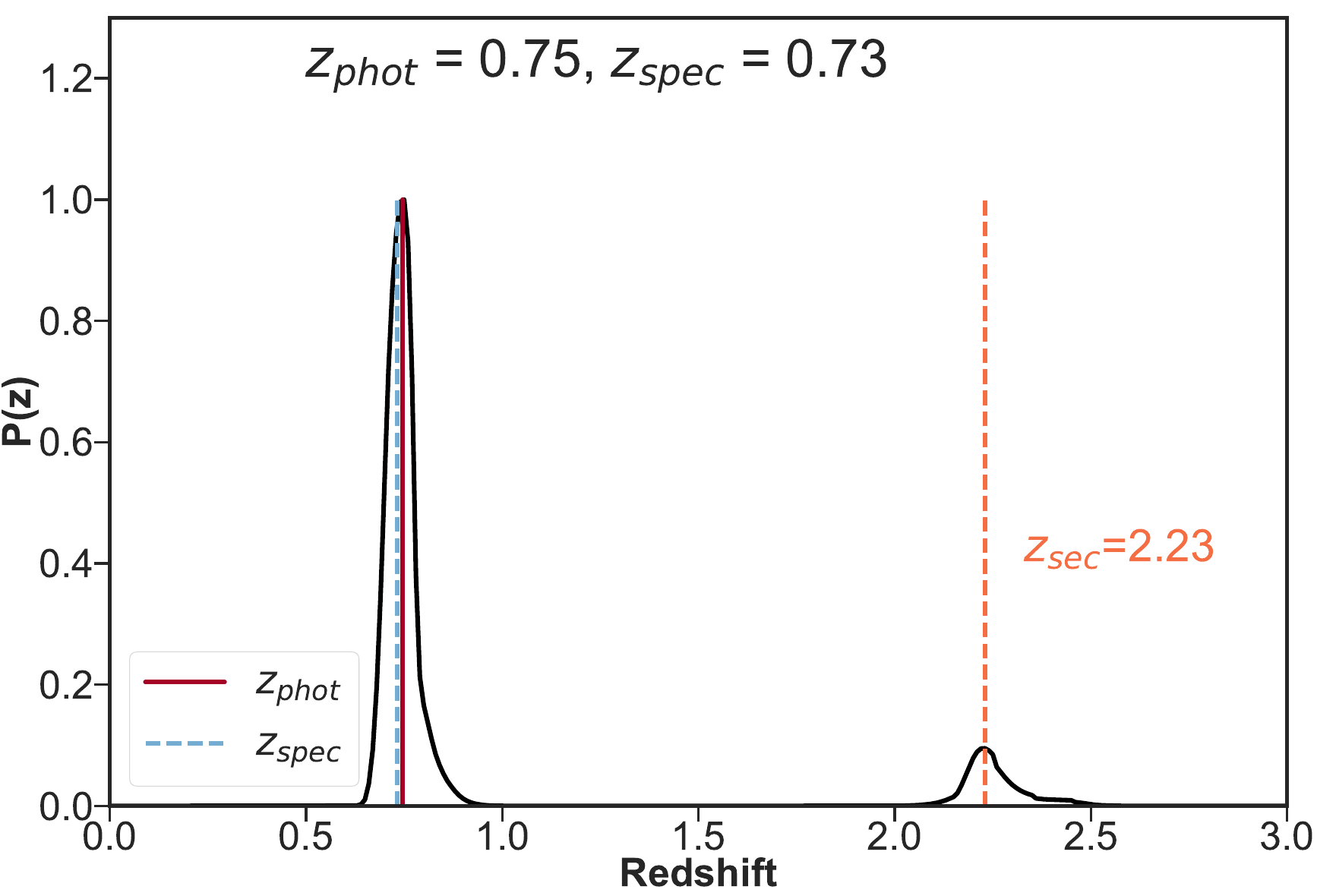}
	\caption{{\it Top panels:} Best-fit templates ({\it blue dashed lines}) for observed SEDs ({\it black points}) of two Stripe~82X AGN, one with a well-defined redshift and a similar one that also has a secondary redshift. {\it Bottom panels:} the probability function of photometric redshifts for these objects. {\it Top left:} Point-like AGN (r-band FWHM $< 1.1^{\prime\prime}$) fitted with a template that is 30\% spiral galaxy and 70\% Type 1 quasar (from the Point Source library; see Table~\ref{table:library}); also shown is the best-fit stellar template ({\it faint solid red line}). {\it Top right:} A point-like AGN fitted with 10\% I22491 starburst galaxy plus 90\% type 1 quasar ({\it blue dashed line}, best fit) or 30\% spiral galaxy plus 70\% Type 1 quasar ({\it orange dash dotted line}, secondary best fit). For the bottom panels, the peak probability value indicates the redshift for the best-fit template, and, when present, a secondary peak indicates a second template that fits the SED well, at a secondary photometric redshift (as shown in bottom right panel).}
	\label{fig:SED_fit}
\end{figure*}

	In order to break this degeneracy, we use secondary information to limit the parameter space of the redshift solution. The full-width at half maximum (FWHM) of sources in ground-based optical images, for example, can tell us whether the source is at low redshift: sources classified as ``extended" in ground-based images cannot be at redshift $z>1$, and their SEDs will have a large host galaxy contribution. 
	
	In contrast, a point-like source will either have a high redshift or be a star or a quasar, and these considerations translate into templates and {luminosity} priors that increase the accuracy of the derived photometric redshifts. Accordingly, we treat point-like and extended sources differently, as shown in S09. 
	
	Photometric catalogs usually provide morphology information, or typical FWHM of stars/point-like objects in different bands. We rely on ground-based morphology provided by FT16, J14 and \citet{vhs}. The exact procedure for the classification into these two subsamples is discussed in \S~\ref{sec:pointlike_extended}.
	Note that the use of  morphological classification done on lower resolution, ground-based images rather than on HST images as in S09, will affect  our results as we explain in \S~\ref{sec:pointlike_extended}.
	
	Photometric redshifts are calibrated using the objects that have spectra (see \S~\ref{sec:redshift_accuracy} for more details). We start by selecting a limited set of SED templates, large enough to represent the full sample yet small enough to avoid degeneracy in $\chi^2$ fitting.
	We do this separately for the two sub-samples (extended and point-like - described in \S~\ref{sec:pointlike_extended}). 
	Using trial and error, we try to achieve photometric redshifts that have small normalized median absolute deviations: 
	\begin{equation}\label{eq:sigma}
	\sigma_{\rm nmad} = 1.48 \times median(\vert (z_{spec}-z_{phot})/(1+z_{spec}) \vert)
	\end{equation}
	and low outlier fraction ($\eta$), where outliers are sources which have 
	\begin{equation}\label{eq:outlier}
	\vert (z_{spec}-z_{phot})/(1+z_{spec}) \vert > 0.15
	\end{equation}
	with respect to the spectroscopic redshifts.
	
	The process of choosing correct luminosity priors, extinction laws and color excesses was very similar to S09.
	We used the same extinction law \citep{prevot1984} and same values for the extinction $E(B-V)$=(0,0.1,0.2 - 0.5) and luminosity priors as in S09, as these settings produced the most reliable results by trial and error. The luminosity priors used for each subset are the same as used in S09: an absolute magnitude range of $-24$~mag $<$ $M_g$ $<$ $-8$~mag for extended, and $-30$~mag $<$ $M_g$ $<$ $-20$~mag for (optically) point-like objects. We fit the SEDs across a redshift range of 0.03-7.0, in steps of 0.01.

\begin{deluxetable}{ll}[h]
	\tablewidth{0pt}
	\tablecaption{\label{table:library} \textsc{Library of Templates\tablenotemark{1} for Point-like\tablenotemark{2} and Extended Sources\tablenotemark{3}}}
	\tablehead{\colhead{\textsc{Point-like SED Templates}} & \colhead{\textsc{Extended SED Templates}}}
	\startdata
	1: I22491\_80\_pl\_TQSO1\_20 & 26: M82 \\
	2: I22491\_90\_pl\_TQSO1\_10 & 27: I22491\_10\_pl\_TQSO1\_90 \\
	3: pl\_QSOH\_template\_norm & 28: I22491\_20\_pl\_TQSO1\_80 \\
	4: pl\_TQSO1\_template\_norm & 29: I22491\_50\_pl\_TQSO1\_50 \\
	5: s250\_10\_pl\_TQSO1\_90 & 30: I22491\_60\_pl\_TQSO1\_40 \\
	6: s180\_30\_pl\_TQSO1\_70 & 31: I22491\_80\_pl\_TQSO1\_20 \\
	7: s800\_40\_pl\_TQSO1\_60 & 32: I22491 \\
	8: fdf4\_40\_pl\_TQSO1\_60 & 33: Ell2 \\
	9: s800\_20\_pl\_TQSO1\_80 & 34: Ell5 \\
	10: I22491\_30\_pl\_TQSO1\_70 & 35: Ell13 \\
	11: I22491\_20\_pl\_TQSO1\_80 & 36: S0 \\
	12: I22491\_10\_pl\_TQSO1\_90 & 37: Sa \\
	13: I22491\_50\_pl\_TQSO1\_50 & 38: Sb \\
	14: Spi4 & 39: Sc \\
	15: Sey2 & 40: Sdm \\
	16: S0\_10\_QSO2\_90 & \\
	17: S0\_20\_QSO2\_80 & \\
	18: S0\_30\_QSO2\_70 & \\
	19: S0\_40\_QSO2\_60 & \\
	20: S0\_50\_QSO2\_50 & \\
	21: S0\_60\_QSO2\_40 & \\
	22: S0\_70\_QSO2\_30 & \\
	23: S0\_80\_QSO2\_20 & \\
	24: S0\_90\_QSO2\_10 & \\
	25: Mrk231 & \\
	\enddata
	\tablenotetext{1}{S150, S180, S800, fdf4, M82, Mrk231, I22491 are starburst galaxies with different levels of star formation. TQSO1, QSO2 are type 1 AGN and type 2 AGN templates. QSOH is high luminosity quasar. Sey2 is a Seyfert 2 template. Spi4, S0, Sa, Sb are spiral galaxies and Ell2, Ell5, Ell13 are ellipticals. I22491\_80\_pl\_TQSO1\_20 is a 80\% I22491 and 20\% QSO1 template. These templetes are described in detail in \citet{Ilbert2009}, S09 and \citet{hsu2014}.}
	\tablenotetext{2}{Point Source templates include hybrids of galaxy, quasar and AGN spectra.}
	\tablenotetext{3}{Extended Source templates include  {elliptical, spiral and starburst galaxies, and starburst-QSO hybrid spectra.}}
\end{deluxetable}

	\begin{figure}[th]
	\centering
	\includegraphics[width=1.0\linewidth]{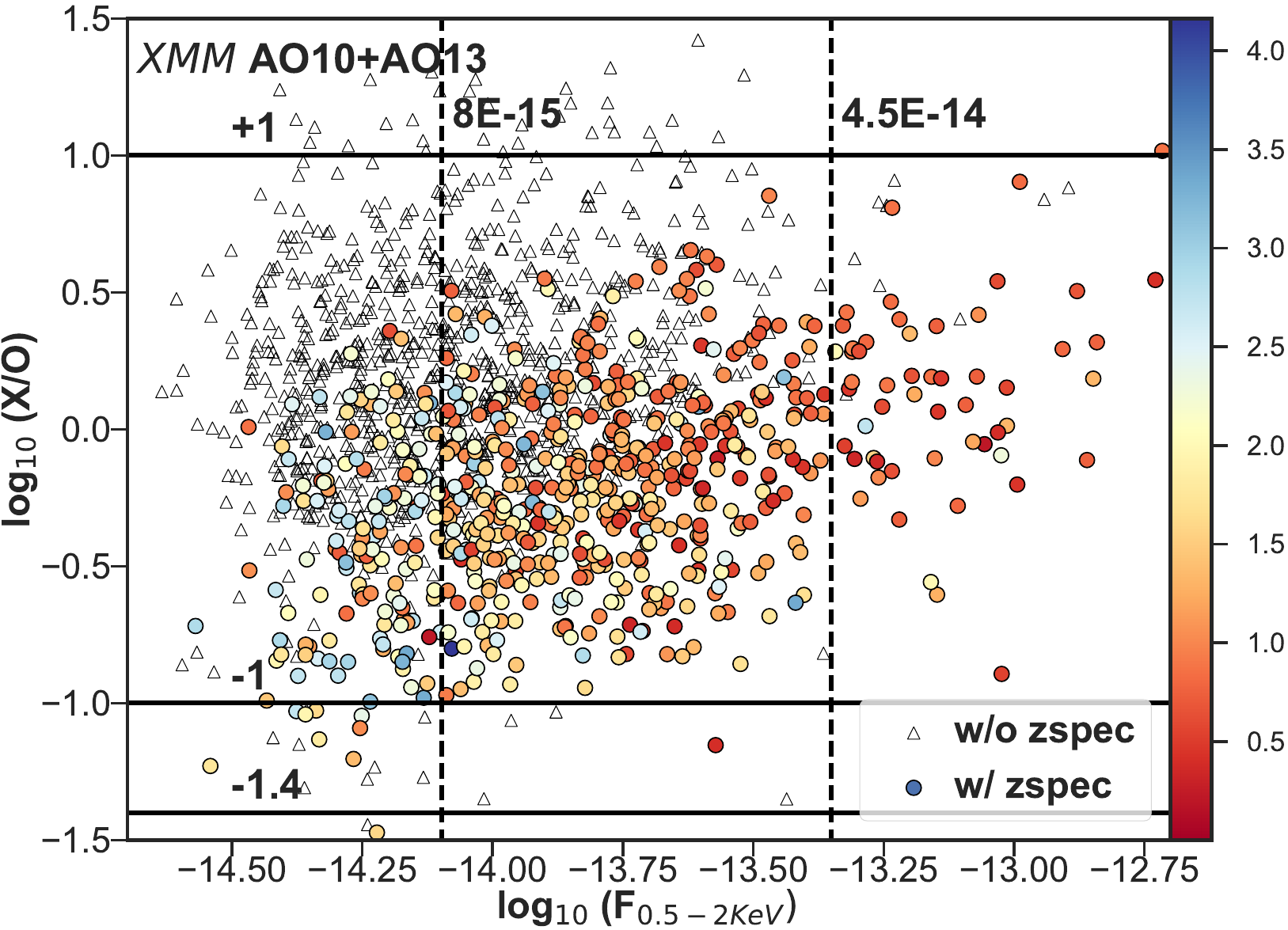}			
	\caption{Log of the X-ray-to-optical flux ratio versus the log of the soft X-ray flux
		{(0.5-2 keV band)} of all point-like objects in $XMM$ AO10 and AO13. 
		The ratio X/O is based on the soft X-ray flux and the SDSS $i$-band flux. The spectroscopic sample is color-coded by redshift. This plot shows 
		that the spectroscopic sample tends to be brighter than the non-spectroscopic sample, although they span the same range in flux and redshift.} 
	\label{fig:xo_plot_pointlike_ao10_ao13} 
\end{figure}

The best-fit photometric redshift is based on the minimum $\chi^2$ of the fit. Two example SED fits are shown in Figure~\ref{fig:SED_fit}. In the first case, the photometric redshift solution is unique, while in the second case there are two possible solutions at different redshifts, obtained with different templates. In addition to the photometric redshift ($z_{\rm phot}$), for each source, LePhare provides the redshift probability distribution function, $P(z)$. We define a quality metric, PDZ, as the probability that the true redshift is within $\pm 0.1(1+z_{phot})$ of $z_{\rm phot}$: 
\begin{equation}
\textit{ PDZ = 100 }\times \frac{\int_{z_{phot}-0.1(1+z_{phot})}^{z_{phot}+0.1(1+z_{phot})} P(z) dz}{\int_{0}^{\infty} P(z) dz}  ,
\end{equation}
\noindent
where  
\begin{equation}\label{eq:pz}
P(z) = \frac{\exp[-\frac{\chi^2min(z)}{2}]}{\exp[-\frac{\chi^2min(zbest)}{2}]} .
\end{equation}
{In LePhare, we set very low values of  MIN\_THRES (0.005) and DZ\_WIN (0.025) (see LePhare manual) so that secondary peaks in the redshift probability distributions could be identified.  Our output catalogs and SED fit plots report these secondary redshifts, as well as the $\chi^2_{sec}$ and PDZ$_{sec}$, as shown in Fig.~\ref{fig:SED_fit} (top right and bottom right panels).}


{The selection of template library and extinction laws was done using all the spectroscopy from SDSS DR13 and DR12. Afterwards, we added in more redshifts from DR12, DR10 and from additional surveys listed in \S~\ref{sec:intro} ($\sim$ 20\% of the whole spectroscopic sample). Therefore, we were treating the initial 80\% as our training set and the 20\% we added in later as a test set. The two subsets (initial data and added data) have similar outlier fractions and overall accuracy. After this paper was submitted, SDSS DR14 was released \citep{sdssdr14}. We present the results for this new spectroscopic sample in \S~\ref{sec:results_discussion}.}
	
	\subsection{Representative Set of Templates}\label{sec:library_selection}
	
	{As discussed in \S~\ref{sec:data_available},} our survey has larger volume and is shallower than fields for which photometric redshifts has previously been calculated. The difference between populations in this work and previous deeper photometric redshift works is summarized in Table~\ref{tab:comparison_object_type_other_fields}. 
	
	In particular, Stripe~82X tends to have more high-luminosity quasars and more AGN-dominated emission, {as evidenced by the higher fraction of high soft flux (flux in 0.5-2 keV band) sources. As a result, we could not use the template set optimized for the deeper fields; instead, we began with a large selection of templates from which we selected a smaller set best suited to Stripe~82X. In Appendix B we describe the process for selecting suitable templates for SED fitting, and we give an example of how we rejected a particular template.}
	
	We started with templates used by \citet{Ilbert2009}, S09, S11 and \citet{hsu2014}, for galaxies and AGN in deep, pencil-beam surveys (CDFS, ECDFS, {\it Chandra}-COSMOS, {\it XMM}-COSMOS). \citet{hsu2014} constructed hybrids combining AGN templates with semi-empirical star-forming galaxy templates from \citet{Noll2009}. {We tested each of these libraries on our spectroscopic sample, noted which templates were frequently providing us with best fits and fewest outliers, and by trial and error selected a small set of templates that minimized outliers fraction and $\sigma_{\rm nmad}$.}
	
	
	{In general, spectroscopy is only possible for brighter objects (see Figure~\ref{fig:xo_plot_pointlike_ao10_ao13}), and photometric redshifts are needed for the faintest sources in a sample. So there is always a possibility that the relatively bright population on which the photometric redshifts are calibrated is systematically different from the faint sources. We took this into account, by including templates that will fit the AGN-dominated sources (mostly Type 1 or starburst spectra) $and$ templates that will fit fainter objects not represented in the spectroscopic sample (Type 2 templates). Newly released SDSS DR14 \citep{sdssdr14} spectra made it possible to test our accuracy using a blind sample of objects fainter than our training set. The results of this test is discussed in \S~\ref{sec:results_discussion}.}
	
{The final list of templates are given in Table~\ref{table:library}. As we mentioned in \S~\ref{sec:data_available}, our template library has also been used to calculate photometric redshifts for $XMM$-XXL \citep{Georgakakis2017}, providing a better accuracy than the original library defined in S09. Thus, we believe this library would be more appropriate for the $eROSITA$ all-sky survey, which has a depth similar to Stripe~82X and $XMM$-XXL (see bottom panel of Figure~\ref{fig:all_xray_histo}). This library should also be considered when computing $z_{\rm phot}$ of AGN in wide surveys such as Dark Energy Survey (DES), Euclid and, more importantly, LSST.}
	 
	\subsection{{Classification as Point-like or Extended Optical/IR Sources}}\label{sec:pointlike_extended}
	
	S09, S11 demonstrated how morphological information, combined with a prior in absolute magnitude, improves the reliability of photometric redshifts for AGN. Ideally, as in COSMOS and other deep, pencil-beam surveys, the morphological information comes from Hubble Space Telescope (HST), without seeing effects deforming the images. For Stripe~82X, we get morphological information from ground-based SDSS images. Besides the lower spatial resolution of the images, we have to deal with the fact that when the seeing is poor, the stars used for defining the PSF are also smeared. This means sources that ought to be classified as extended are wrongly classified as point-like. A quantification of this effect was given in \citet{hsu2014}.
	
	As mentioned in \S~\ref{sec:assoc}, FT16 created deep stacked images from which low quality data were removed, improving on the overall photometry and morphological classification. We determined that the FT16 classification as point-like or extended is the most reliable compared to other available morphological information, because it gave the smallest number of outliers in the comparison of photometric and spectroscopic redshifts. Using the J14 morphological information, our results had around 25\% outliers, whereas using FT16 helped us reduce that to around 14\%.

\begin{figure*}[th]
		\centering
		\includegraphics[width=0.7\linewidth]{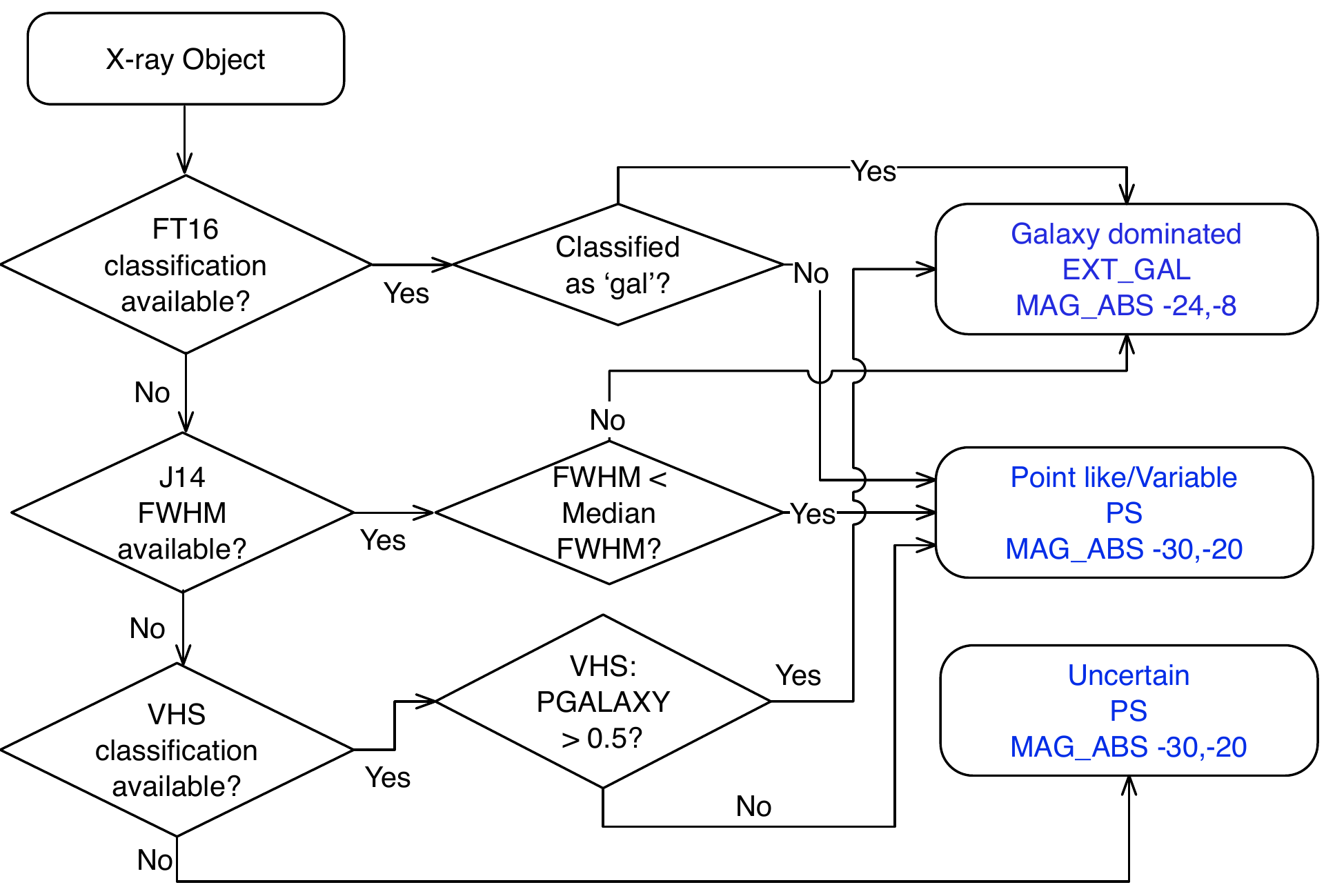}
		\caption{Procedure used to classify AGN as point-like or extended, based on optical and NIR morphology. Based on this classification, different {luminosity} priors and template libraries are applied on these objects to calculate photometric redshifts.}
		\label{fig:final_classification}
	\end{figure*}

	We summarize the classification process in Fig.~\ref{fig:final_classification}.
	When FT16 classification was not available (18\% of objects), 
	we used the information from J14, 
	ideally in the $r$ band (for 6\% of the objects); failing that, we used the information from the $i$, $z$, $g$, and $u$ bands, in that order (4\%). 
	If no optical information was available, we used the NIR morphology from VHS data (2\%); specifically, if PGAL$> 0.5$ \citep{vhs}, we consider the object extended, otherwise we consider it point-like. 
	
	Objects that are not extended in SDSS or VHS data are grouped with the point-like objects but flagged as uncertain. In the end, 40\% (2486) 
	of the Stripe~82 X-ray sources are clearly extended, 53\% (3289) 
	are clearly point-like, and  5\% (269) 
	are tentatively point-like but flagged uncertain. The remaining 2\% (137) do not have a multi-wavelength counterpart. 
	
	We indicate the final morphology assigned to a source using ``classification" column in the final output table, as described in Appendix A. Once objects have been classified as point-like or extended, we apply a different set of templates (summarized in Table~\ref{table:library}) and priors to these two morphological classes to calculate photometric redshifts.
	
	For 1156 
	objects with saturated photometry or bright nearby neighbors ($\sim16$\%),
	we calculate photometric redshifts but flag them as uncertain. We discuss the accuracy and outlier fractions of objects with and without bright nearby neighbors separately, as the photometry of the former is not reliable.
	
	The comparison of photometric and spectroscopic redshifts for point-like and extended AGN in the $XMM$ AO10 and AO13 samples (to differentiate from the $XMM$ archival sample) is shown in Fig.~\ref{fig:zspec_zphot} in \S~\ref{sec:results_discussion}. 
	A census of all the sources is given in Table~\ref{table_classif}. 
	
 	\begin{deluxetable*}{lcccc}[th]
		\tablewidth{0pt}
		\tablecaption{\label{table_classif} \textsc{Summary of Available Photometric Redshifts, by Sub-sample, for the 6181 Stripe~82X Sources.\tablenotemark{1}}}
		\tablehead{\colhead{\textsc{Type}} & \colhead{\textsc{$XMM$ AO10 + AO13}} & \colhead{\textsc{Archival $XMM$}} & \colhead{\textsc{$Chandra$}} & \colhead{\textsc{Total}}}
		\startdata
		Total & 3613 & 1607 & 1146 & 6366 \\
		After removing duplicates & 3541 & 1496 &  1144 & 6181 \\
		Objects with associations & 3516 (99\%) & 1450 (97\%) & 1061 (93\%) & 6027 (98\%) \\ 
		All point-like & 1976 (56\%) & 757 (51\%) & 546 (48\%) & 3279 (53\%) \\
		With $z_{spec}$ &  793 (23\%) & 292 (20\%) & 246 (22\%) & 1331 (22\%) \\
		All extended & 1404 (40\%) & 641 (43\%) & 434 (38\%) & 2479 (40\%) \\
		With $z_{spec}$ & 288 (8\%) & 138 (9\%) & 108 (9\%) & 534 (9\%) \\
		Cannot classify\tablenotemark{2} & 136 (4\%) & 52 (3\%) & 81 (7\%) &  269 (5\%)\\
		No associations & 25 (1\%) & 46 (3\%) & 83 (7\%) &  154 (2\%)\\
		Reporting $z_{\rm phot}$ for\tablenotemark{3} & 3501 (100\%) & 1418 (95\%) & 1025 (91\%) &  5944 (96\%) \\
		\enddata
		\tablenotetext{1}{All percentages relative to total number of non-duplicate sources, and rounded to nearest integer.}
		\tablenotetext{2}{There are multi-wavelength associations in at least one band for these sources, but we do not have enough SDSS or VHS information for proper classification. We calculate photometric redshifts for these sources assuming they are point-like.}
		\tablenotetext{3}{Number of X-ray sources for which we are reporting photometric redshifts. Photometric redshifts calculated with less than 2 bands of data should not be considered reliable. To assist the user, we provide the ``num\_filt" column in our final catalog with the number of bands of data available for a source.}
	\end{deluxetable*}
	
	\subsection{{Photometric Inhomogeneities Among Wavebands}}\label{sec:systematicshifts}
	
	The success of photometric redshifts, especially for AGN, is determined to a large extent by the homogeneity of the optical and infrared photometry from different surveys. While photometry in pencil-beam areas is relatively homogeneous - e.g., CANDELS fields \citep{hsu2014} the entire COSMOS (S09, S11), AEGIS surveys \citep{nandra2015} - the situation is worse for wide field areas. 
	
	In particular, Stripe~82X has NIR photometry reported for different apertures, in images of different resolution 
	i.e., sampling different amounts of the host galaxies. 
	SpIES and SHELA provide aperture and SEXTRACTOR \citep{sextractor} AUTO fluxes (a Kron-like flux) for {\it Spitzer} IRAC sources, while the VHS catalog reports Kron, PSF and Petrosian magnitudes. The photometry reported in {\it GALEX} also provides AUTO magnitude, whereas the {\it AllWISE} catalog provides a total flux (though this is affected by source blending).
	Optical photometry is available for different aperture sizes, as well as Petrosian and AUTO magnitude. 
	
	Under these circumstances, computing and correcting for systematic zero-point effects (e.g., \citealp{Ilbert2009}, \citealp{hsu2014}, S09) is at best difficult and at minimum unreliable.
	{For this reason, we came up with a compromise approach: instead of applying systematic shifts, we tested different types of photometry,} in order to find the combination that provides the highest accuracy and the smallest fraction of outliers. 
	
	Accordingly, after some experimentation, we decided to use AUTO magnitudes (where available) for both extended and point-like sources. Only for the NIR, we decided on 2.8$^{\prime\prime}$ radius aperture magnitude for point-like sources and 5.7$^{\prime\prime}$ radius aperture magnitude for extended sources after testing several aperture sizes for each sample. {For {\it AllWISE}, we provide $mpro$ magnitudes in all four bands, although we only use $w1mpro$ and $w2mpro$ to construct SEDs}.
	
	\subsection{Improving Accuracy Using UKIDSS and AllWISE}\label{ssec:ukidss_wise}
	In \S~\ref{ssec:sed_construction}, we noted that we tried including UKIDSS and {\it AllWISE} photometry in the fitted SEDs even though VHS/IRAC bands cover the same 
	wavebands
with slightly better accuracy, because we wanted to determine whether having additional bands of data over the same range of wavelengths lead to better photometric redshifts. We found that excluding UKIDSS and {\it AllWISE} in cases where we have VHS/IRAC marginally improves results by lowering outliers by $\sim 1$ \% in the overall sample, and also leads to fewer cases where we are completely unable to compute photometric redshifts. 

Therefore, we do not use UKIDSS photometry if we have VHS data available. Similarly, we ignore $AllWISE$ data if we have IRAC data available. Other than these exceptions, we use all available bands from FUV to W2/CH2 for SED fitting. We report {\it AllWISE} W3 and W4 magnitudes in the final catalog but do not use it for z$_{\rm phot}$ calculation. Note that we do use UKIDSS and/or {\it AllWISE} bands to compute photometric redshifts for objects that do not have VHS/IRAC data.
	
	\subsection{{Accuracy of spectroscopic redshifts}}\label{sec:redshift_accuracy}
	The accuracy of photometric redshifts is based on a sample of secure spectroscopic redshifts. In practice, the fitting procedure, classifications, templates --- all the aspects of determining photometric redshifts --- are adjusted to produce the most accurate results with respect to spectroscopic sample.
	Therefore, to ensure the most accurate results, we excluded any spectroscopic redshifts flagged as uncertain.

	SDSS DR12Q \citep{paris2016} presents quasar redshifts that have been visually inspected and are therefore reliable. For cases where we do not find any redshifts from DR12Q, we look for redshifts in SDSS DR13 \citep{dr13_2016}. For DR13, no public visual inspection information is yet available, and the pipeline sometimes returns multiple redshifts for the same object. This occurs because the reliability of a redshift is dependent upon the signal-to-noise ratio of the spectrum (discussed in detail in \citep{menzel2016}) and the nature of the source. 
	
	We found that for 300 cases, the same object was associated with more than one redshift in DR13, with no warning flag (zwarning=0). Among these, 289 sources have multiple redshifts that agree down to 2 decimal places. A difference on the 3rd decimal place corresponds to a velocity of about 1000 km/s, which is below our photometric redshift sensitivity of 0.01 (due to limited number of bands). For these cases, we select the redshift with lowest error. For the 11 cases where the difference is significant, or two different object types (AGN/star) are identified by two different spectra, we proceeded to a visual inspection and determined the correct redshift. {We made use of \citep{dwelly2017} to visually inspect all DR13 spectra reported in this work.}
	
	SDSS DR10, DR12Q, DR13 provide redshifts for 1577 cases, and additional spectroscopy (sources listed in output table and in LM16) provides redshifts for another 288 sources; the latter have also all been visually inspected. 
	

	\section{{Results and Discussion}}\label{sec:results_discussion}
	
	\begin{figure*}[th]
		\centering
		\includegraphics[width=0.5\linewidth]{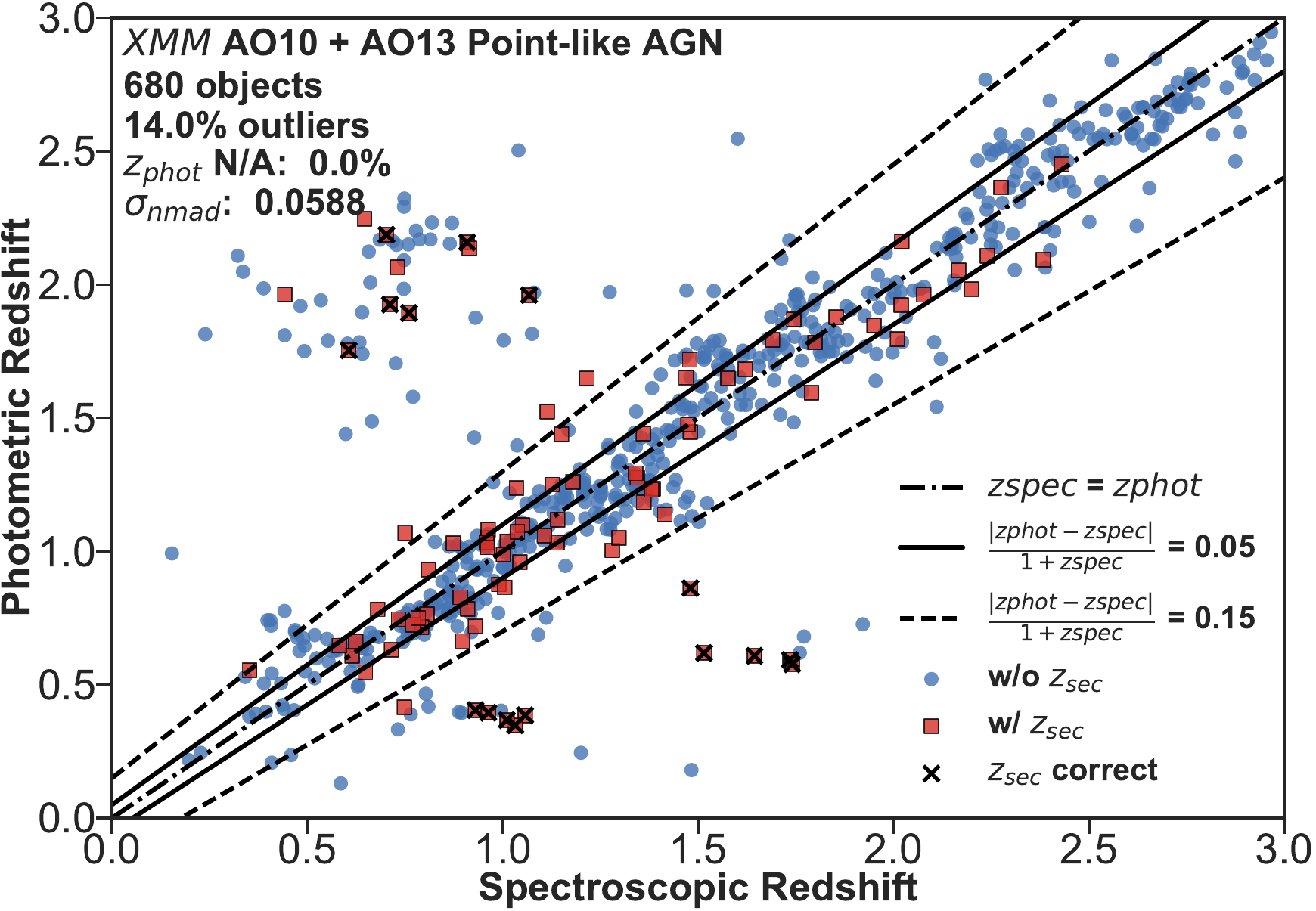}~
		\includegraphics[width=0.5\linewidth]{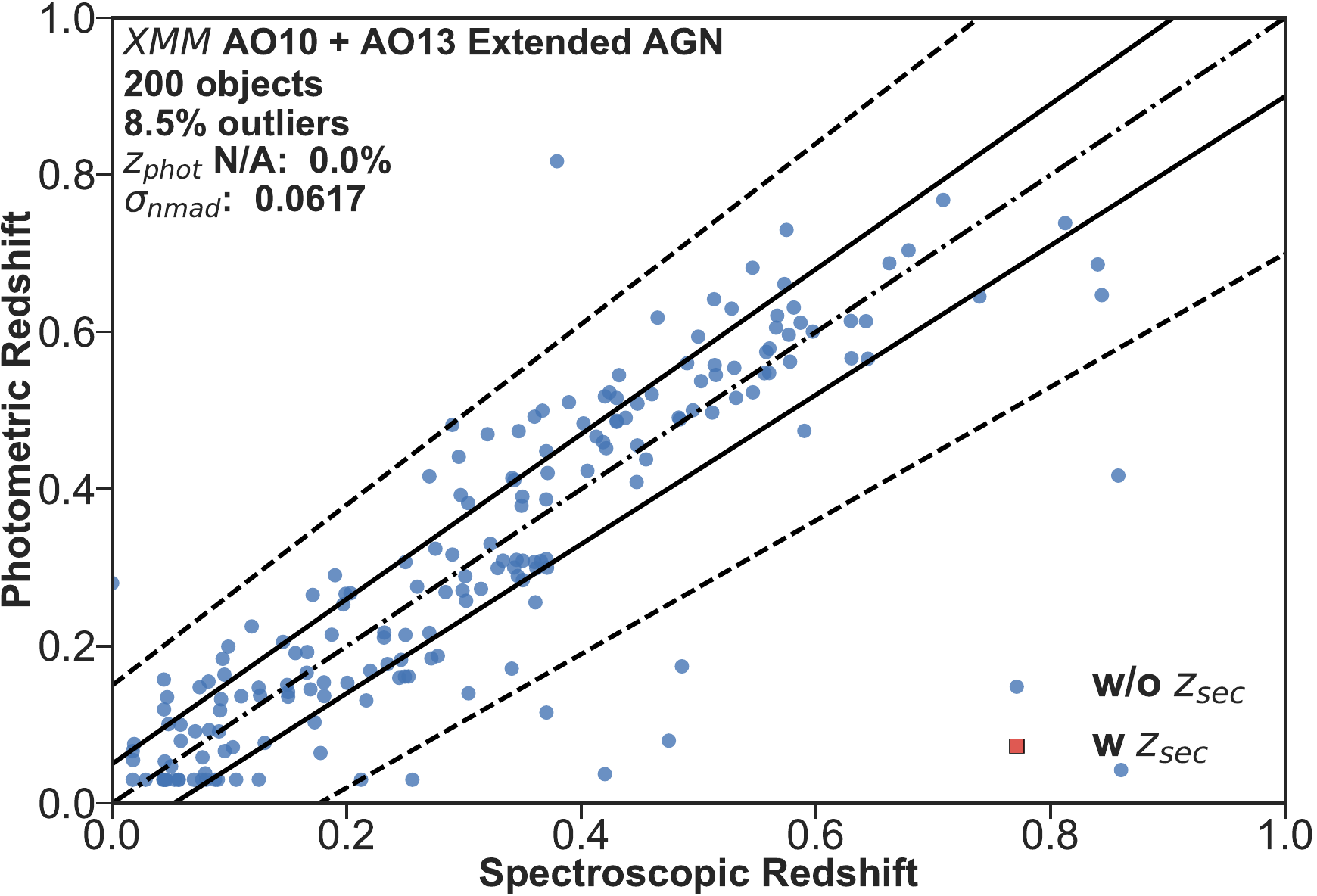}
		\caption{Spectroscopic versus photometric redshift for point-like ({\it left}) and extended ({\it right}) sources in the uniform $XMM$-Newton fields. The results in the other fields are very similar, as can be seen in Table~\ref{tab:overall_results}. Black crosses indicate cases where the primary redshift is incorrect and the secondary redshift is correct (i.e., it lies within dashed black lines). {The quantity ``zphot N/A'' reports the percentage of sources for which we are unable to calculate photometric redshifts altogether due to data in very few bands, or not having any an appropriate SED template in our template library.}} 
		\label{fig:zspec_zphot}
	\end{figure*}
	
	We calculated photometric redshifts following the methods explained in (\S~\ref{sed_fitting}) for 96\% of the Stripe~82X X-ray sources, and achieved an overall accuracy ($\sigma_{\rm nmad}$) of 0.06 and an outlier fraction ($\eta$) of 13.69\%. The $\sigma_{\rm nmad}$ and $\eta$ of each subset of the data are given in Table~\ref{tab:overall_results}. These results do not take into account objects with bright nearby neighbors. As photometry for such cases is less reliable due to contamination, their $\eta$ and $\sigma_{\rm nmad}$ is calculated separately. There are 352 spectroscopic objects with bright nearby neighbors, and this sample has $\eta$ = 19.32\% and $\sigma_{\rm nmad}$ = 0.0882. 
	
	The population that we are probing is brighter in X-ray than populations in fields such as CDFS, and so typical AGN in our population is composed of a power-law plus emission lines. Objects that have a significant galaxy component tends to have emission lines, and with narrow or intermediate band photometry these lines would be identifiable, making their photometric redshifts more accurate. However, with broad-band photometry, the correct model is harder to pinpoint, which leads to a higher $\eta$ and $\sigma_{\rm nmad}$.	

	Similar to previous AGN photometric redshift works (S09, \citealp{hsu2014}), we find more accurate redshifts for point-like objects, but with a higher percentage of outliers, than for extended objects. As in \citet{hsu2014}, most of the outliers in the point-like samples have an over-estimated photometric redshift. The problem is related to the wrong morphological classification, which is done in this work using ground-based images. Figure~\ref{fig:zspec_zphot} shows a plot of $z_{\rm phot}$ against $z_{spec}$. The plot on the left is for point-like objects, where a fraction (18\%) of the outliers have secondary photometric redshifts that are very close to the spectroscopic redshifts (both primary and secondary photometric redshifts are given in Appendix A).
	
	\begin{deluxetable*}{lccccc}[th]
		\tablewidth{0pt}
		\tablecaption{\label{tab:overall_results} \textsc{Photometric Redshift Outlier Fractions and Accuracy for All Three Fields.\tablenotemark{1}}}
		\tablehead{\colhead{\textsc{Object type}} & & \colhead{\textsc{$XMM$ AO10 + AO13}} & \colhead{\textsc{Archival $XMM$}} & \colhead{\textsc{$Chandra$}} & \colhead{\textsc{Total}}}
		\startdata
		Point-like AGN & Total number with $z_{spec}$ & 797 & 292 & 246 &  1335\tablenotemark{2} \\
		& W/o bright nearby neighbor & 680 & 247 & 208 & 1135 \\ 
		&  Outlier &  14.0\%  & 14.6\% &  19.7\%  &  15.18\%  \\ 
		&  $\sigma_{\rm nmad}$ &  0.0588 & 0.0605 &  0.0573 &   0.0589 \\ 
		Extended & Total number with $z_{spec}$ & 288  & 139  & 109 & 536 \\
		&  W/o bright nearby neighbor & 200 & 94 & 79 & 373 \\ 
		&  Outlier &  8.5\%  &  9.6\%  &  10.3\%  & 9.14\%   \\
		& $\sigma_{\rm nmad}$ & 0.0617 & 0.0618 & 0.0733 &  0.064 \\	
		Overall & Total number with $z_{spec}$ & 1081  & 430  & 354 & 1865 \\
		&  W/o bright nearby neighbor & 878 & 341 & 286 & 1505 \\ 
		&  Outlier &  12.76\%  &  13.20\%  &  17.13\%  & 13.69\%   \\
		& $\sigma_{\rm nmad}$ & 0.0595 & 0.0609 & 0.0617 &  0.0602 \\
		\enddata
		\tablenotetext{1}{The outlier percentage and accuracy values reported here only apply to objects without SDSS photometry flags indicating contamination due to bright nearby neighbors.}
		\tablenotetext{2}{This number includes stars, which we eliminate using the method described in \S~\ref{section_star_id}.}
	\end{deluxetable*}

\begin{figure}[th]
		\centering
		\includegraphics[width=1.0\linewidth]{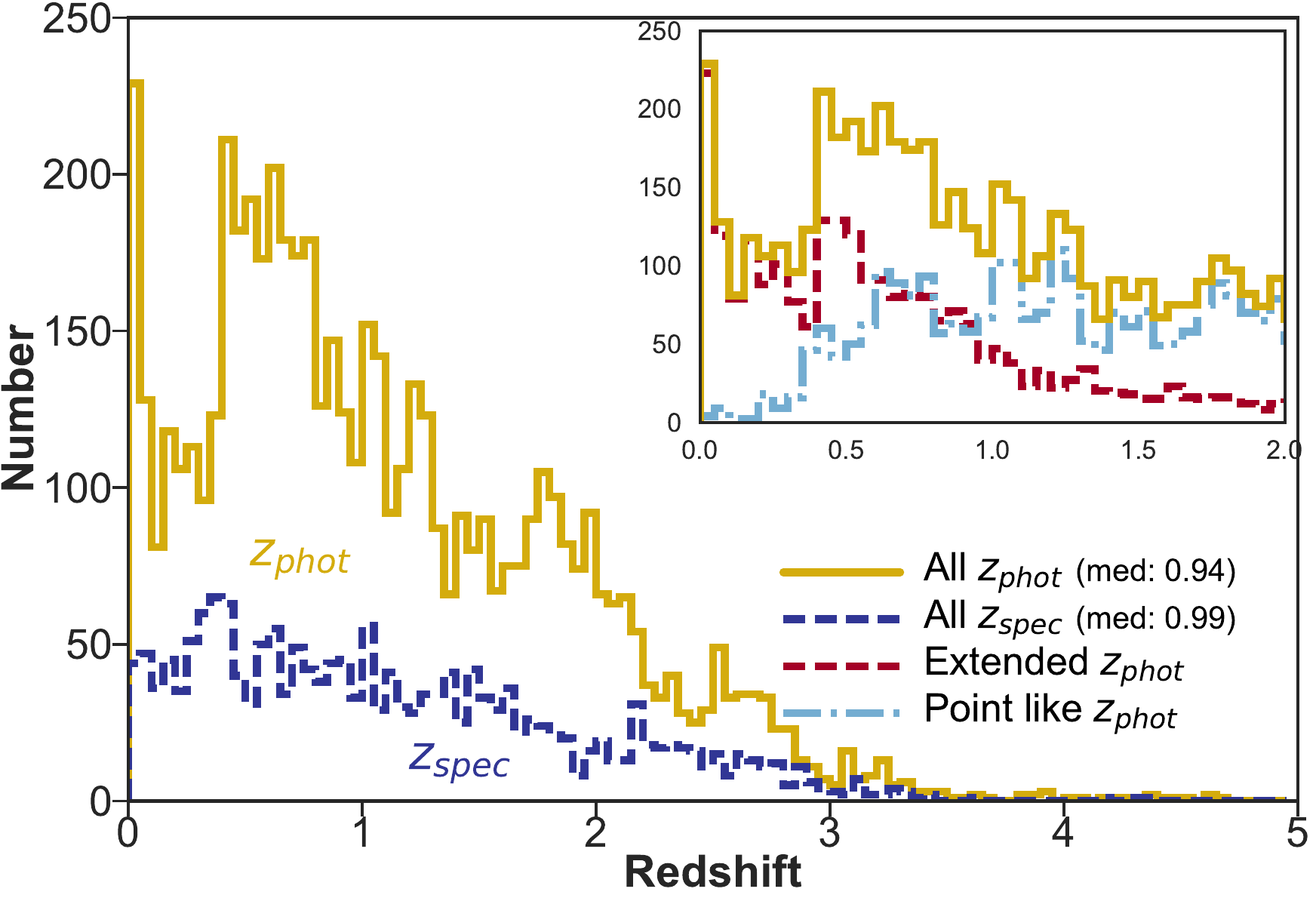}

		\caption{Best fit photometric ({\it gold}) and spectroscopic ({\it blue}) redshift distributions for all 5716 sources with good photometry. The overdensity at z$\sim$0.75 in the distribution of photometric redshifts is not echoed in the spectroscopic distribution, suggesting fainter sources are more clustered at $z\sim0.7$ than bright sources. The inset shows that the populations of extended sources and point sources cross at this approximate redshift; we checked that point sources were not mistakenly identified as extended (see text for details).}
		\label{fig:redshift_distribution}
	\end{figure}

	The redshift distribution for our objects are given in Figure~\ref{fig:redshift_distribution}. Spectroscopic and photometric redshifts lie within the same range, but the photometric redshift distribution show an overdensity at z$\sim$0.75 (between 0.5 and 1). We tested whether this peak could be due to a systematic effect in our computation process. In particular, we tested the impact of switching priors and libraries on the two subsamples (extended and point-like). 
	
	We found that the extended sources work equally well with the point-like sample's priors and libraries, but not vice-versa. This is because the point-like sources' library includes AGN-galaxy hybrids which also apply to extended objects. Additionally, extended sources host an AGN with absolute magnitudes that are within the range allowed by the point-like sample's prior\footnote{Note that there cannot be many extended non-active galaxies in our sample because the template fits would have failed.}.
	Point-like sources are not fitted well by the extended sample's templates. The high absolute magnitude priors for extended sources force LePhare to assign the lowest possible redshift ($z\sim0.03$) to a significant fraction (25\%) of the point-like objects whose photometric redshifts lie in the overdensity region. 
	Thus the overdensity is not due to systematic effects from the SED fitting.
	
	Overdensities at redshift $\sim$ 0.8 are a common feature in all photometric redshift distributions for deep surveys (e.g. \citealp{Gilli2004,Barger2003,Szokoly2004}) and similar overdensity has been seen for star-forming galaxies and emission-line galaxies \citep{sobral2013,favole2016,Iovino2016}.

	In the context of the LCDM cosmological model, characterized by the observationally determined values for cosmological parameters like $\Lambda$ and $H_0$, z $\sim$ 0.8 is the redshift at which the Universe switches from being matter-dominated to dark energy dominated. At this epoch, the Hubble distance (which is a function of cosmological parameters) changes slope \citep{inflectionpoint2015}. 
	
	Another potential explanation is that the competition between the increase in accessible volume with redshift and the decreasing number of detectable objects crosses over at this cosmic epoch leaving the imprint of a peak in the redshift distribution. This makes sense if the objects with spectroscopic redshifts are luminous enough that none have been excluded by the flux limit. And indeed, it also marks the fact that all the surveys considered here are definitely complete out to this redshift.
	
	\begin{deluxetable*}{lcccccccc}[th]
	\tablewidth{0pt}
	\tablecaption{\label{compare_fields} \textsc{Comparison of Photometric Redshifts for X-ray Surveys with Different Depths (of X-ray and Multi-Wavelength Data) and Number of Photometric Bands.}}
	\tablehead{\colhead{\textsc{Field}} & \colhead{\textsc{Ref}} & \colhead{\textsc{Area}} & \colhead{\textsc{Number of Sources}} & \colhead{\textsc{Mean Exp time}} & \colhead{\textsc{Soft flux limit}} & \colhead{\textsc{Bandwidth}} & \colhead{\textsc{Accuracy}} & \colhead{\textsc{Outliers}}}
	\startdata
	& & (deg$^2$) & & (ks) & ($F_{\rm 0.5-2~keV}$) & N/I/B\tablenotemark{1} & ($\sigma_{\rm nmad}$) &  $\eta$ (\%) \\
	S82 Archival $Chandra$ &  & 6 & 1146 & 5.7-71.2 &  $8 \times 10^{-16}$ & B & 0.0627 & 17.1\\				
	S82 Archival $XMM$ & & 7.4 & 1607 &  17-65 & $7 \times 10^{-17}$ & B & 0.0609 & 13.2 \\
	S82 $XMM$ AO10+AO13 & & 20.2 & 3613 & 4-8 & $2 \times 10^{-15}$ & B & 0.0595 & 12.8 \\
	Lockman Hole & \citet{fotopoulou2012} & 0.20 & 388\tablenotemark{2} & 185 & $1.9 \times 10^{-16}$ & B & 0.069 & 18.3 \\ 
	CDFS & \citet{hsu2014} & 0.13 & 740 &  ~2000 & $9.1 \times 10^{-18}$ & I & 0.014 & 6.73\\  				
	ECDFS & \citet{hsu2014} & 0.3 & 762 & 250 & $1.1 \times 10^{-16}$ & I & 0.016 & 10.14 \\ 				
	AEGIS-XD & \citet{nandra2015} \tablenotemark{3} & 0.29 & 1325 & 200 & $5.3 \times 10^{-17}$ & N/B & 0.022 & 2.8\\
	$Chandra$-COSMOS & \citet{Marchesi2016} & 0.90 & 4016 & 200 & $1.9 \times 10^{-16}$ & N/I/B & 0.015 & 6 \\ 
	$XMM$-COSMOS & S09 & 2.13 & 1887 & 60 & $1.7 \times 10^{-15}$ & N/I/B & 0.015 & 6.3 \tablenotemark{4} \\
	\enddata
	\tablenotetext{1}{N, I and B stand for narrow-, intermediate- and broad-band, respectively.}
	\tablenotetext{2}{for X-ray detected sources (3.9$\sigma$ significance), which are more similar to our sample}
	\tablenotetext{3}{this paper only includes photometric redshifts for point sources, but we report total number of sources in this table}
	\tablenotetext{4}{for the QSOV sample}
	\end{deluxetable*}
	
	{Since this paper was submitted, SDSS DR14 spectra were made public \citep{sdssdr14}, including 810 Stripe~82X X-ray sources that did not previously have spectroscopic redshifts. This allowed a new test of the photometric redshifts for a population of fainter sources (the median r-band magnitude of this blind sample is two magnitudes fainter than our training sample), with systematically lower spectroscopic signal-to-noise ratio. We visually inspected all DR14 spectra reported in this work using \citep{dwelly2017}.} 
	{For these 810 redshifts, we obtained $\sigma_{\rm nmad}$ = 0.0626, essentially the same as for the training and test data sets, and 19\% outliers, which is higher than the 13.7\% for the training set.}
	
	\subsection{{Variability}}
	
	AGN vary on time scales of hours to years. 
	In Stripe~82, the SDSS survey includes 80-100 repeated images over a period of three years, so it is sensitive to variability on time scales of months. \citet{variability} used color cuts and morphology (point-like) from SDSS DR12 to select quasar-like objects from Stripe~82, and  calculate their strength of variability using a neural network code. As all of our objects are not necessarily quasars, only 3052 of our X-ray objects (49\%) overlap with their sample. 
	
	The neural network output parameter from \citet{variability}, $nnv$, which reflects the strength of quasar-like variability, varies between 0 and 1. (A value of $-1$ indicates that not enough epochs of data were available to draw reliable conclusions.) Most of these sources --- 2432 objects, which is 39\% of our sample --- have $nnv > 0.5$, meaning highly variable. Of these, 204 
	have extended morphology and are probably nearby Type 1 AGN, while 2228 
	have point-like morphologies and so could be variable stars or quasars. Unlike S09, we divided variable sources according to their morphology. For the extended variable sources, we have spectra for 45 sources, and we achieved an outlier fraction of 29\% and $\sigma_{\rm nmad}$ = 0.0669. For the point-like variable objects, we have spectra for 1013 sources, and we get 14.9\% outliers and $\sigma_{\rm nmad}$ = 0.0561. 
	
	{AGN variability can have an effect on the photometric redshifts, particularly when data at different wavelengths were obtained at different times. However, for the sources dominated by stellar light (3387 sources; 54.8\% overall, of which 1044 objects, or 16.9\%, are fainter than $r\gtrsim22.5$~mag), AGN variablity is unlikely to affect the photometric redshifts. For the most luminous AGN (2329 objects; 37.7\%), variability can affect the shape of the SED, but most of these (1216 objects, 19.7\%) have spectroscopic redshifts. So at most AGN variability could affect up to 1113 objects (18\% of the sample); of these, 
	471 (7.6\%) have  $nnv>$0.5 and 188 (3\%) do not have sufficient information to calculate $nnv$. These sources can be easily identified in the final catalog.} 
	
	\subsection{Comparison With Other Fields Using This Approach For Photometric Redshifts}
	
	Table~\ref{compare_fields} compares the quality of our photometric redshifts with those in other X-ray surveys. 
	Not surprisingly, the surveys with intermediate- or narrow-band photometry generally have more accurate photometric redshifts and fewer outliers, because narrow-band photometry can easily detect the emission lines of the galaxy component. 
	In addition, narrow- and intermediate-band photometry fill the gaps between the broad-band filters. 
	This allows the detection of characteristic features like the 4000$\AA$ and Ly-$\alpha$ breaks that can be missed in broad-band photometry.
	
	Our results are most similar to the \citet{fotopoulou2012} results for the Lockman Hole, where only a limited number of broad-band filters were used. Our results are slightly better due to a template library that is better optimized for bright AGN. It is worth noting that, contrary to other surveys, the published photometry for Stripe~82X was not analyzed homogeneously, as explained in \S~\ref{sec:systematicshifts}; reprocessing the data uniformly across multi-wavelength catalogs---in particular, defining consistent total magnitudes---would likely reduce the photometric uncertainties and thus improve the accuracy of the photometric redshifts.  
	
	\subsection{{Soft X-ray flux and photometric redshift accuracy}}\label{sec:soft_flux_and_accuracy}
	
	This is the first photometric redshift study of a large, bright, X-ray-selected AGN sample. In particular, this is the largest investigation of AGN-appropriate templates for sources with $F_{\rm 0.5-2~keV} >$ 10$^{-14}$ ergs/cm$^2$ (38\%, 2426 objects). The well-studied pencil-beam surveys have far fewer objects at these bright fluxes; for example, CDFS has 7 sources (1\%) and AEGIS has 22 sources (2\%) brighter than this limit. Even $XMM$-COSMOS, with 2~deg$^2$ of coverage, has only 221 (6\%) sources of this type. 
	We need to quantify the fitting quality at these bright fluxes because a point-like object with high soft flux is most likely a quasar with a nearly perfect power-law spectrum. Such SEDs lack features such as the 4000~\AA~break, which are important for template fitting. As such, we expect that our results should deteriorate with increasing soft X-ray flux. 
	
\begin{figure}[th]
		\centering
		\includegraphics[width=0.5\linewidth]{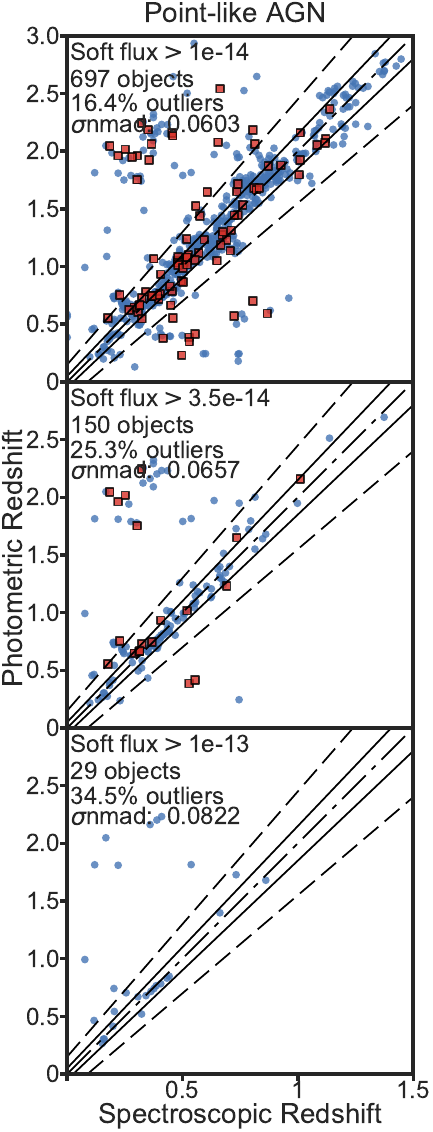}~
		\includegraphics[width=0.47\linewidth]{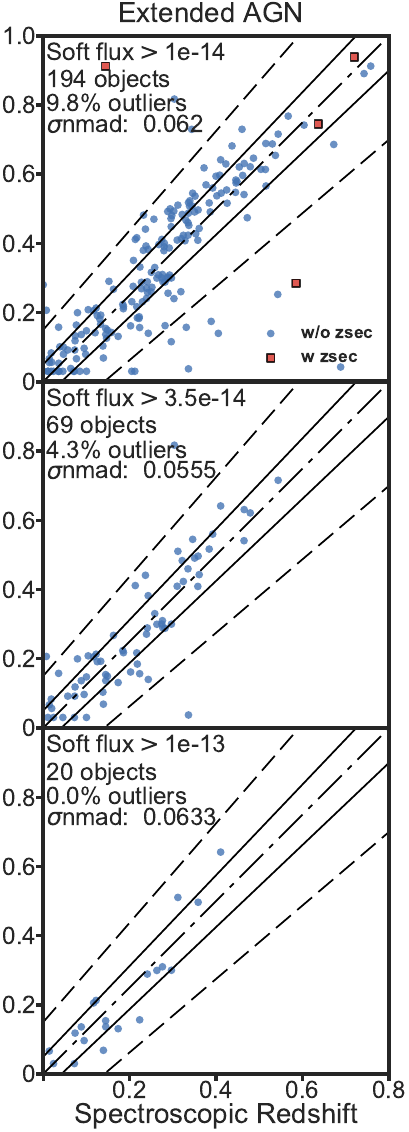}
		\caption{Difference in accuracy and outlier numbers for spectroscopic samples selected by limits on soft flux (0.5-2 keV). {These values are also summarized in Table~\ref{table:soft_flux_cuts}.}} 
		\label{fig:soft_flux_cuts}
	\end{figure}

	We tested the accuracy of photometric redshifts for three X-ray bins brighter than this limit, as shown in Figure~\ref{fig:soft_flux_cuts} and summarized in Table \ref{table:soft_flux_cuts}. For point-like objects, the fraction of outliers and $\sigma_{\rm nmad}$ increases for higher soft flux samples, while the opposite happens for extended samples. This is likely because the X-ray-brightest extended objects are nearby, and their SEDs are dominated by host galaxy light, whereas the brightest point-like objects are quasars dominated by pure power laws. 
	Note that it is the outlier fraction that sees the biggest change with flux limit; the accuracy of the photometric redshifts changes much less because it is the median value.
	
	\begin{deluxetable*}{lccc}[th]
		\tablewidth{0pt}
		\tablecaption{\label{table:soft_flux_cuts} \textsc{Outlier Fractions for Different Soft Flux Cuts\tablenotemark{1}.}}
		\tablehead{\colhead{\textsc{Object Type}} & \colhead{\textsc{$F_{0.5-2~keV} >$ $10^{-14}$ ergs/cm$^2$ s}} & \colhead{\textsc{$F_{0.5-2~keV} >$ 3.5$\times10^{-14}$ ergs/cm$^2$ s}}  & \colhead{\textsc{$F_{0.5-2~keV} >$10$^{-13}$ ergs/cm$^2$ s}}}
		\startdata
		{\bf Point-like} & 16.4\% outlier & 25.5\% outlier & 34.3\% outlier \\
		& $\sigma_{\rm nmad}$ = 0.0605  & $\sigma_{\rm nmad}$ = 0.0665 & $\sigma_{\rm nmad}$ = 0.0822  \\
		& 695 objects & 149 objects & 29 objects \\			
		{\bf Extended} & 9.8\% outlier  & 4.3\% outlier  & 0\% outlier\\	
		& $\sigma_{\rm nmad}$ = 0.0619 &  $\sigma_{\rm nmad}$ = 0.0555  & $\sigma_{\rm nmad}$ = 0.0633  \\
		& 193 objects & 69 objects &  20 objects \\
		\enddata 
		\tablenotetext{1}{Only for objects without bright nearby neighbors.}
	\end{deluxetable*}
	
	\subsection{Comparisons with Other Methods Of Calculating Photometric Redshift}\label{subsec:comparisons}

	{Since Stripe~82 is fully covered by SDSS, others have calculated photometric redshifts for subsets of optically or spectroscopically selected quasars using a variety of techniques. Here we compare results for the Stripe~82X sources that are in common with those studies, as summarized in Table~\ref{table:photoz_compare}. The samples that exclusively select quasars from optical data are expected to (and in some cases do) perform better than our approach. This is because of the difference in population type: our sample includes Type 1 and Type 2 AGN, starburst galaxies, some spiral and elliptical galaxies as well. The diversity in object type requires a large number of models in our template library, which leads to degeneracy and increases our outlier numbers and inaccuracy.}
	
	\citet{xdqso_photoz_2012} (hereafter XDQSO model) uses spectroscopically confirmed quasars from SDSS DR7, with $z>0.3$, to model color-redshift relationship. The XDQSO model uses this color-redshift relationship, along with an apparent-magnitude-dependent redshift prior to calculate a probability distribution of photometric redshift for each object. 
	They find accurate photometric redshifts ($\vert \triangle z \vert < 0.3$) for 97\% of the quasars (we do not have their spectroscopic sample available to calculate $\eta$ and $\sigma_{\rm nmad}$). There are 945 objects in Stripe~82X sample which overlap with the XDQSO work, and also have spectra, so we compared our results. 
	
	\citet{peters2015} used a Bayesian technique to select quasar candidates from SDSS and Stripe~82 and used the method described in \citet{weinstein2004} to find photometric redshifts for these objects. This work also optically selects quasars only. This approach uses astrometric parameters and optical photometry and reaches an accuracy of ($\vert \triangle z \vert < 0.3$) for 76.9\% of the quasars. Using the spectroscopic sample of that work, this translates to $\eta$ = 23.3\% and $\sigma_{\rm nmad}$ = 0.035. 
	
	{\citet{richards2015} uses the same approach as \citet{peters2015}, but with more bands (adding $GALEX$ and IRAC to SDSS, VHS and $AllWISE$ data). More bands improves results: this work has an accuracy of ($\vert \triangle z \vert < 0.3$) 93\% of the time (when NIR is present), which translates to $\eta$ = 7.03\% and $\sigma_{\rm nmad}$ = 0.018 for the entire spectroscopic sample of that work.} 

	
	A machine learning approach to calculate SDSS quasar luminosity was taken by \citep{mlpqna_photoz_2013}. However, the spectroscopic redshifts used to train the data were from SDSS DR7, and while trying to compare our results, we found that with the updated pipeline in SDSS DR12/DR13, many of these objects have changed spectroscopic redshift. As a consequence, our results were not comparable.

	\begin{deluxetable}{lccc}[th]
		\tablewidth{0pt}
		\tablecaption{\label{table:photoz_compare} \textsc{Comparison to Photometric Redshifts in the Literature for Sources Overlapping with Our Work\tablenotemark{1}.}}
		\tablehead{ \colhead{\textsc{Photoz catalog}} & \colhead{\textsc{Source Counts}} & \colhead{\textsc{Outliers}} & \colhead{\textsc{$\sigma_{\rm nmad}$}}}
		\startdata
		\hline
		\citet{peters2015} (Quasars) & 565 & 16.11\% & 0.0475  \\
		This work &  & 10.8\% & 0.0529 \\
		\hline
		\citet{richards2015} (Quasars) & 920 & 6.8\% & 0.0236 \\
		This work &  & 16\% & 0.0601 \\
		\hline
		\citet{xdqso_photoz_2012} (Quasars) & 945 & 8.15\% & 0.0312 \\
		This work &  & 13.76\% & 0.0559 \\
		\enddata
		\tablenotetext{1}{The values $\sigma_{\rm nmad}$ and fraction of outliers are from a comparison of photometric and spectroscopic redshifts for the sources in common to both works.}
	\end{deluxetable}
	
	\subsection{{Identification of Stars and Comparison with Color-Based Star Identification}}\label{section_star_id}
	
	\begin{figure}[th]
		\centering
		\includegraphics[width=1.0\linewidth]{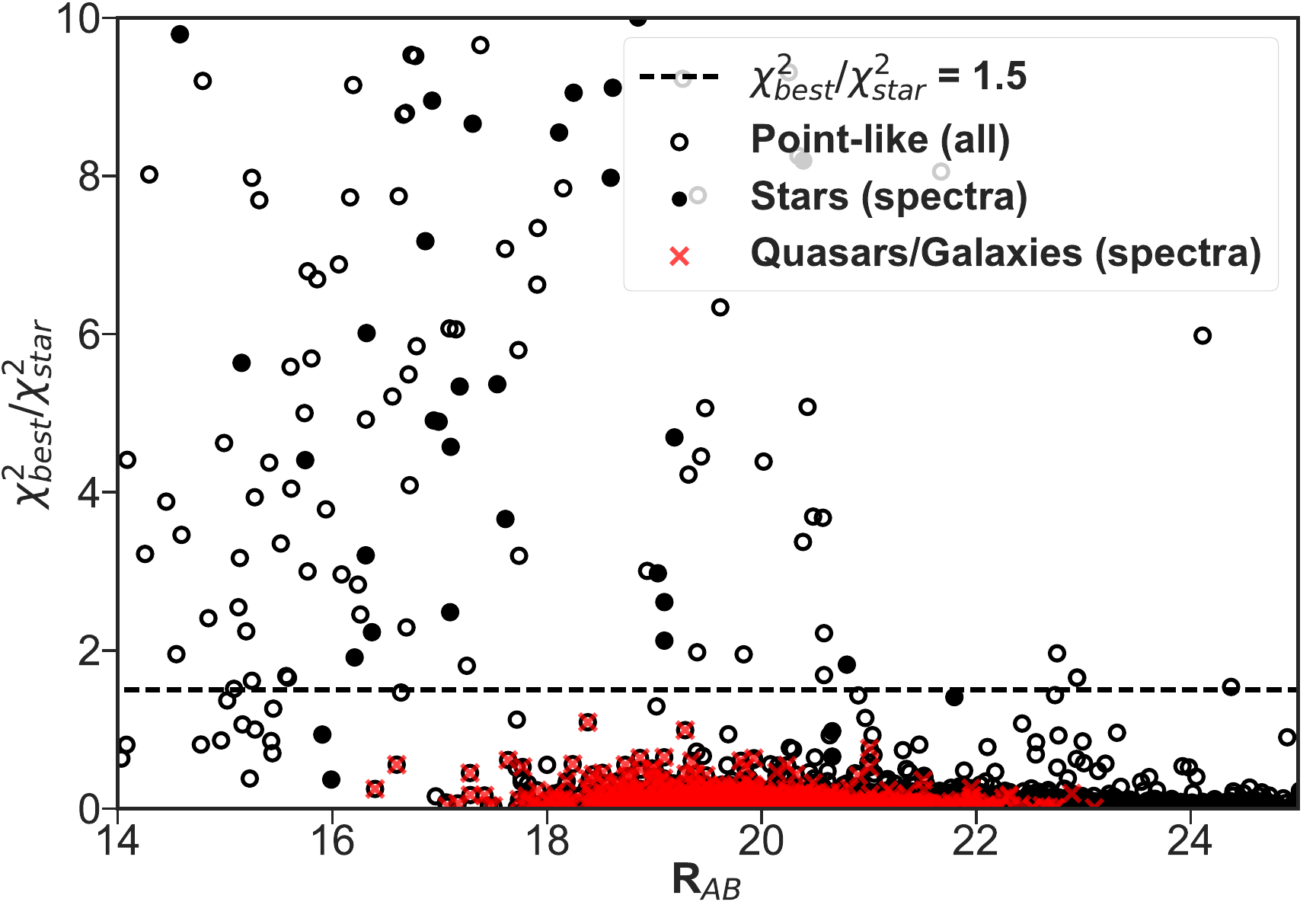}
		\caption{Ratio of $\chi^2$ values for best-fit quasar/Galaxy model ($\chi^2_{best}$) to best-fit star ($\chi^2_{star}$), against $R_{AB}$ magnitude. 
			Low ratios indicate secure quasar/galaxy identifications; {\it red crosses} show spectroscopically confirmed quasar/AGN. {\it Black filled points} show spectroscopically confirmed stars, only 5 of which (all cataclysmic variables) have ratios $<1.5$ and thus are misclassified (10\% of the 57 stars with spectra).			
			}
		\label{fig:chistar_chibest_r}
	\end{figure}
	
	Following the method explained in S09, we identified stars by SED fitting using the following criteria on point-like sources: $
	\chi^2_{best} \ge 1.5 \times \chi^2_{star}$. Here, $\chi^2_{best}$ corresponds to $\chi^2$ value returned by the best-fit AGN/galaxy template, and $\chi^2_{star}$ corresponds to the best-fit star template.
	The appropriate multiplicative factor depends on photometry and usually falls between 1 and 2. 
	Figure~\ref{fig:chistar_chibest_r} compares the quality of fit for galaxy/quasar and stellar templates. 
	The spectroscopically confirmed AGN are shown as red points, all of which fall below a ratio of 1.5.	
	
	Generally, if a star has a close neighbor, it could mistakenly be classified as extended, in which case a galaxy template will fit about as well as the stellar template. This is the biggest cause of misclassification of stars. 
	Truly point-like objects can easily be distinguished as stars or AGN/quasars, as shown in Fig.~\ref{fig:SED_fit} (and as confirmed by their spectra). 
	
	
	To investigate if we can improve the accuracy of classification, we looked into a color-based method of stellar identification presented in \citet{steph2016Rw1}: $R-W1 = (0.998 \pm 0.02) (R-K) + 0.18$, where all magnitudes are in the Vega magnitude system. We found that the SED fitting method is more accurate in correctly identifying stars, based on objects for which we have spectroscopic information. We summarize our findings in Table~\ref{table:star_compare}. 
		
	One caveat of finding stars using template fitting is that if the object is misclassified as extended due to close nearby neighbors, we use incorrect priors and cannot correctly classify such objects. \citet{scranton2002} discusses some challenges that arise in morphological classification due to close nearby neighbors. We found five cases where a close nearby object causes a star to appear extended. These stars were misclassified into the extended sample. Without better resolution data, we cannot correct for nearby neighbors distorting FWHM information. Nearby elliptical galaxies are very well fitted by stellar templates, so we cannot apply the $\chi^2_{best}/\chi^2_{star} > 1.5$ threshold on the extended sample without misclassifying a lot of galaxies as stars. This is why we search for stars only among the point-like sources. One of these five stars is recognized as a star by the color-color test, one is mis-categorized as ``extragalactic" and for the rest we do not have color diagnostic information. 
	
	\begin{deluxetable}{lcc}[h]
	\tablewidth{0pt}
	\tablecaption{\label{table:star_compare} \textsc{Comparison of Accuracy of SED Fitting Method and Color-color Method for Identifying Stars, for Objects with Spectra}}
	\tablehead{\colhead{\textsc{Classification type}} & \colhead{\textsc{SED}} & \colhead{\textsc{Color}}}
	\startdata
	Stars correctly classified & 52 & 25 \\
	Stars misclassified as extragalactic & 5 & 18 \\
	Extragalactic objects misclassified as stars & 0 & 0 \\
	\enddata
	\end{deluxetable}		
	
	\subsection{{Characterization of the Sources}}
	\begin{figure}[th]
		\centering
		\includegraphics[width=1.0\linewidth]{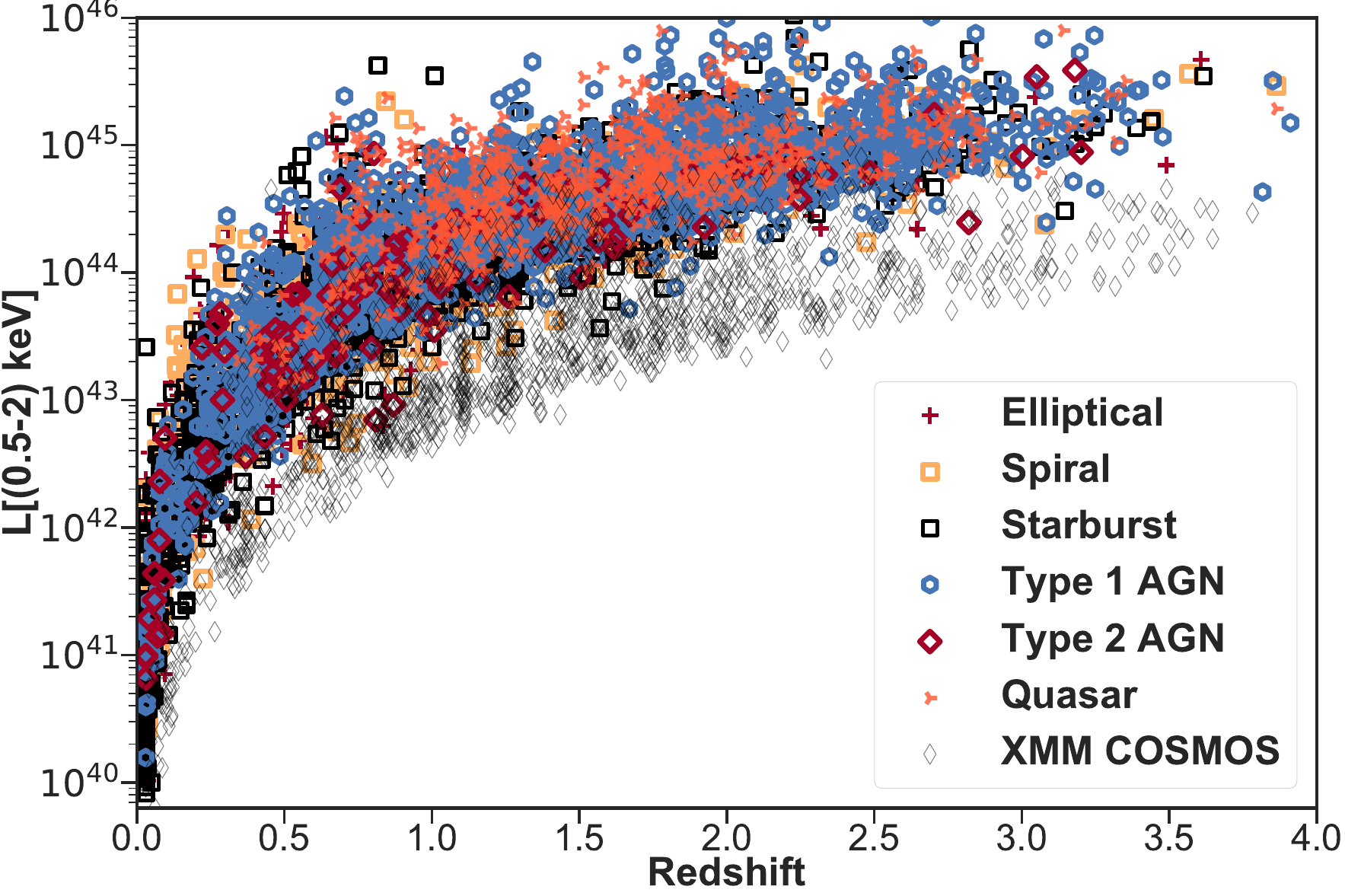}
		\caption{X-ray luminosity versus redshift for X-ray sources in Stripe~82X and $XMM$-COSMOS. 
		Because of its larger volume, the Stripe~82X AGN/quasars are more luminous than the $XMM$-COSMOS AGN. 
		The Type~1 AGN/quasars in Stripe~82X sample are the brightest class of objects in X-ray, as expected, and stars line up at the bottom of the distribution. $XMM$-COSMOS objects are less luminous in X-ray.}
		\label{fig:xray_characteristics}
	\end{figure}
	As we discussed in the \S~\ref{sec:intro}, Stripe~82X provides information on a population that was barely represented in COSMOS due to its smaller volume. 
Figure~\ref{fig:xray_characteristics} shows the luminosity-redshift distributions of these two surveys, demonstrating that the larger Stripe~82X volume samples $\sim15$ times higher luminosities.

\begin{deluxetable*}{lccc}[th]
\tablewidth{0pt}
\tablecaption{\label{table:object_types} \textsc{Comparison of Accuracy of SED Fitting Method and Color-color Method for Identifying Stars, for Objects with Spectra.}}
\tablehead{\colhead{\textsc{Best fit SED type}} & \colhead{\textsc{Template Number}} & \colhead{\textsc{Description}} & \colhead{\textsc{Number of Objects}}}
\startdata
	Stars & $\chi^2_{best} \ge 1.5 \times \chi^2_{star}$ & As explained in \S~\ref{section_star_id}  & 230 (3.7\%) \\
	Ellipticals & 33-35 (From Table~\ref{table:library}) & Elliptical galaxies & 797 (12.9\%; 4 at z$>$3) \\
	Spirals & 14, 36-40 & Spiral galaxies & 1000 (16.2\%; 6 at z$>$3) \\
	Starbursts & 1, 2, 13, 25, 26, 29-32 & M82, Mrk231, I22491-type 1 hybrids & 1474 (23.8\%; 24 at z$>$3) \\
	& & with  $>=$50\% I22491 contributions & \\
	Type 2 & 15-24 & Any type-2 hybrid & 123 (2\%; 9 at z$>$3) \\
	Type 1 hybrid & 5-12, 27, 28 & Type~1 hybrids with $>$50\% Type 1 contribution & 1655 (26.8\%; 56 at z$>$3) \\
	High luminosity quasar & 2, 3 & 100\% quasars & 682 (11\%; 17 at z$>$3) \\
	{\bf Total} & & & 5961 (96.4\%; 114 at z$>$3) \\
	\enddata 
\end{deluxetable*}

\begin{figure*}[th]
	\centering
	\includegraphics[width=0.5\linewidth]{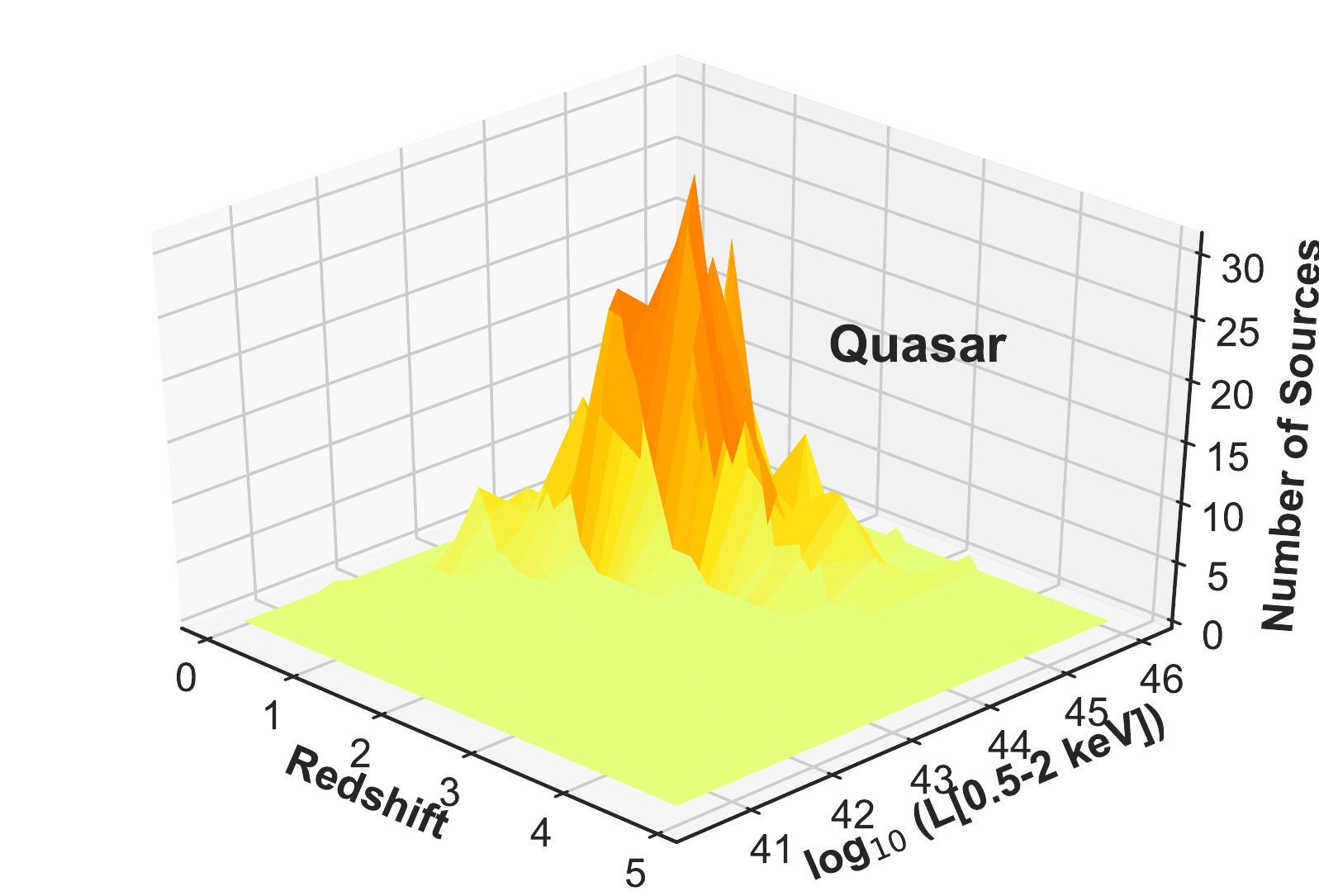}~
	\includegraphics[width=0.5\linewidth]{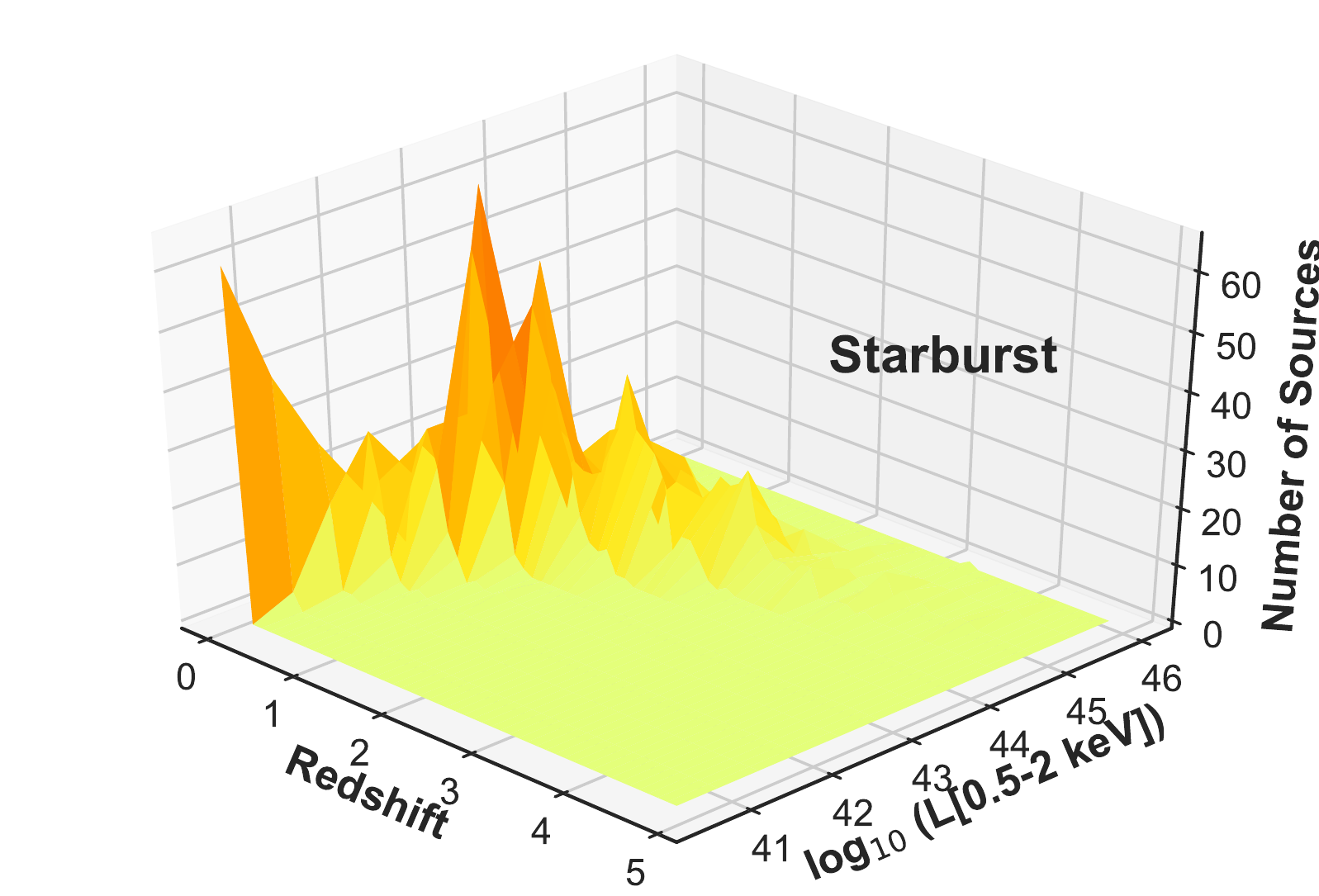}
	\includegraphics[width=0.5\linewidth]{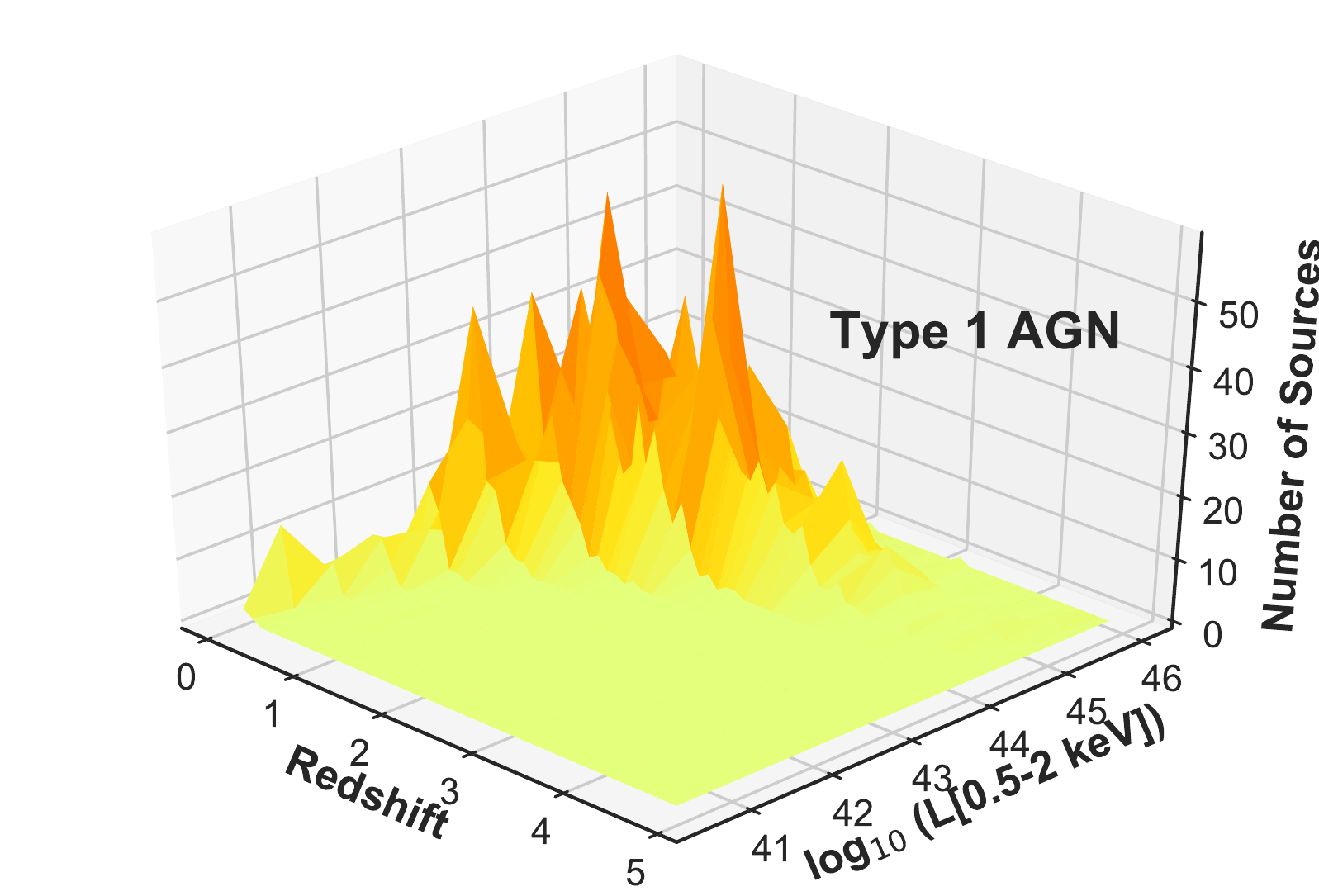}~
	\includegraphics[width=0.5\linewidth]{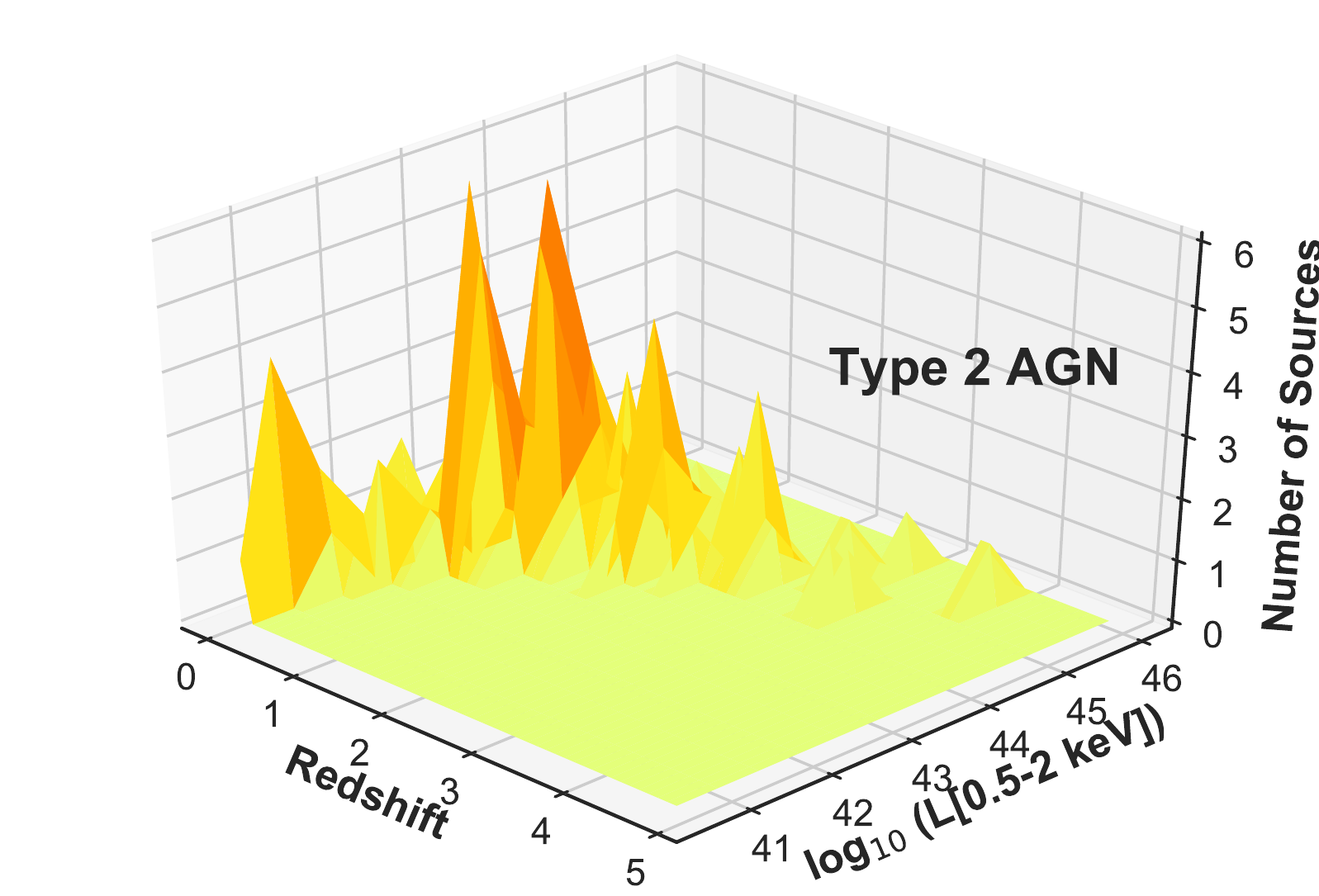}
	\includegraphics[width=0.5\linewidth]{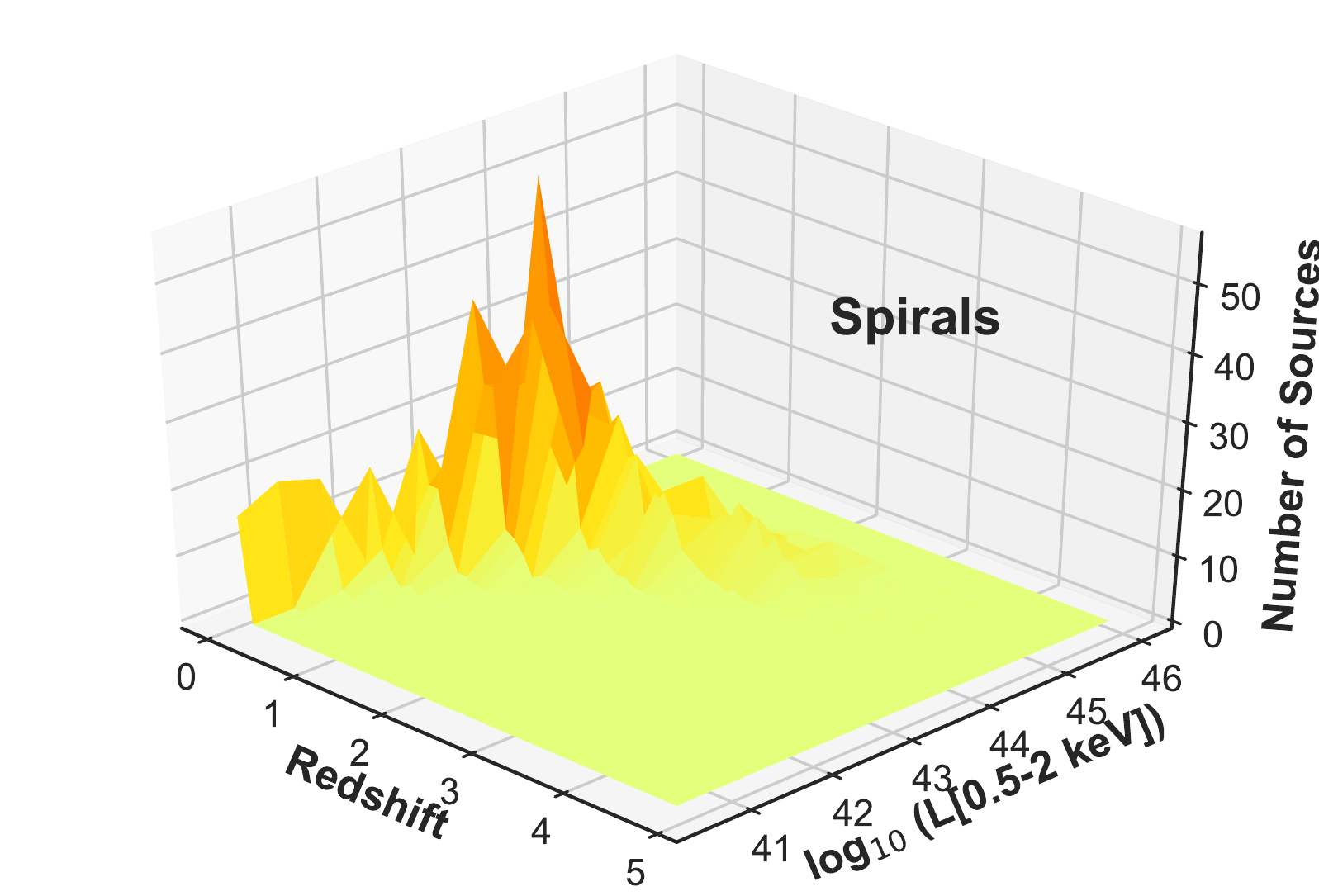}~
	\includegraphics[width=0.5\linewidth]{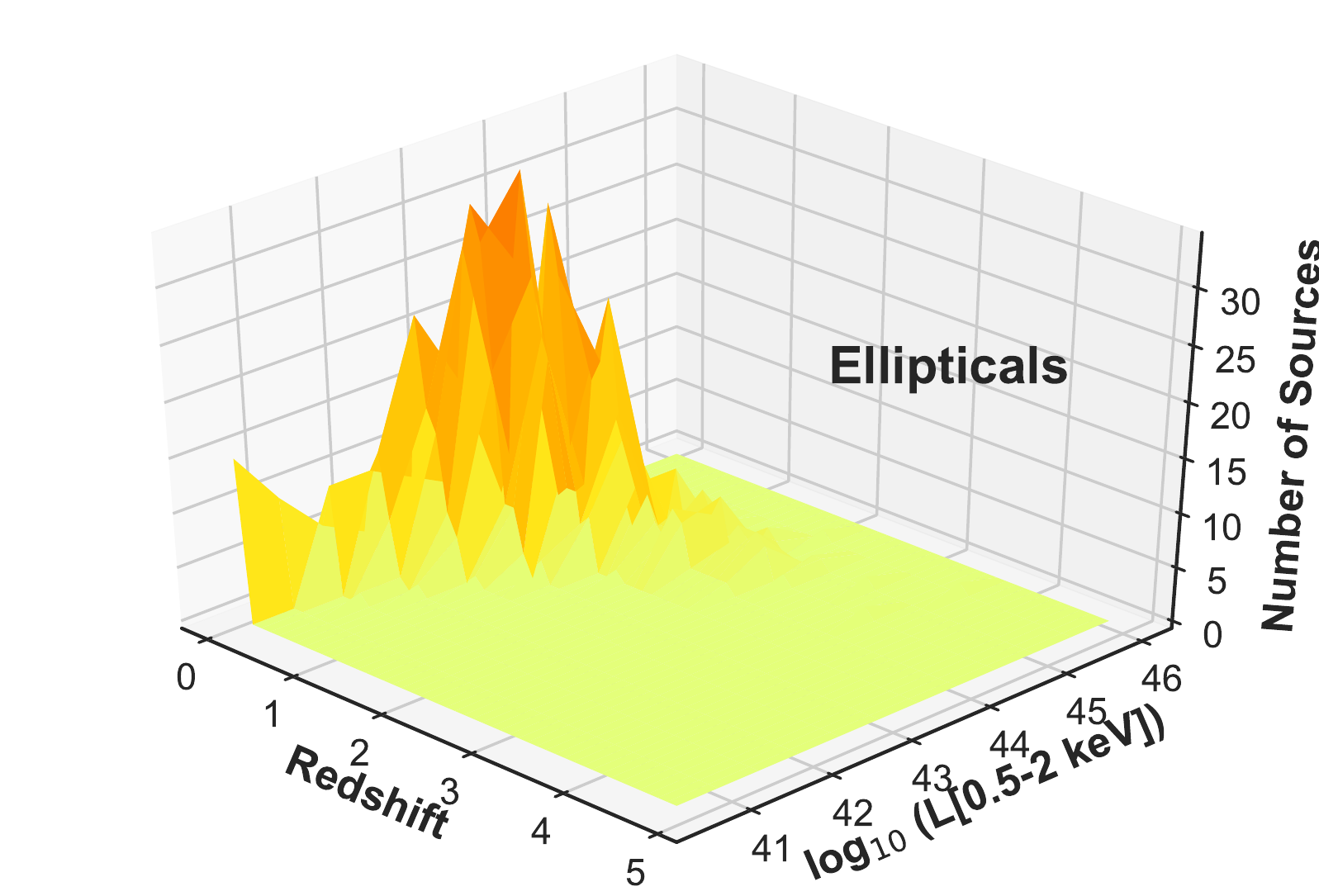}
	\caption{Luminosity and redshift distribution of \textit{top left:} quasars (QSOs), \textit{top right:} starbursts, \textit{middle left:} type 1 AGNs, \textit{middle right:} type 2 AGN, \textit{bottom left:} spiral galaxies and \textit{bottom right:} elliptical galaxies in Stripe~82X survey, using best-fit SED templates.}
	\label{fig:3D}
	\end{figure*}

	Template fitting gives us a rough idea of the demographics of Stripe~82 X-ray sources. Table~\ref{table:object_types} presents numbers of each type of object based on best fit SED template.
	Of course, since most of the ``galaxies'' have X-ray luminosities in excess of 10$^{42.5}$~erg/s, these are really obscured AGN. Together with the explicitly identified Type~2s, obscured AGN make up roughly 1/3 of the extragalactic objects. The observed ratio of Type~2 to Type~1 AGN is roughly 4:5; the intrinsic ratio can be much higher because obscured AGN are fainter than unobscured AGN (for the same underlying luminosity). 
	In Figure~\ref{fig:3D} we present the overall distribution in luminosity and redshift for each type of object.
	
	\begin{figure}[th]
	\centering
	\includegraphics[width=1.0\linewidth]{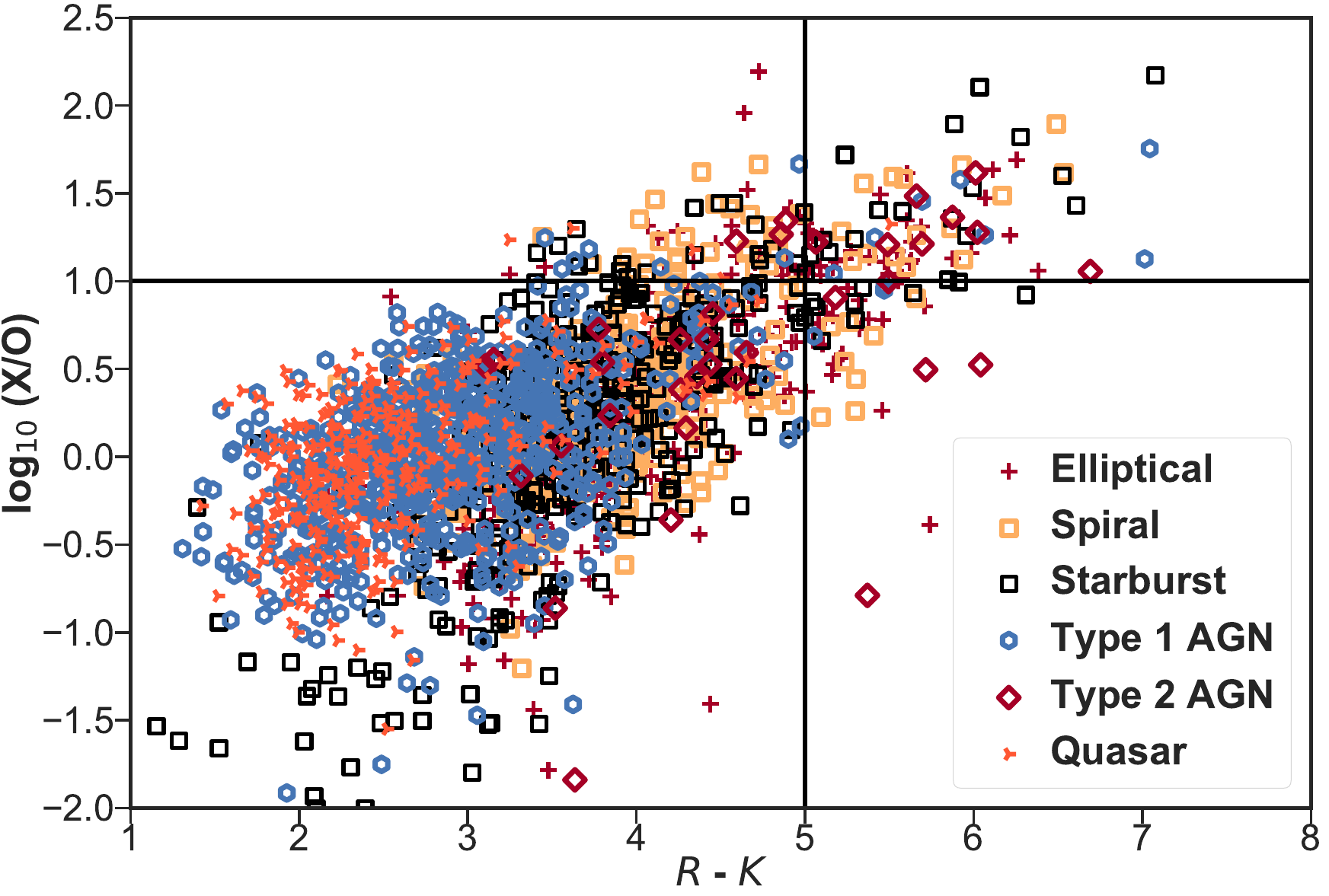}
	\caption{Log of ratio of hard band X-ray flux to $r$ band flux plotted against $R-K$ Vega magnitude. Objects with $\log (X/O) > 1$ and R-K $>$ 5 are candidates for heavily obscured AGN; most of the 82 objects that fit these criteria have SEDs best fit by starburst, spiral, elliptical and Type 2 templates (see legend for symbols), as expected for obscured AGN.}
	\label{fig:wise_characteristics_2}
\end{figure}

	The most interesting question is whether there are many obscured sources at high luminosity.
	Following \citet{Brusa2010}, we look for an obscured population using R-K Vega magnitudes and the 2-10~keV X-ray-to-$r$ band flux ratio (X/O), as shown in Figure~\ref{fig:wise_characteristics_2}. 
	Objects with $\log (X/O) > 1$ and R-K $>$ 5 are obscured AGN candidates. 
	In the Stripe~82X sample, 368 objects have $\log (X/O) > 1$. Of the 368 objects, 78 also have (R-K)$_{Vega} > 5$ ($\sim$ 2.5 deg$^{-2}$), and their SEDs are well fitted by starburst, spiral, elliptical and Type~2 templates, consistent with what we expect for heavily obscured AGN/quasars.
	
	For comparison, \citet{Brusa2010} found 105 objects in this region of color-color space in the $XMM$-COSMOS survey ($\sim$ 52.5 deg$^{-2}$).
	Nine of our 78 candidate obscured objects have redshift $z>2$, and 1 has redshift $z>3$; independent of color, 815 (129) objects have $z>2$ ($z>3$). 
	{Of the 129 sources (2\% of the entire sample)  that have redshift $z>3$,  only 32 ($<$1\%) are spectroscopically confirmed. Among the 97 without spectroscopic confirmation, only 25 have $PDZ>$ 90\% and can be considered reliable. The low PDZ of the remaining sources is due to the limited number of photometric points and/or their large photometric errors.}

\begin{figure*}[th]
	\centering
	\includegraphics[width=0.5\linewidth]{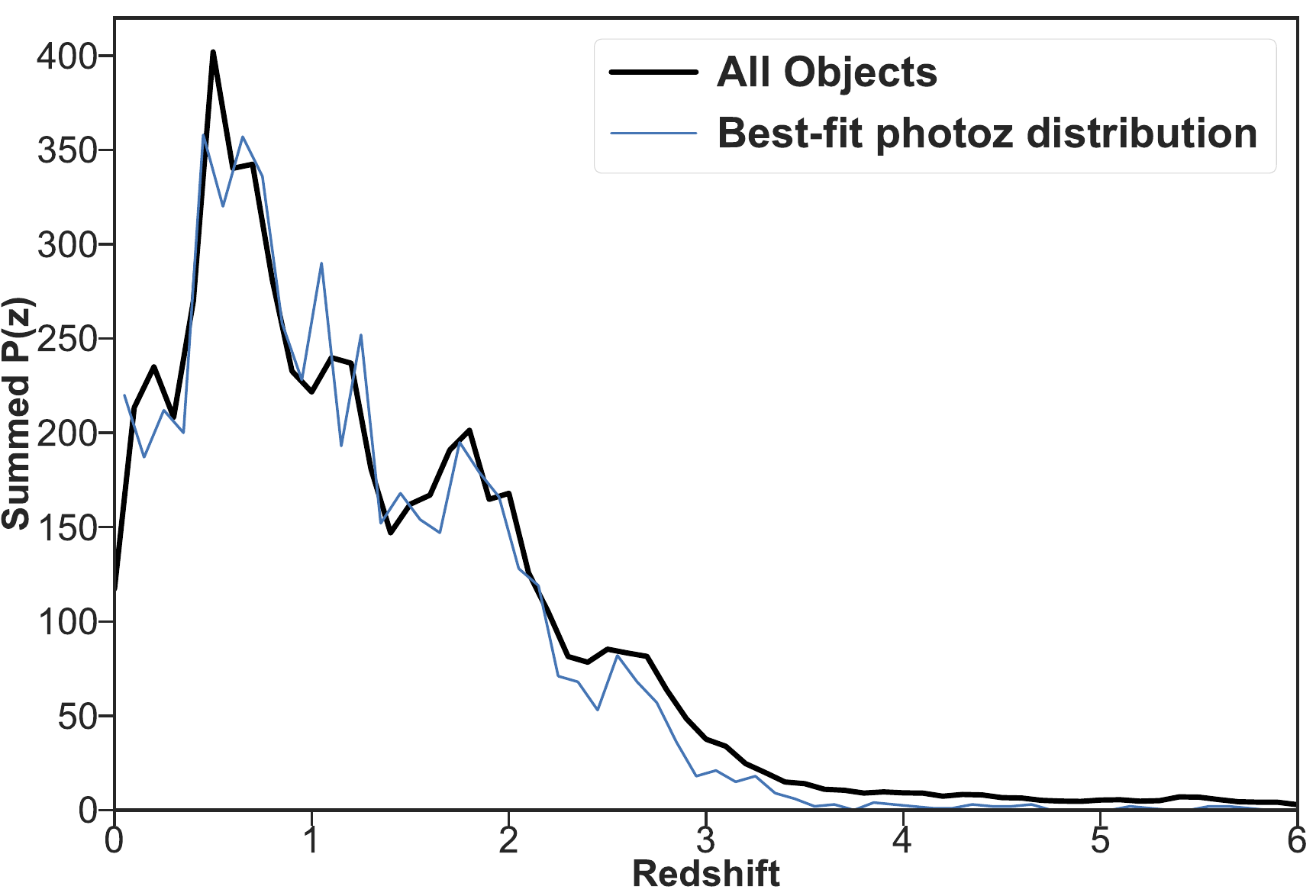}~
	\includegraphics[width=0.5\linewidth]{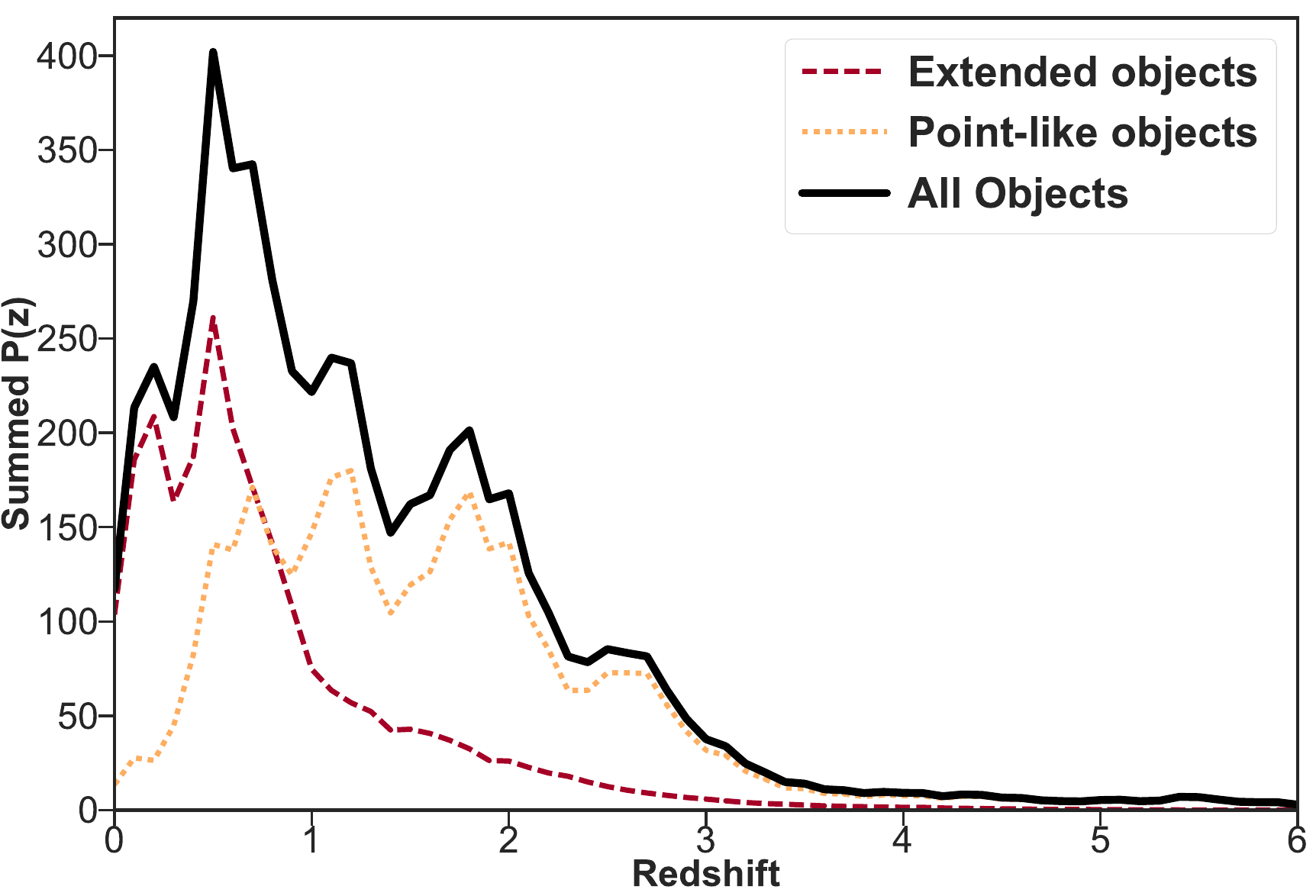}
	\caption{Summing the normalized probability distributions of photometric redshifts for each object. The probability distribution, P(z), for each object is described in Eq.~\ref{eq:pz}. The {\it left panel} shows how summed P(z) compares to distribution of best-fit photometric redshifts, and the {\it right panel} shows the summed P(z) of all objects, along with the contribution of extended and point-like fractions.}
	\label{fig:pdz_redshift}
\end{figure*}
	
	Using slightly different criteria,
	\citet{Perola2004,Civano2005} found that objects with $\log (X/O) > 1$ and F$_{\rm (2-10) keV} \ge 10^{-14}$ erg/(cm$^{2}$ s) are high-luminosity obscured quasar candidates. 
	375 Stripe~82X sources meet these criteria (11.981 deg$^{-2}$). 
	We have a dedicated spectroscopic follow-up program targeting candidate obscured quasars (LaMassa et al., submitted), and we are focusing particularly on obscured and/or high-redshift candidates.

\begin{figure}[th]
	\centering
	\includegraphics[width=1.0\linewidth]{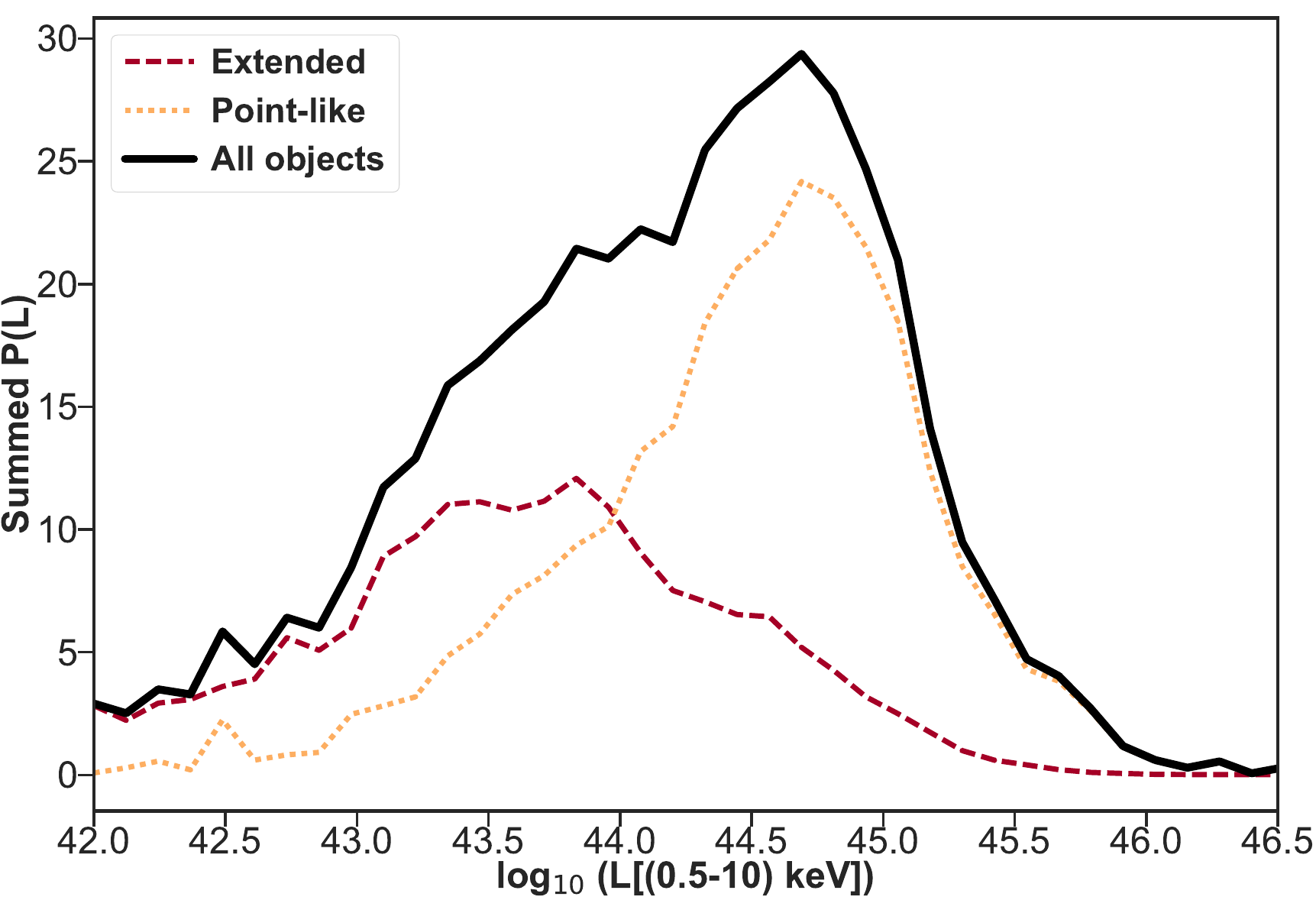}
	\caption{X-ray full-band luminosity (0.5-10 keV) distribution by taking into account normalized probability distribution of photometric redshift of each object.}
	\label{fig:pdz_luminosity}
	\end{figure}

	Finally, we report the total distribution of photometric redshifts and X-ray full-band luminosities (0.5-10 keV) by taking into account the probability distribution of redshifts for each object, and not just the best-fit value, in Figure~\ref{fig:pdz_redshift} and \ref{fig:pdz_luminosity}, respectively. Typical probability distributions of redshift of individual objects look similar to the bottom two panels of Figure~\ref{fig:SED_fit}. The total redshift distribution as shown in Figure~\ref{fig:pdz_redshift} was calculated by summing over probability distributions of all objects. 
	
	As expected, the Stripe~82X luminosity distribution is skewed to higher luminosities, as shown in Figure~\ref{fig:pdz_luminosity} (calculated using the full photometric redshift probability distributions). {These probability distribution of individual objects can also be used to derive luminosity functions, clustering and other redshift-dependent results, as shown by \citet{allevato2016}, which will be the subject of future work.}

	\section{{Conclusion}}\label{sec:conclusions}
	We fitted SED templates to 5961 X-ray objects (96.4\%) in the Stripe~82X survey \citep[LM16]{steph2013a, steph2013b}, and calculated photometric redshifts with an outlier fraction of 13.69\% and a $\sigma_{\rm nmad}$ of 0.06. This is the largest volume X-ray survey to date for which photometric redshifts have been calculated, and as such, has given us important insights, such as optimizing template selection for the higher flux population found in large area, shallow surveys. We have described our attempts to identify a set of templates that would be representative of all objects, but be small enough in size to minimize degeneracy. These template libraries may be a helpful starting point for other large volume surveys that would require photometric redshifts, such as XXL \citep{pierre2016, menzel2016}, 3$XMM$ \citep{xmm_ra_dec_err}, eROSITA \citep{merloni2012} and LSST \citep{ivezic2008}.
	
	{With this paper we release a catalog containing information about the X-ray sources, their correct associations to the multi-wavelength ancillary data, the multi-wavelength photometry used for template fitting, visually inspected spectroscopic redshifts, variability information, the photometric redshifts, including the $P(z)$ probability distributions. We also provide cutout images at various bands, the best-fit SED model (as in Figure~\ref{fig:SED_fit}) for each source and new libraries of SED optimized for wide and shallow X-ray surveys.}
	
	We find that in the case of point-like objects, 18\% of the outliers have a secondary redshift that are very close to the spectroscopic redshifts of those objects. This lends support to the argument that a more accurate approach in using photometric redshifts is to consider the probability distribution of redshifts for one object, rather than just one best fit value (e.g. \citealp{Georgakakis2014,buchner2015,Miyaji2015}).
	
	We were able to determine that most outliers tend to have important spectral features such as the 4000$\AA$ break and Ly-$\alpha$ break that fall in between filter response curves, or in the UV region, where we do not have data for 80$\%$ of the sources. We also found that we have comparable/slightly better results than other studies that only have broad-band data, possibly because our library contains more Type 1 templates. Accuracies can be improved significantly when at least some narrow or intermediate band filters are present.
	
	One interesting result of the study is that for point-like objects that are bright in the soft X-ray band ($>$ 1e$^{-13}$ ergs/cm$^2$/s), we get a large fraction of outliers ($\sim$ 33\%). The point-like morphology and high soft flux indicates that these objects are luminous unobscured quasars. Without the ability to see strong emission lines with narrow- or intermediate-band filters, it is difficult to get the correct redshift for such objects using template fitting. {As for relatively X-ray bright objects below 1e$^{-13}$ ergs/cm$^2$ s, the template libraries presented in this work should be appropriate for fields with similar depth.}
	
	This work can be improved if intermediate- or narrow-band photometry becomes available for the region, or with deeper optical photometry (e.g., HyperSuprimeCam imaging).
	Higher-resolution imaging would be useful for better distinguishing between point-like and extended samples.
	

	\acknowledgements
	{We are very grateful to the anonymous referee for a careful reading of this work and providing comments that helped us improve a number of areas that were not sufficiently clear in our first submission.}
	
	TA wishes to thank her parents, M. A. Quayum and Shamim Ara Begum, and her husband, Mehrab Bakhtiar, for their constant support under all conditions. She is very grateful to Ms. Birgit Boller of Max Planck, Garching, for supporting her visit there. The authors thank Ms. Geriana Van Atta of Yale University. 
This project was supported by Yale University and the $XMM$-Newton analysis was partially supported by NASA grant NNX15AJ40G. 
	
	\software{Astropy \citep{astropy}, Matplotlib \citep{matplotlib}, Topcat \citep{topcat}, STILTS \citep{stilts}}

	\appendix

	\section{Output catalog columns}
	The final catalog can be downloaded from
	\href{https://yale.app.box.com/s/6o4dtks5tmxuq7j083s0wujf2ke8slvx}{here}.
	Description of the output table columns are in this section. Note that this catalog supersedes LM16, because we explicitly resolved conflicting identifications (previously left to the reader), and used deeper optical and mid-infrared (MIR) catalogs.
	\begin{longtable}[th]{p{0.3\linewidth} p{0.6\linewidth}}
		\caption{Column Descriptions of Final Output Table.} \label{table:final_output_table}\\
		\hline
		\textbf{Column} & \textbf{Description} \\
		\hline
		\endfirsthead
		\multicolumn{2}{l}{{\bfseries \tablename \thetable{} -- continued from previous page}}\\
		\noalign{\smallskip}\hline\noalign{\smallskip}
		Column & Description \\
		\noalign{\smallskip}\hline\noalign{\smallskip}
		\endhead
		\noalign{\smallskip}\noalign{\smallskip} \multicolumn{2}{r}{{Continued on next page}} \\ \noalign{\smallskip}\hline\noalign{\smallskip}
		\endfoot

		\hline 
		\endlastfoot
		
		Rec\_no & ID number for each source as assigned in LM16. \\
		
		\hline \textbf{X-ray data:} & \\
		
		\hline Catalog & Name of the original X-ray survey from LM16 (AO10, AO13, $Chandra$ or Archival-$XMM$) \\
		Xray RA, DEC, RA\_DEC\_ERR & RA, DEC and error in position of the X-ray source \\
		
		Soft\_flux & X-ray flux in 0.5-2~keV band. \\ 
		
		Soft\_detml & {Significance of detection in the 0.5-2 keV band: detml = -ln $P_{\rm random}$}  \\
		
		Hard\_flux & X-ray flux in 2-10~keV band. \\ 
		
		Full\_flux & Total X-ray flux in 0.5-10~keV band. \\ 

		\hline \textbf{Association data:} & \\

\hline Association & This field indicates how many catalogs agree that the association we selected to the X-ray object is the correct association. Sp, Sh, V F and J stand for Spies, Shela, VHS, SDSS co-added catalogs FT16 and J14 respectively. ``SpVJr" indicates that IRAC, VHS and J14 r band agree and there is no conflict on the association of this X-ray object. ``V" might indicate we only have VHS data for this object, or that we found VHS counterpart to be the most likely counterpart. ``LM16" indicates that these sources were added from the LM16 catalog - we did not find them through our MLE matches but these sources exist in SDSS DR13, in at least one singe epoch frame. Details of how we arrived at this results is discussed in \S~\ref{ssec:correct_assoc}.\\ 

RADEC\_from & When we select RA, DEC for the best X-ray association, we look at all the catalogs that agree, and choose coordinates from the one with the best astrometric accuracy (SDSS, than VHS, than IRAC). \\ 

CTP\_RA, CTP\_DEC & Right ascension and declination of the best associations to the X-ray object.\\ 

QF & Quality flag - this field indicates whether or not there was a conflict in association. Described in detail in Figure~\ref{fig:assoc_flowcharts}. \\ 

Xray\_ctp\_dist & Distance (in arseconds) between X-ray coordinates and counterpart. \\




SDSS\_rel\_class, VHS\_rel\_class, IRAC\_rel\_class & Reliability class (Secure/Ambiguous/Sub-threshold) for SDSS, VHS and IRAC counterparts respectively.\\

Manual\_Check & For each visually checked source (all sources with conflicting associations), we added a manual comment describing the issue that might lead to difficulty in counterpart identification or photometry contamination. Some examples:  `2150: two very close sources', `2152: extended source, so spies coordinates is a little off', `2155: looks like one very bright source in optical, and one bright and one nearby faint source in spies'\\

		\hline \textbf{Photometric data:} & \\

		\hline mag\_fuv, magerr\_fuv, mag\_nuv, magerr\_nuv & Extinction corrected {\it GALEX} Far UV (FUV) and near UV (NUV) SEXTRACTOR auto magnitude and magnitude errors, all magnitudes are in the AB system \\ 
		
		u, uerr, g, gerr, r, rerr, i, ierr, z, zerr & SDSS co-added (SEXTRACTOR AUTO) magnitudes and corresponding errors. If the field ``Coadded" is set to ``FT16" the data comes from FT16 and if it is set to ``J14", the photometry comes from J14. We allow all photometry data from the former catalog, and when no photometry is available, we look at the latter and take all data with SEXTRACTOR flags set to 0 or 2. We tried to correct for galactic extinction whenever extinction information was available. If not extinction was not available sdss\_ext\_corr is set to ``N/A".\\ 
		
		j, jerr, h, herr, k, kerr & Extinction corrected magnitudes and magnitude errors from VISTA VHS survey in J, H and K bands (2.8$^{\prime\prime}$ and 5.6$^{\prime\prime}$ radius aperture magnitude for point-like and extended sources, respectively) \\ 
		
		juk, juk\_err, huk, huk\_err, kuk, kuk\_err & Extinction corrected magnitudes from UKIDSS Large Area Survey catalog in J, H and K bands (2.8$^{\prime\prime}$ and 5.6$^{\prime\prime}$ aperture magnitude for point-like and extended sources respectively) \\ 
		
		ch1\_spies, ch1err\_spies, ch2\_spies, ch2err\_spies & IRAC CH1 and CH2 data (SEXTRACTOR AUTO mag;  not extinction corrected) from \citet{spies}. \\ 
		
		ch1\_shela, ch1err\_shelas, ch2\_shela, ch2err\_shela & IRAC CH1 and CH2 (SEXTRACTOR AUTO mag) data (not extinction corrected) from \citet{sheladata}. \\ 
		
		w1, w1err, w2, w2err, w3, w3err, w4, w4err &  {\it AllWISE} (mpro) magnitudes and errors (not extinction corrected). W3 and W4 data appear later in the table (after spectroscopic redshift data) because we do not use it to construct SEDs. W3 and W4 magnitude errors are blank if the profile-fit magnitude is a 95\% confidence upper limit or if the source is not measurable (\url{http://wise2.ipac.caltech.edu/docs/release/allwise/expsup/sec2_1a.html}). \\ 
		
		Context & LePhare needs this field to understand which filters to use to calculate photometry. For example, if we just need to use filters FUV (filter number 0), SDSS i (5) and CH1 SPIES (13), the context is set to   $2^0+2^5+2^{13}$. In the catalogs, we use this column to indicate which bands we used for final template fitting by LePhare.\\ 
		
		\hline \textbf{Spectroscopic data:} & \\
		
		\hline Redshift & Spectroscopic redshift \\ 
		
		Redshift\_Source & Source of spectroscopic data (e.g. SDSS DR13, DR12Q) \\
				
		SPEC\_class & Spectroscopic classification of this object \\ 
		
		SPECOBJID & SDSS spectra object ID DR12Q, DR13, DR14 \\ 
		
		SPEC\_subclass & Additional spectroscopic data on object type \\ 
		
		Redshift\_err & Spectroscopic redshift error \\ 
		
		zwarning & Warning on redshift. Only ZWARNING = 0 was used in our training sample. \\ 
		
		\hline \textbf{Photometry/Variability flags:} & \\
		
		
		\hline Fliri\_flag & SEXTRACTOR extraction flag from FT16 catalog.\\ 
				
		EXT\_CORR\_FROM & The co-added SDSS catalogs do not come with extinction information. At the same time, if we can find an extinction match in SDSS DR7, this field is set to ``SDSS". If we cannot, we use EBV from VHS catalog to calculate extinciton, in which case this flag is set to ``VHS".\\ 
		
		artifact\_fuv, artifact\_nuv & {\it GALEX} artifact flags as described here: \url{http://www.galex.caltech.edu/DATA/gr1_docs/GR1_Pipeline_and_advanced_data_description_v2.htm} \\ 
		
		fwhm\_fuv, fwhm\_nuv & FWHM for {\it GALEX} data, measured in pixels \\ 
		
		Flags\_nuv, Flags\_fuv & SEXTRACTOR flags for NUV and FUV. \\ 
		
		{\it GALEX}\_extended & {\it GALEX} object classification \\ 
		
		{\it GALEX}\_manflag & {\it GALEX} manual flag \\ 
		
		u\_fwhm, g\_fwhm, r\_fwhm, i\_fwhm, z\_fwhm & The FWHM for each SDSS band from J14. \\ 
		
		nnv & Variability information for optical sources, varies from 0 to 1 (-1 for completely non-varying sources). \\ 
		
		VHS\_mergedclass &  Class flag from available measurements (1|0|-1|-2|-3|-9=galaxy|noise|stellar|probableStar|probableGalaxy|saturated)\\ 
		
		VHS\_pstar, VHS\_pgal & VHS pipeline probability that an object is a star or a galaxy, respectively\\ 
		
		ch1\_flag\_spies, ch2\_flag\_spies, ch1\_flag\_shela, ch2\_flag\_shela & SEXTRACTOR flags for IRAC data \\ 
		
		spies\_class\_star & SPIES object classification \\ 
		
		high\_rel1\_spies, high\_rel2\_spies & Reliability of CH1 and CH2 SPIES \citep{spies} data. \\ 
		
		XO & X-ray to optical ratio, as explained in Appendix C. This ratio is for X-ray hard band (0.5-2 keV) to SDSS r band. \\ 
		
		Coadded & Indicates whether the photometry we use is from FT16 or J14. We prefer FT16 and use it when available. \\ 
		
		Jiang\_flag & J14 r-band SEXTRACTOR flag \\ 
		
		Nearby\_Neighbor\_SEXTRACTOR & This is the remainder of Jiang\_flag - it is 1 that means there is a nearby neighbor bright enough to significantly impact photometry. We exclude these objects from the training set. \\
		
		Classification & Classification according to Figure~\ref{fig:final_classification}. \\
		
		
		
		\hline \textbf{Photometric Redshift Data:} & \\
		
		\hline Photoz & Best fit photometric redshift\\
		Photoz\_best68\_low, Photoz\_best68\_high  & 1-$\sigma$ lower and upper bound on photometric redshift, respectively. \\
		Chi\_best & $\chi^2$ value for our best fit SED  \\
		Mod\_best & Best fit SED \\
		Extlaw\_Best & Extinction law used for best fir SED (1 for \citet{prevot1984}) and 0 for none. \\
		EBV\_best & Color excess for host galaxy of AGN. \\
		PDZ\_best & Described in Section 5. Can also be taken as LePhare's confidence on $z_{\rm phot}$. \\
		Photoz\_sec & Secondary redshift for SED with $\chi^2$ value very close to best fit $\chi^2$ value.  \\
		Chi\_sec & $\chi^2$ value for second best fit. \\
		Mod\_sec & SED for second best fit. \\
		EBV\_sec & Color excess used in second best fit. \\
		PDZ\_sec & Check section 3. \\
		Chi\_star & $\chi^2$ value for best star template fit. \\
		Nband\_used & Number of bands used by Lephare to determine $z_{\rm phot}$  \\
		Morphology & Morphology according to template fitting: 1: Stars, 2: Ellipticals, 3: Spirals, 4: Type 2, 5: Starbursts, 6: Type 1, 7: QSOs. \\
		Color\_morphology & \citet{steph2016Rw1} color morphology: extragalactic or star. \\
		SED\_name & Name of the best fit template. \\
		AstrometricJHK\_2015\_photoz & \citet{peters2015} photoz. \\
		XDQSO\_2012\_photoz & \citet{xdqso_photoz_2012} photoz. \\
		Richards\_2015\_photoz & \citet{richards2015} photoz. \\
		Luminosity\_distance\_phot, Luminosity\_distance\_spec & Luminosity distance calculated using photoz and specz respectively. \\
		Luminosity\_phot, Luminosity\_spec & Full flux X-ray luminosity calculated using photoz and specz respectively. \\
		\hline \textbf{X-ray source duplication data:} & \\
		\hline Duplication\_flag & 0: the object is unique, 1, -1: one of a duplicate pair of X-ray sources. We identify the same counterpart for both X-ray observations. 2, -2: duplicate pairs - these sources find different counterparts for X-ray sources. The positive flags (1, 2) indicate more reliable X-ray positions - we use only values of 0 or above to calculate outlier fractions/accuracy.\\
		
		Duplication ID 1 & {REC NO of duplicate X-ray source}.\\
		Duplication ID 2 & {Duplicate X-ray source REC NO if there is a second duplicate source.}\\
		
	\end{longtable}

\section{Selecting template SEDs}

\begin{figure*}[th]
	\centering
	\includegraphics[width=0.5\linewidth]{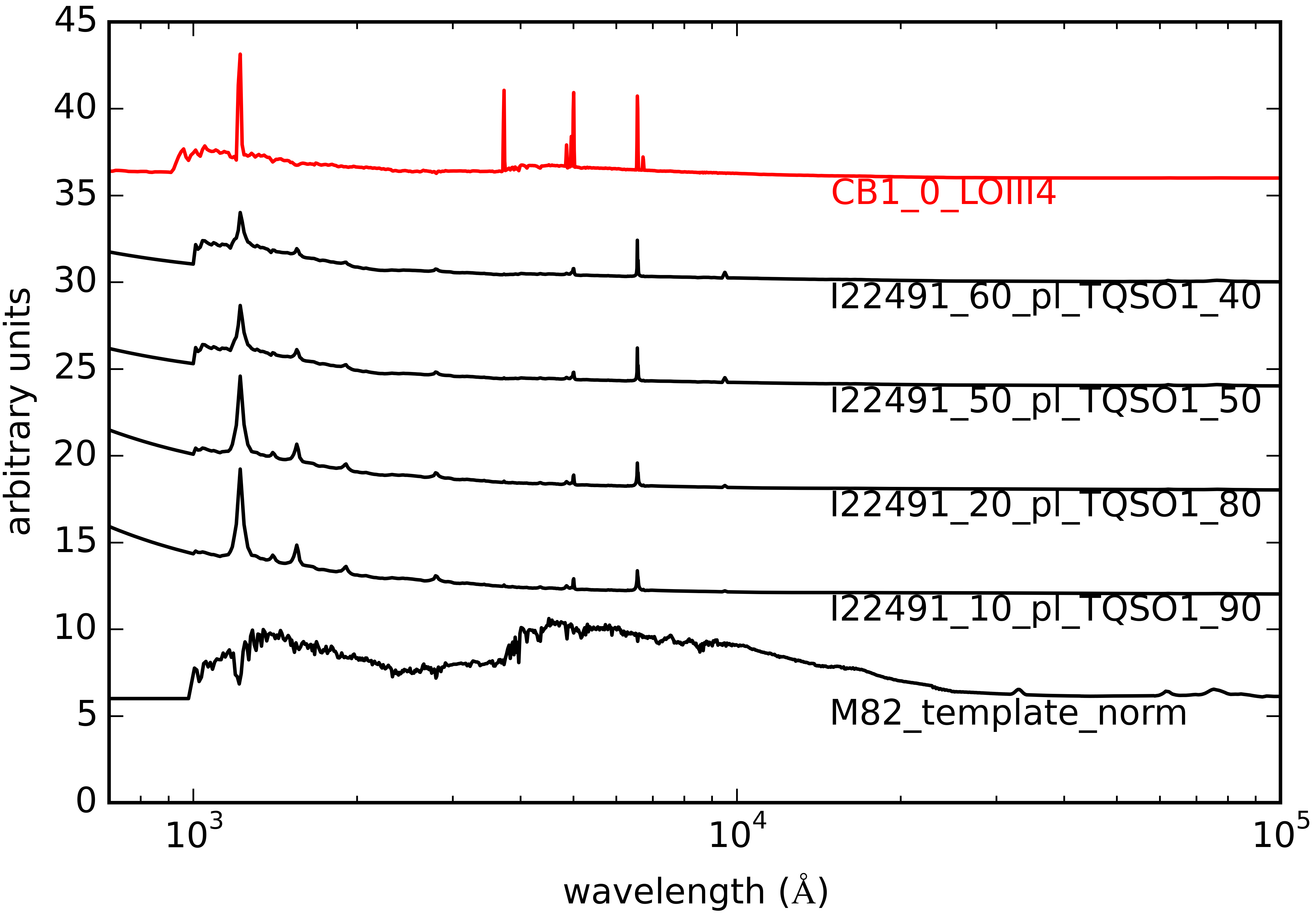}~
	\includegraphics[width=0.5\linewidth]{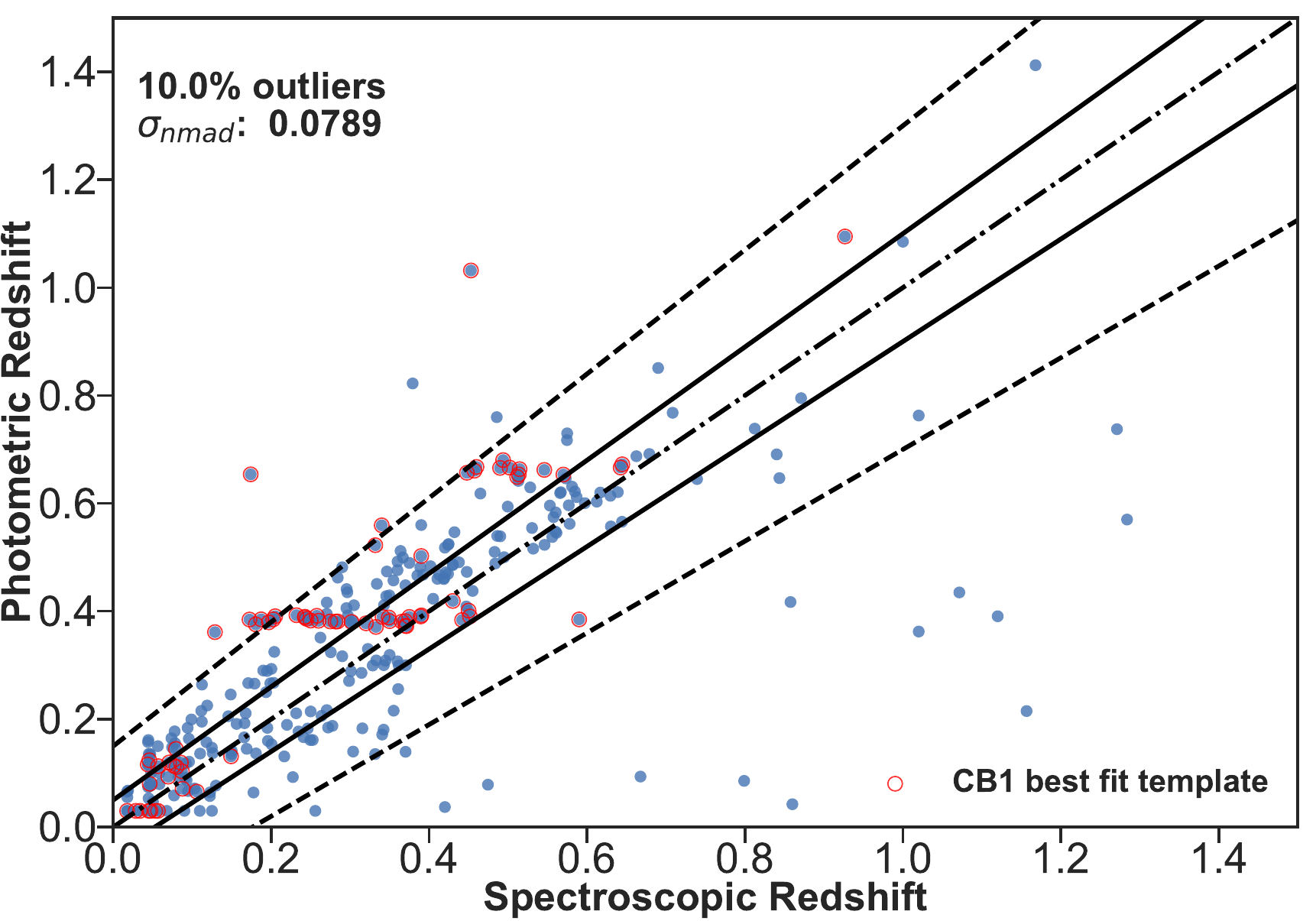}
	\caption{{\it Left:} All the templates in black are from the extended library. The red template is an extra SED that belongs to a young blue starforming galaxy from \citet{Ilbert2009}, and has sharp features similar to hybrids of I22491, also a starburst galaxy. The four templates below the red template are I22491 hybrids. {\it Right:} This figure shows an instance of degeneracy in fitting. The template library and objects used to fit are identical to the one used to fit the objects in Figure~\ref{fig:zspec_zphot} (right), with one exception: this library additionally contains the template indicated in red in top image. This template has many of the same features as the I22491 starburst/type 1 QSO hybrids, and so some objects which are more accurately fitted by I22491 are now degenerately fitted by the red template, giving rise to the horizontal line at $z_{phot} \sim$ 0.39 in this plot. Note that even though the addition of this template improves the the fits for some very low redshift objects ($z_{spec}<0.2$), which appear in a horizontal line in Figure~\ref{fig:zspec_zphot} (right) at $z_{phot}\sim$0.03. So, we identify that this horizontal lines are caused by template degeneracy. We cannot improve on this without increasing inaccuracy of overall work, so the optimum decision is to leave out the red template. Note that the objects which are in the $z_{phot}\sim$0.03 line have acceptable $\sigma_{\rm nmad}$, and the photometric redshifts calculated by this work is not largely different from the actual redshift of these objects.}
	\label{fig:extended_seds}
\end{figure*}		

One example of minimizing a library is discussed here. 
Ideally, when constructing a library we need to make sure that all types of sources are represented. However, an excessive number of templates applied to sources with a SED defined by a limited number of photometric points, introduces degeneracies in the redshift solution. Thus, the compromise is to make sure that key SED features  are represented in the library, keeping the number of templates as small as possible. In this way, even if the template is not exactly the best template fitting the data, it is sufficiently close to it for allowing a reliable redshift determination.

 For the extended objects sample, the template colored in red in Figure~\ref{fig:extended_seds} (left) closely fits most of the objects that are well fitted by the I22491 hybrids (plotted below the red template), although it produces inaccurate redshift as shown in Figure~\ref{fig:extended_seds} (right). The horizontal lines at  $z_{phot}\sim0.39$ is caused by this template. The photometric redshift code is very dependent on sharp features in the spectra, such as the Ly-$\alpha$ break. Therefore, the degeneracy is not surprising, because the I22491 hybrids and this red template has many of the same features. However, some very low redshift objects that appear in a horizontal line at $z_{phot}\simeq 0.03$ in Figure~\ref{fig:zspec_zphot} (right) are better fitted by the red template. This template was not included in the final library, because the photometric redshifts of those low redshift objects are within acceptable $\sigma_{\rm nmad}$ without it, and because this template was causing bigger inaccuracies at higher redshifts.

It is not possible to remove templates when outliers are caused by essential models which accurately represent a significant fraction of the objects. For point-like objects, this work arrived at a smaller library than \citet{MS09}, \citet{hsu2014}. But even with a small number of templates the photometric data is not sufficient to avoid degeneracy entirely. This is shown in Figure~\ref{fig:zspec_zphot}, where some of the outliers (red squares) have secondary redshifts that are closer to the true spectroscopic redshifts. 

	Each template in the point-like library that causes outliers/degeneracy also produces the correct photometric redshifts for many other objects, so none can be removed. Note that these outliers with correct secondary redshifts demonstrate that it is more accurate to consider the entire probability distribution of redshifts for one object. Overall, the choice of templates was driven by the quality of SED fits, the frequency of use of a particular template, and the absence of systematics in the redshift results when each template is used (see S09 and \citealp{hsu2014} for details).

	\section{X/O conversions}
To calculate X/O (introduced by \citet{maccacaro1988}) for our sample, we use soft flux (ergs/cm$^2$ s) and SDSS i-band AB magnitude. We convert i-band flux to the correct units as follows:
\begin{equation}
	f_{\nu} = 3631 \times 10^{-\frac{m_{AB}}{2.5}}\textit{ Jy}
\end{equation}
To convert to $f_{\lambda}$:
\begin{equation}
	f_{\lambda} = \frac{f_{\nu}}{3.34 \times 10^4} (\frac{\textit{\AA}}{\lambda})^2 \frac{erg}{cm^2\textit{ \AA} \textit{ s}}
\end{equation}
To get total i-band flux, multiplying by the bandwidth of i-band (1300 \AA):
\begin{equation}
	F_{\lambda} = \delta \lambda\textit{ \AA}\textit{ }f_{\lambda}
\end{equation}
Taking log of flux:
\begin{equation}
	\log_{10} F_{\lambda} = - \frac{m_{AB}}{2.5} - 5.61425
\end{equation}
So X/O:
\begin{equation}
	\log_{10} (\frac{F_X}{F_{\lambda}}) = \log_{10} (F_{\textit{0.5-2~keV}}) + \frac{m_{AB}}{2.5} + 5.61425
\end{equation}

\clearpage

\end{document}